\documentclass{lmcs}
\pdfoutput=1

\usepackage{lastpage}
\renewcommand{\dOi}{14(3:8)2018}
\lmcsheading{}{1--\pageref{LastPage}}{}{}%
{Jul.~22,~2016}{Aug.~20,~2018}{}

\keywords{higher types, recursion, fixed point operator, definability, PCF}

\usepackage{hyperref}
\usepackage{amssymb}
\usepackage{graphicx}

\def\N{\mathbb{N}}

\def\lsem{[\![}
\def\rsem{]\!]}
\newcommand\sem[1]{\lsem #1 \rsem}

\def\iso{\cong}

\def\KKar{{\mathbf{K}}}

\def\arrow{\rightarrow}
\def\parrow{\rightharpoonup}

\def\darrow{\Rightarrow}

\def\reducesto{\rightsquigarrow}

\def\Set{{\mathcal{S}\!\!et}}

\def\id{{\mathit{id}}}

\def\lev{{\mathrm{lv}}}
\newcommand\pure[1]{\overline{#1}}
\newcommand\num[1]{\widehat{#1}} 

\def\dom{{\mathrm{dom}}}

\def\nat{{\mathtt{N}}}  
\def\bool{{\mathtt{B}}} 
\def\unit{{\mathtt{U}}}

\def\ttrue{{\mathit{tt}}}
\def\ffalse{{\mathit{ff}}}

\def\iif{{\mathit{if}}}

\def\ccase{{\mathtt{case}}}
\def\oof{{\mathtt{of}}}

\newcommand\caseof[2]{\ccase\;#1\;\oof\;(#2)}

\newcommand\dang[1]{\ll #1 \gg}

\newcommand\myvec[1]{\vec{#1}}

\def\suc{{\mathit{suc}}}
\def\pre{{\mathit{pre}}}

\def\rec{{\mathit{rec}}}

\def\ifzero{{\mathit{ifzero}}}

\def\mmin{{\mathit{min}}}
\def\byval{{\mathit{byval}}}

\def\por{{\mathit{por}}}
\def\pif{{\mathit{pif}}}

\def\eexists{{\mathit{exists}}}

\def\catch{{\mathit{catch}}}

\def\Set{{\mathsf{S}}}

\def\Ct{{\mathsf{Ct}}}

\def\HEO{{\mathsf{HEO}}}

\def\PC{{\mathsf{PC}}}

\def\SP{{\mathsf{SP}}}

\def\SF{{\mathsf{SF}}}

\def\Klex{{\mathsf{Klex}}}

\def\eff{{\mbox{\scriptsize \rm eff}}}

\def\fin{{\mbox{\scriptsize \rm fin}}}

\def\lwf{{\mbox{\scriptsize \rm lwf}}}

\def\smallmin{{\mbox{\scriptsize \rm min}}}

\def\PCF{{\mathrm{PCF}}}
\def\sysT{{\mathrm{T}}}

\def\LLL{{\mathcal{L}}}

\newcommand\details[1]{}
\newcommand\pending[1]{}

\def\up{{\mathit{up}}}
\def\down{{\mathit{down}}}
\def\iter{{\mathit{iter}}}

\def\Klexmin{{\Klex^\smallmin}}

\def\mydot{\prime}

\begin{document}

\title[Recursion hierarchy for PCF]{The recursion hierarchy for PCF is strict}
\titlecomment{Revised, corrected and expanded version of Informatics Research Report EDI-INF-RR-1421, University of Edinburgh, 2015}
\author{John Longley}
\address{School of Informatics, University of Edinburgh, 10 Crichton Street, Edinburgh EH8 9AB, UK}
\email{jrl@staffmail.ed.ac.uk}

\begin{abstract}
We consider the sublanguages of Plotkin's PCF obtained by imposing some bound \texorpdfstring{$k$}{k}
on the levels of types for which fixed point operators are admitted.
We show that these languages form a strict hierarchy,
in the sense that a fixed point operator for a type of level \texorpdfstring{$k$}{k} can never be defined
(up to observational equivalence) using fixed point operators for lower types.
This answers a question posed by Berger.
Our proof makes substantial use of the theory of nested sequential procedures
(also called PCF B\"ohm trees) as expounded in the recent book of Longley and Normann.
\end{abstract}

\maketitle

\section{Introduction}
\label{sec-intro}

In this paper we study sublanguages of Plotkin's functional programming language $\PCF$,
which we here take to be the simply typed $\lambda$-calculus over a single base type $\nat$,
with constants 
\[ \begin{array}{rlcrl}
\num{n} & : \nat \mbox{~~for each $n \in \N$} \;, & \hspace*{1.0em} &
\suc,\pre & : \nat\arrow\nat \;, \\
\ifzero & : \nat\arrow\nat\arrow\nat\arrow\nat \;, & &
Y_\sigma & : (\sigma\arrow\sigma)\arrow\sigma \mbox{~~for each type $\sigma$} \;.
\end{array} \]
As usual, we will consider this language to be endowed with a certain (call-by-name) operational semantics,
which in turn gives rise to a notion of \emph{observational equivalence} for $\PCF$ programs.

We define the \emph{level} $\lev(\sigma)$ of a type $\sigma$ inductively by
\[ \lev(\nat) ~=~ 0 \;, \hspace*{3.0em} 
   \lev(\sigma\arrow\tau) ~=~ \max\,(\lev(\sigma)+1,\lev(\tau)) \;, \]
and define the \emph{pure type} $\pure{k}$ of level $k \in \N$ by
\[ \pure{0} ~=~ \nat \;, \hspace*{3.0em}
   \pure{k+1} ~=~ \pure{k} \arrow \nat \;. \] 
Modifying the definition of $\PCF$ so that the constants $Y_\sigma$ are admitted only
for types $\sigma$ of level $\leq k$, we obtain a sublanguage $\PCF_k$ for any $k \in \N$.
Our main result will be that for each $k$, the expressive power of $\PCF_{k+1}$ strictly
exceeds that of $\PCF_k$: in particular, there is no closed term of $\PCF_k$ that is observationally equivalent to $Y_{\pure{k+1}}$.
(Fortunately, `observational equivalence' has the same meaning for all the 
languages in question here, as will be explained in Section~\ref{sec-background}.)
This answers a question posed explicitly by Berger in \cite{Berger-min-recursion}, 
but present in the folklore at least since the early 1990s.
It is worth remarking that the situation is quite different for various extensions of $\PCF$
considered in the literature, in which one may restrict to recursions at level~1 types
without loss of expressivity (see Subsection~\ref{subsec-embed}).

We can phrase our result more denotationally in terms of the type structure
$\SF$ of \emph{(PCF-)sequential functionals} or its effective substructure $\SF^\eff$.
As will be reviewed in Section~\ref{sec-background},
the latter may be conveniently characterized up to isomorphism as the closed term model for $\PCF$
modulo observational equivalence.
Our result can therefore be understood as saying that more elements of $\SF^\eff$ (and hence of $\SF$)
are denotable in $\PCF_{k+1}$ than in $\PCF_k$.
From this we may easily infer that there is no finite `basis'
$B \subseteq \SF^\eff$ relative to which all elements of $\SF^\eff$ are $\lambda$-definable 
(see Corollary~\ref{no-finite-basis-cor} below).

The models $\SF$ and $\SF^\eff$ are \emph{extensional}:
elements of type $\sigma \arrow \tau$ can be considered as mathematical functions mapping 
elements of type $\sigma$ to elements of type $\tau$.
Whilst our theorem is naturally stated in terms of these extensional models, 
its proof will make deep use of a more intensional model which yields $\SF$
as its extensional quotient.
This intensional model of PCF has been considered many times before in the literature,
for instance as the \emph{$\PCF$ B\"ohm tree} model of Amadio and Curien \cite{Amadio-Curien},
and it is known to be isomorphic to the game models of PCF given by 
Abramsky, Jagadeesan and Malacaria \cite{AJM} and by Hyland and Ong \cite{Hyland-Ong}.
In this paper, we choose to work with the \emph{nested sequential procedure} (NSP) 
presentation of this model, as studied in detail in the recent book of Longley and Normann 
\cite{Longley-Normann}. 
(We will touch briefly on the possible use of other presentations for this purpose
in Section~\ref{sec-further-work}.)
We shall denote the NSP model by $\SP^0$; 
its construction will be reviewed in Section~\ref{sec-background},
but in the meantime, let us offer a high-level overview of our proof method
without assuming detailed knowledge of this model.

As a motivating example, fix $k \in \N$, and consider the $\PCF$ term
\[  \Phi_{k+1} ~:~ (\nat \arrow \pure{k+1} \arrow \pure{k+1}) \arrow \nat \arrow \pure{k+1}  \]
given informally by
\[ \Phi_{k+1}\;g\;n ~=~ g\;n\;(\Phi_{k+1}\;g\;(n+1)) ~=~ g\;n\;(g\;(n+1)\;(g\;(n+2)\;(g\;\cdots))) \;,
\]
or more formally by
\[ \Phi_{k+1} ~=~ \lambda g n.\; Y_{\nat \arrow \pure{k+1}} \,
   (\lambda f m.\,g\,m\,(f(m+1)))\,n  \;, \]
where $f$ has type $\nat \arrow \pure{k+1}$.
Clearly, $\Phi_{k+1}$ is a term of $\PCF_{k+1}$; however, we will be able to
show that the element of $\SF$ that $\Phi_{k+1}$ defines is not denotable in $\PCF_k$. 

What is the essential feature of this function that puts it beyond the reach of $\PCF_k$?
To get a hint of this, we may observe from the informal definition above that
$\Phi_{k+1}$ seems implicitly to involve an infinite nested sequence of calls to its argument $g$, 
and indeed the NSP model makes this idea precise.
Furthermore, each call to $g$ involves an argument of type level $k+1$ resulting from another such call.
Broadly speaking, we shall refer to such a sequence of nested calls (subject to certain other conditions)
as a \emph{$k\!+\!1$-spine} in the NSP associated with $\Phi_{k+1}$.
As a second example, we can see from the natural recursive definition of $Y_{\pure{k+1}}$ itself 
that this too involves a spine of this kind:
\[  Y_{\pure{k+1}} \; h ~=~ h(Y_{\pure{k+1}}\;h) ~=~ h(h(h(\cdots))) \;.  \]
where $h$ has type $\pure{k+1} \arrow \pure{k+1}$.
In fact, we may view $Y_{\pure{k+1}}$ as a `special case' of $\Phi_{k+1}$, since
$Y_{\pure{k+1}} \,h$ is observationally equivalent to $\Phi_{k+1} \,(\lambda n.h) \,\num{0}$.

A suitable general definition of $k\!+\!1$-spine turns out to be quite delicate to formulate;
but having done this, it will be possible to prove that no $\PCF_k$-denotable NSP contains a
$k\!+\!1$-spine (Theorem~\ref{no-gremlin-thm}).
This will be proved by induction on the generation of such NSPs: in essence, we have to show that none of
the generating operations for the interpretations of $\PCF_k$ terms are capable of manufacturing 
spinal NSPs out of non-spinal ones.  This already suffices to show that within the
intensional model $\SP^0$, the evident procedure for $Y_{\pure{k+1}}$ is denotable in $\PCF_{k+1}$
but not in $\PCF_k$.

However, this does not yet establish our main theorem, which concerns not $\SP^0$ but its extensional quotient $\SF$. For this purpose, we undertake a closer analysis of the function $\Phi_{k+1}$ defined above: 
we show that
not only the NSP arising from the above definition, but any \emph{extensionally equivalent} NSP, 
must necessarily involve a $k\!+\!1$-spine. 
This shows that, within $\SF$ (or $\SF^\eff$), the functional given by $\Phi_{k+1}$ is
not denotable in $\PCF_k$.
This establishes Berger's conjecture that the languages $\PCF_k$ form a strict hierarchy.

Since $\Phi_{k+1}$ is definable from $Y_{\nat \arrow \pure{k+1}}$, the above shows that
the element $Y_{\nat \arrow \pure{k+1}} \in \SF$ is not $\PCF_k$-denotable.
To complete the picture, however, we would also like to know that the simpler element 
$Y_{\pure{k+1}} \in \SF$ is not $\PCF_k$-denotable.
We show this via a more refined version of the above analysis which is of some interest in its own right.
Just as $\PCF$ is `stratified' into sublanguages $\PCF_k$, we show that each $\PCF_k$ may be further
stratified into sublanguages $\PCF_{k,1}, \PCF_{k,2}, \ldots$ on the basis of the `width' of the types $\sigma$
for which $Y_\sigma$ is permitted (see Definition~\ref{width-def}).
An easy adaptation of our earlier proofs then shows that, for any $l$, 
there are operators $Y_\sigma$ in $\PCF_{k,l+2}$ that are not denotable in $\PCF_{k,l}$
(the appearance of $l+2$ is admittedly a curiosity here).
Since all these $Y_\sigma$ are themselves readily definable from $Y_{\pure{k+1}}$,
it follows that $Y_{\pure{k+1}} \in \SF$ itself is not denotable in $\PCF_{k,l}$ for any $l$,
and hence not in $\PCF_k$.
This finer analysis illustrates the remarkable richness of structure that $\SF$ has to offer.

The paper is organized as follows.
In Section~\ref{sec-background} we recall the necessary technical background on $\PCF$ 
and on the models $\SP^0$ and $\SF$, fleshing out many of the ideas outlined above.
In Section~\ref{sec-denotations} we obtain a convenient inductive characterization 
of the class of procedures denotable in (the `oracle' version of) $\PCF_k$,
framed in terms of constructions on the procedures themselves.
In Section~\ref{sec-Y-k+1} we introduce the central concept of a $k\!+\!1$-spinal procedure,
and show using our inductive characterization that no $\PCF_k$-denotable procedure can be 
$k\!+\!1$-spinal (this is the most demanding part of the proof).
As noted above, this already shows that $\PCF_{k+1}$ denotes more elements of $\SP^0$ 
than $\PCF_k$ does.
In Section~\ref{sec-extensional} we obtain the corresponding result for $\SF$, showing that the element 
$\Phi_{k+1} \in \SF$, and hence $Y_{\nat \arrow \pure{k+1}} \in \SF$, is not $\PCF_k$-denotable.
In Section~\ref{sec-pure-type} we adapt our methods to the more fine-grained hierarchy of languages
$\PCF_{k,l}$ that takes account of the widths of types; this enables us to show also that
$Y_{\pure{k+1}} \in \SF$ is not $\PCF_k$-denotable.
We conclude in Section~\ref{sec-further-work} with a discussion of related and future work.

\section{Background}  \label{sec-background}

We here summarize the necessary definitions and technical background from 
\cite{Longley-Normann}, especially from Chapters~6 and 7.

\subsection{The language PCF}  \label{subsec-PCF}

In \cite{Scott-TCS}, Scott introduced the language LCF for computable functionals of simple type.
This language is traditionally called $\PCF$ when equipped with a standalone
operational semantics as in Plotkin \cite{LCF-considered}.
We will work here with the same version of $\PCF$ as in \cite{Longley-Normann},
with the natural numbers as the only base type.
Our types $\sigma$ are thus generated by
\[  \sigma ~::=~ \nat ~\mid~ \sigma \arrow \sigma \;, \]
and our terms will be those of the simply typed $\lambda$-calculus
constructed from the constants
\[ \begin{array}{rcll}
  \num{n} & : & \nat & \mbox{for each $n \in \N$} \;, \\
  \mathit{suc},\;\mathit{pre} & : & \nat\arrow\nat \;, \\
  \mathit{ifzero} & : & \nat\arrow\nat\arrow\nat\arrow\nat \;, \\
  Y_\sigma & : & (\sigma\arrow\sigma) \arrow \sigma &
  \mbox{for each type $\sigma$} \;.
\end{array} \]
We often abbreviate the type $\sigma_0 \arrow\cdots\arrow \sigma_{r-1} \arrow \nat$
to $\sigma_0,\ldots,\sigma_{r-1} \arrow \N$ or just $\vec{\sigma} \arrow \nat$.
As usual, we write $\Gamma \vdash M:\sigma$ to mean that $M$ is a well-typed
term in the environment $\Gamma$ (where $\Gamma$ is a finite list of typed variables).
Throughout the paper, we shall regard the type of a variable $x$ as intrinsic to $x$, and will
often write $x^\sigma$ to indicate that $x$ carries the type $\sigma$.
For each $k \in \N$, the sublanguage $\PCF_k$ is obtained by admitting the constants
$Y_\sigma$ only for types $\sigma$ of level $\leq k$.

We endow the class of closed $\PCF$ terms with the following small-step reduction rules:
\[ \begin{array}{rclcrcl}
  (\lambda x.M)N & \reducesto & M[x \mapsto N]   \;, & &
  \mathit{ifzero}\;\num{0} & \reducesto & \lambda xy.x \;, \\
  \mathit{suc}\;\num{n} & \reducesto & \num{n+1} \;, & &
  \mathit{ifzero}\;\num{n+1} & \reducesto & \lambda xy.y \;, \\
  \mathit{pre}\;\num{n+1} & \reducesto & \num{n} \;, & &
  Y_\sigma M & \reducesto & M(Y_\sigma M) \;. \\
  \mathit{pre}\;\num{0}   & \reducesto & \num{0} \;,
\end{array} \]
We furthermore allow these reductions to be applied in certain term contexts.
Specifically, the relation $\reducesto$ is inductively generated by the rules above along with the clause:
if $M \reducesto M'$ then $E[M] \reducesto E[M']$,
where $E[-]$ is one of the contexts
\[ [-]N \;, \hspace*{2.0em} \suc\,[-] \;, \hspace*{2.0em}
   \pre\,[-] \;, \hspace*{2.0em} \ifzero\,[-] \;. \]
We write $\reducesto^*$ for the reflexive-transitive closure of $\reducesto$.
If $Q$ is any closed $\PCF$ term of type $\nat$, it is easy to check that either
$Q \reducesto^* \num{n}$ for some $n \in \N$ 
or the (unique) reduction path starting from $Q$ is infinite.

This completes the definition of the languages $\PCF$ and $\PCF_k$.
Whilst the language $\PCF_0$ is too weak for programming purposes
(it cannot even define addition), it is not hard to show that even $\PCF_1$ is Turing-complete:
that is, any partial computable function $\N \parrow \N$ is representable 
by a closed $\PCF_1$ term of type $\nat \arrow \nat$.

We will also refer to the non-effective language $\PCF^\Omega$
(or \emph{oracle $\PCF$}) obtained by extending the definition of $\PCF$ with a
constant $C_f: \nat\arrow\nat$ for every set-theoretic partial function $f : \N \parrow \N$,
along with a reduction rule $C_f \num{n} \reducesto \num{m}$ for every $n,m$ such that
$f(n)=m$. (In $\PCF^\Omega$, the evaluation of a closed term $Q:\nat$ may fail to
reach a value $\num{n}$ either because it generates an infinite computation, or because
it encounters a subterm $C_f(n)$ where $f(n)$ is undefined.)
The languages $\PCF^\Omega_k$ are defined analogously.

If $M,M'$ are closed $\PCF^\Omega$ terms of the same type $\sigma$, 
and $\LLL$ is one of our languages $\PCF_k$, $\PCF_k^\Omega$, $\PCF$, $\PCF^\Omega$,
we say that $M,M'$ are \emph{observationally equivalent in $\LLL$}, 
and write $M \simeq_\LLL M'$,
if for all closed program contexts $C[-^\sigma] : \nat$ of $\LLL$ and all $n \in \N$, we have
\begin{eqnarray*}
C[M] \reducesto^* \num{n} & \mbox{iff} & C[M'] \reducesto^* \num{n} \;. 
\end{eqnarray*}
(It makes no difference to the relation $\simeq_\LLL$ whether we take $C[-]$ to range over
single-hole or multi-hole contexts.)

Fortunately, it is easy to show that all of the above languages give rise to exactly the
same relation $\simeq_\LLL$.
Indeed, it is immediate from the definition that if $\LLL,\LLL'$ are two of our languages
and $\LLL \supseteq \LLL'$, then $\simeq_{\LLL} \,\subseteq\, \simeq_{\LLL'}$;
it therefore only remains to verify that $M \simeq_{\PCF_0} M'$ implies
$M \simeq_{\PCF^\Omega} M'$.
We may show this by a `syntactic continuity' argument,
exploiting the idea that any of the constants $Y_\sigma$ or $C_f$ in $\PCF^\Omega$
can be `approximated' as closely as necessary by terms of $\PCF_0$.
Specifically, let us write $\bot$ for the non-terminating program $Y_0(\lambda x^0.x) : \nat$
(a term of $\PCF_0$), and for any type $\sigma$ write $\bot_\sigma$ for the term of type $\sigma$
of the form $\lambda \vec{x}.\bot$.
For any $j \in \N$, we may then define $\PCF_0$ terms
\begin{eqnarray*} 
   Y_\sigma^{(j)} & = & \lambda f^{\sigma\arrow\sigma}.\,f^j(\bot_\sigma) \;,  \\
   C_f^{(j)} & = & \lambda n.\,\mathit{case}~n~\mathit{of}~
   (0 \darrow \num{f(0)} \mid \cdots \mid j-1 \darrow \num{f(j-1)}) \;,
\end{eqnarray*}
where we use some evident syntactic sugar in the definition of $C_f^{(j)}$.
For any $\PCF^\Omega$ term $M$, let $M^{(j)}$ denote the `approximation' obtained from
$M$ by replacing all occurrences of constants 
$Y_\sigma, C_f$ by $Y_\sigma^{(j)}, C_f^{(j)}$ respectively. 
It is then not hard to show that for closed $Q:\nat$, we have
\begin{eqnarray*}
Q \reducesto^* \num{n} & \mbox{iff} & \exists j.\; Q^{(j)} \reducesto^* \num{n} \;. 
\end{eqnarray*}
From this it follows easily that if $C[-]$ is an observing context of $\PCF^\Omega$
that distinguishes $M,M'$, then some approximation $C^{(j)}[-]$ (a context of $\PCF_0$)
also suffices to distinguish them. This establishes that 
$\simeq_{\PCF_0} \,\subseteq\; \simeq_{\PCF^\Omega}$.
We may therefore write $\simeq$ for observational equivalence without ambiguity.

In fact, an even more restricted class of observing contexts suffices for ascertaining
observational equivalence of $\PCF^\Omega$ terms. 
The well-known \emph{(equational) context lemma}, due to Milner \cite{Milner-PCF},
states that $M \simeq M' : \sigma_0,\ldots,\sigma_{r-1} \arrow \nat$
iff $M,M'$ have the same behaviour in all \emph{applicative contexts} of $\PCF$---%
that is, if for all closed $\PCF$ terms $N_0:\sigma_0,\ldots,N_{r-1}:\sigma_{r-1}$, we have
\begin{eqnarray*}
M N_0 \ldots N_{r-1} \reducesto^* n & \mbox{iff} &
M' N_0 \ldots N_{r-1} \reducesto^* n \;. 
\end{eqnarray*}
Furthermore, using the above idea of approximation, it is easy to see that we obtain
exactly the same equivalence relation if we allow the $N_i$ here to range only over
closed $\PCF_0$ terms---this gives us the notion of 
\emph{$\PCF_0$ applicative equivalence}, which we denote by $\sim_{0}$.

We have concentrated so far on giving a purely operational description of $\PCF$.
We are now able to express the operational content of our main theorems as follows.
As in Section~\ref{sec-intro}, we define the type $\pure{k}$ by $\pure{0} = \nat$,
$\pure{k+1} = \pure{k} \arrow \nat$;
we shall write $\pure{k}$ simply as $k$ where there is no risk of confusion.

\begin{thm}  \label{operational-main-thm}
For any $k \geq 1$, there are functionals definable in $\PCF_{k+1}$ but not in $\PCF^\Omega_k$.
More specifically: 

\begin{enumerate}[label=(\roman*)]
\item There is no closed term $M$ of $\PCF^\Omega_k$ such that 
$M \simeq Y_{\nat \arrow (k+1)}$ (or equivalently $M \sim_{0} Y_{\nat \arrow (k+1)}$).
\item There is even no closed $M$ of $\PCF^\Omega_k$ such that
$M \simeq Y_{k+1}$.
\end{enumerate}
\end{thm}

\noindent
We shall obtain part~(i) of this theorem at the end of Section~\ref{sec-extensional},
then in Section~\ref{sec-pure-type} resort to a more indirect method to obtain the stronger 
statement (ii).
The theorem also holds when $k=0$, but this rather uninteresting case does not require
the methods of this paper; it will be dealt with in Subsection~\ref{subsec-power-pcf1}.

Theorem~\ref{operational-main-thm} can be construed as saying that in a suitably 
pure fragment of a functional language such as Haskell,
the computational strength of recursive function definitions increases strictly as the admissible
type level for such recursions is increased.
The point of the formulation in terms of $\sim_{0}$ is to present our result in 
as manifestly strong a form as possible:
there is no $M \in \PCF_k$ that induces the same partial function as $Y_{k+1}$ 
even on closed $\PCF_0$ terms.

A more denotational formulation of our theorem can be given in terms of
the model $\SF$ of \emph{sequential functionals}, which we here define as the
type structure of closed $\PCF^\Omega$ terms modulo observational equivalence.
Specifically, for each type $\sigma$, let $\SF(\sigma)$ denote the set of closed
$\PCF^\Omega$ terms $M:\sigma$ modulo $\simeq$.
It is easy to check that application of $\PCF^\Omega$ terms
induces a well-defined function 
$\cdot : \SF(\sigma\arrow\tau) \times \SF(\sigma) \arrow \SF(\tau)$
for any $\sigma,\tau$;
the structure $\SF$ then consists of the sets $\SF(\sigma)$ along with these application operations.
Using the context lemma, it is easy to see that $\SF(\nat) \iso \N_\bot = \N \sqcup \{ \bot \}$,
and also that $\SF$ is \emph{extensional}:
if $f,f' \in \SF(\sigma\arrow\tau)$ satisfy
$f \cdot x = f' \cdot x$ for all $x \in \SF(\sigma)$, then $f = f'$.
Thus, up to isomorphism, each $\SF(\sigma\arrow\tau)$ may be considered simply as a 
certain set of functions from $\SF(\sigma)$ to $\SF(\tau)$.

Any closed $\PCF^\Omega$ term $M:\sigma$ naturally has a denotation $\sem{M}^\SF$
in $\SF(\sigma)$, namely its own equivalence class. We may therefore restate
Theorem~\ref{operational-main-thm} as:

\begin{thm}  \label{SF-main-thm}
Suppose $k \geq 1$.
\begin{enumerate}[label=(\roman*)]
\item The element $\sem{Y_{\nat\arrow(k+1)}}^\SF$ is not denotable in $\PCF^\Omega_k$.
\item Even $\sem{Y_{k+1}}^\SF$ is not $\PCF^\Omega_k$-denotable.
\end{enumerate}
\end{thm}

\noindent
It follows immediately that in any other adequate, compositional model of $\PCF^\Omega$
(such as Scott's continuous model or Berry's stable model), the element
$\sem{Y_{k+1}}$ is not $\PCF^\Omega_k$-denotable,
since the equivalence relation on $\PCF^\Omega$ terms induced by such a model must be contained
within $\simeq$.

By taking closed terms of $\PCF$ rather than $\PCF^\Omega$ modulo observational
equivalence, we obtain the type structure $\SF^\eff$ of 
\emph{effective sequential functionals}, which can clearly be seen as a substructure of $\SF$.
Although the above constructions of $\SF$ and $\SF^\eff$ are syntactic,
there are other more mathematical constructions 
(for instance, involving game models \cite{AJM,Hyland-Ong})
that also give rise to these structures, 
and experience suggests that these are mathematically natural classes of 
higher-order functionals.
We now see that Theorem~\ref{SF-main-thm}(i)
implies an interesting absolute property of $\SF^\eff$,
not dependent on any choice of presentation for this structure or any selection of language primitives:

\begin{cor}[No finite basis]   \label{no-finite-basis-cor}
There is no finite set $B$ of elements of $\SF^\eff$ such that all
elements of $\SF^\eff$ are $\lambda$-definable relative to $B$.
In other words, the cartesian category of $\PCF$-computable functionals is not finitely generated.
\end{cor}

\begin{proof}
Suppose $B = \{ b_0, \ldots, b_{n-1} \}$ were such a set.
For each $i$, take a closed $\PCF$ term $M_i$ denoting $b_i$.
Then the terms $M_0,\ldots,M_{n-1}$ between them contain only finitely many
occurrences of constants $Y_\sigma$, 
so these constants are all present in $\PCF_k$ for large enough $k$.
But this means that $b_0,\ldots,b_{n-1}$, and hence all elements of $\SF^\eff$, 
are $\PCF_k$-denotable, contradicting Theorem~\ref{SF-main-thm}(i).
\end{proof}

\subsection{The model \texorpdfstring{$\SP^0$}{SP0}}  \label{sec-SP0}

We turn next to an overview of the \emph{nested sequential procedure} (or NSP) model, denoted by $\SP^0$.
Further details and motivating examples are given in \cite{Longley-Normann}.
In some respects, however, our presentation here will be more formal than that of \cite{Longley-Normann}:
in particular, our discussion of bound variables and $\alpha$-conversion issues will be somewhat 
more detailed, in order to provide a solid foundation for the delicate syntactic arguments that follow.

The ideas behind this model have a complex history.
The general idea of sequential computation via nested oracle calls 
was the driving force behind Kleene's later papers
(e.g.\ \cite{Kleene-revisited-IV}), although the concept did not receive a 
particularly transparent or definitive formulation there.
Many of the essential ideas of NSPs can be found in early work of
Sazonov \cite{Sazonov-early}, in which a notion of 
\emph{Turing machine with oracles} was used to characterize the 
`sequentially computable' elements of the Scott model.
NSPs as we study them here were first explicitly introduced
in work on game semantics for $\PCF$---%
both by Abramsky, Jagadeesan and Malacaria \cite{AJM} 
(under the name of \emph{evaluation trees}) 
and by Hyland and Ong \cite{Hyland-Ong} (under the name of \emph{canonical forms}).
In these papers, NSPs played only an ancillary role;
however, it was shown by Amadio and Curien \cite{Amadio-Curien} how 
(under the name of \emph{$\PCF$ B\"ohm trees}) they
could be made into a model of $\PCF$ in their own right.
Similar ideas were employed again by Sazonov
\cite{Sazonov-PCF} to give a standalone characterization of
the class of sequentially computable functionals.
More recently, Normann and Sazonov \cite{Normann-Sazonov} 
gave an explicit construction of the NSP model in 
a somewhat more semantic spirit than \cite{Amadio-Curien}, 
using the name \emph{sequential procedures}.
As in \cite{Longley-Normann}, we here add the epithet `nested'
to emphasize the contrast with other flavours of sequential computation.%
\footnote{A major theme of \cite{Longley-Normann} is that NSPs serve equally well to 
capture the essence of PCF computation and that of Kleene's S1--S9 computability;
this is one reason for preferring a name that is not biased towards PCF.}

As in \cite{Longley-Normann}, our NSPs are generated by means of the following
infinitary grammar, interpreted coinductively.
Here $\bot$ is a special atomic symbol and $n$ ranges over natural numbers.
\[ \begin{array}{rrcl}
\mbox{\it Procedures:} &
p,q & ::= & \lambda x_0 \cdots x_{r-1}.\,e \\
\mbox{\it Expressions:} &
d,e & ::= & \bot ~\mid~ n ~\mid~ \ccase~a~\,\oof~ 
            (i \darrow e_i \mid i \in \N) \\
\mbox{\it Applications:} &
a   & ::= & x\, q_0 \cdots q_{r-1}
\end{array} \]

Here we write $(i \darrow e_i \mid i \in \N)$ to indicate an infinite sequence of `branches':
$(0 \darrow e_0 \mid 1 \darrow e_1 \mid 2 \darrow e_2 \mid \cdots)$.

We will use vector notation to denote finite (possibly empty) lists of variables or procedures:
$\vec{x}$, $\vec{q}$.
Our convention will be that a list $\vec{x}$ must be non-repetitive, though a list $\vec{q}$ need not be.
We may use $t$ to range over \emph{NSP terms} of any of the above three kinds;
note that a `term' is formally a (possibly infinite) syntax tree as generated by the above grammar. 
A procedure $\lambda \vec{x}.\bot$ will often be abbreviated to $\bot$.

For the most part, we will be working with terms modulo (infinitary) $\alpha$-equivalence,
and most of the concepts we introduce will be stable under renamings of bound variables.
Thus, a statement $t=t'$, appearing without qualification, will mean that $t,t'$ are $\alpha$-equivalent
(although we will sometimes write $=$ as $=_\alpha$ if we wish to emphasize this).
When we wish to work with terms on the nose rather than up to $=_\alpha$, we shall refer to them
as \emph{concrete terms}.

If each variable is assigned a simple type over $\nat$, then 
we may restrict our attention to \emph{well-typed}\index{well-typed} terms.
Informally, a term will be well-typed unless a typing violation occurs
at some particular point within its syntax tree. 
Specifically, within any term $t$, occurrences of procedures $\lambda \vec{x}.e$
(of any type), applications $x \vec{q}$ (of the ground type $\nat$) and expressions $e$
(of ground type) have types that are related to the types of their constituents and of variables
as usual in type $\lambda$-calculus extended by $\ccase$ expressions of type $\nat$.
We omit the formal definition here since everything works in the expected way;
for a more precise formulation see \cite[Section~6.1.1]{Longley-Normann}.

If $\Gamma$ is any environment (i.e.\ a finite non-repetitive list of variables),
we write $\Gamma \vdash e$ and $\Gamma \vdash a$ to mean that 
$e,a$ respectively are well-typed with free variables in $\Gamma$. 
We also write $\Gamma \vdash p:\tau$ when $p$ is well-typed in $\Gamma$ and
of type $\tau$. 

We shall often refer to variable environments that arise from combining several lists of variables,
which may be represented by different notations, e.g.\ $\Gamma,V,\vec{x}$.  
Since such environments are required to be non-repetitive, we take it to be
part of the content of a typing judgement such as 
$\Gamma,V,\vec{x} \vdash t \,(:\tau)$ that the entire list $\Gamma,V,\vec{x}$ is non-repetitive.
However, the \emph{order} of variables within an environment will typically be of much less concern to us 
(clearly our typing judgements $\Gamma \vdash T \,(:\tau)$ are robust under permutations of $\Gamma$),
and we will sometimes abuse notation by identifying a finite \emph{set} $Z$ of variables
with the list obtained from some arbitrary ordering of it.

It will also be convenient to place another condition on concrete
well-typed terms (not imposed in \cite{Longley-Normann}) in order to exclude \emph{variable hiding}.
Specifically, we shall insist that if $\Gamma \vdash t \,(:\tau)$
then no variable of $\Gamma$ appears as a bound variable within $t$,
nor are there any \emph{nested} bindings within $t$ of the same variable $x$.
(Clearly any concrete term not satisfying this restriction is $\alpha$-equivalent to one that does.)
This will help to avoid confusion in the course of some delicate arguments in which questions of
the identity of variables are crucial.

With these ideas in place, we may take $\SP(\sigma)$ to be the set of
well-typed procedures of type $\sigma$ modulo $=_\alpha$, 
and $\SP^0(\sigma) \subseteq \SP(\sigma)$ the subset constituted by the \emph{closed} procedures
(i.e.\ those that are well-typed in the empty environment).
By inspection of the grammar for procedures, it is easy to see that $\SP^0(\nat) \iso \N_\bot$.

As in \cite{Longley-Normann}, we shall need to work not only with NSPs
themselves, but with a more general calculus of NSP \emph{meta-terms}
designed to accommodate the intermediate forms that arise in the course of computations:
\[ \begin{array}{rrcl}
\mbox{\it Meta-procedures:} &
P,Q & ::=   & \lambda \vec{x}.\,E \\
\mbox{\it Meta-expressions:} &
D,E & ::=   & \bot ~\mid~ n ~\mid~ 
      \caseof{G}{i \darrow E_i \mid i \in \N} \\
\mbox{\it Ground meta-terms:} &
G   & ::=   & E ~\mid~ x\, \vec{Q} ~\mid~ P \vec{Q}
\end{array} \]

Here again, $\vec{x}$ and $\vec{Q}$ denote finite lists.
We shall use $T$ to range over meta-terms of any of the above three kinds;
once again, a meta-term is formally a syntax tree as generated by the above grammar.
(Unless otherwise stated, we use uppercase letters for general meta-terms and lowercase
ones for terms.)
Once again, we will normally work with meta-terms up to (infinitary) $\alpha$-equivalence,
but may also work with concrete meta-terms when required.

The reader is warned of an ambiguity arising from the above grammar: a surface form
$\lambda.E$ may be parsed either as a meta-procedure $\lambda \vec{x}.E$ with $\vec{x}$ empty,
or as a ground meta-term $(\lambda \vec{x}.E)\vec{Q}$ with both $\vec{x},\vec{Q}$ empty.
Formally these are two quite distinct meta-terms, bearing in mind that meta-terms are officially syntax trees.
To remedy this ambiguity, we shall therefore in practice write `$()$' to indicate the presence of 
an empty argument list $\vec{Q}$ to a meta-procedure $P$, 
so that the ground meta-term above will be written as $(\lambda.E)()$.
In the absence of `$()$', a surface form $\lambda.E$ should always be interpreted as a meta-procedure.

Meta-terms are subject to the expected typing discipline, leading to typing judgements
$\Gamma \vdash P:\sigma$, $\Gamma \vdash E$, $\Gamma \vdash G$
for meta-procedures, meta-expressions and ground meta-terms respectively.
Again we omit the details: see \cite[Section~6.1.1]{Longley-Normann}.
We shall furthermore require that well-typed concrete meta-terms are subject to the 
no-variable-hiding condition.
We will sometimes write e.g.\ $\Gamma \vdash P$ to mean that $P$ is a well-typed meta-procedure
in environment $\Gamma$, if the type itself is of no particular concern to us.

There is an evident notion of simultaneous capture-avoiding \emph{substitution} 
$T[\vec{x}\mapsto\vec{Q}]$ for well-typed concrete terms. 
Specifically, given $\Gamma,\vec{x} \vdash T \,(:\tau)$ and 
$\Gamma,\vec{y} \vdash Q_i : \sigma_i$ for each $i<r$,
where $\vec{x} = x_0^{\sigma_0},\ldots,x_{r-1}^{\sigma_{r-1}}$, 
we will have $\Gamma,\vec{y} \vdash T[\vec{x}\mapsto\vec{Q}] \,(:\tau)$.
Note that this may entail renaming of bound variables both within $T$ (in order to avoid capture of
variables in $\vec{y}$) and in the $Q_i$ (in order to maintain the no-hiding condition for variables
bound within $T$).
The details of how this renaming is performed will not matter, provided that for each $T,\vec{x},\vec{Q}$ as above we have a determinate choice of a suitable concrete term $T[\vec{x}\mapsto\vec{Q}]$,
so that multiple appearances of the same substitution will always yield the same result.
We also note that substitution is clearly well-defined on $\alpha$-equivalence classes.
Finally, we will say a substitution $[\vec{x}\mapsto\vec{Q}]$ \emph{covers}
a set $V$ of variables if $V$ consists of precisely the variables $\vec{x}$.

As a mild extension of the concept of meta-term, we have an evident notion of a
\emph{meta-term context} $C[-]$: essentially a meta-term containing a `hole' $-$,
which may be of meta-procedure, meta-expression or ground meta-term type
(and in the case of meta-procedures, will carry some type $\sigma$).
Our convention here is that a meta-term context $C[-]$ is permitted to contain only a single 
occurrence of the hole $-$.
Multi-hole contexts $C[-_0,-_1,\ldots]$ will occasionally be used, but again, each hole $-_i$ may
appear only once.

By the \emph{local variable environment} associated
with a concrete meta-term context $\Gamma \vdash C[-]$,
we shall mean the set $X$ of variables $x$ bound within $C[-]$ whose scope includes the hole, 
so that the environment in force at the hole is $\Gamma,X$.
(The no-variable-hiding convention ensures that $X$ and indeed $\Gamma,X$ is non-repetitive.)
Although in principle local variable environments pertain to particular choices of concrete contexts,
most of the concepts that we define using such environments will be easily seen to be invariant
under renamings of bound variables.

Next, there is a concept of \emph{evaluation} 
whereby any concrete meta-term $\Gamma \vdash T\, (:\sigma)$ evaluates to 
an ordinary concrete term $\Gamma \vdash\, \dang{\!T\!} (:\sigma)$.
To define this, the first step is to introduce a \emph{basic reduction}\index{basic reduction} relation $\reducesto_b$ for concrete ground meta-terms, which we do by the following rules:
\begin{description}
\item[(b1)] \( (\lambda \vec{x}.E) \vec{Q} \reducesto_b 
         E [\myvec{x} \mapsto \myvec{Q}] ~~~\) (\emph{$\beta$-rule}). 
\item[(b2)] \( \caseof{\bot}{i \darrow E_i} \reducesto_b \bot \).
\item[(b3)] \( \caseof{n}{i \darrow E_i} \reducesto_b  E_n \).
\item[(b4)] \( \caseof{(\caseof{G}{i \darrow E_i})}{j \darrow F_j} \reducesto_b 
                    \caseof{G}{i \darrow \caseof{E_i}{j \darrow \!F_j}} \).
\end{description}
Note that the $\beta$-rule applies even when $\vec{x}$ is empty: 
thus, $(\lambda.2)() \reducesto_b 2$.

From this, a \emph{head reduction} relation $\reducesto_h$ 
on concrete meta-terms is defined inductively:
\begin{description}
\item[(h1)] If $G \reducesto_b G'$ then $G \reducesto_h G'$.
\item[(h2)] If $G \reducesto_h G'$ and $G$ is not a $\ccase$ meta-term, then 
\[ \caseof{G}{i \darrow E_i} ~\reducesto_h~ \caseof{G'}{i \darrow E_i} \;. \]
\item[(h3)] If $E \reducesto_h E'$
then $\lambda \myvec{x}.E ~\reducesto_h~ \lambda \myvec{x}.E'$.
\end{description}

Clearly, for any meta-term $T$, there is at most one $T'$ with $T \reducesto_h T'$.
We call a meta-term a \emph{head normal form} if it cannot be further 
reduced using $\reducesto_h$. 
The possible shapes of head normal forms are 
$\bot$, $n$, $\caseof{y \vec{Q}}{i \darrow E_i}$
and $y \vec{Q}$, the first three optionally prefixed by $\lambda \myvec{x}$
(where $\vec{x}$ may contain $y$).

We now define the \emph{general reduction} relation $\reducesto$ inductively as follows:
\begin{description}
\item[(g1)] If $T \reducesto_h T'$ then $T \reducesto T'$.
\item[(g2)] If $E \reducesto E'$ then 
$\lambda \myvec{x}.E ~\reducesto~ \lambda \myvec{x}.E'$.
\item[(g3)] If $Q_j = Q'_j$ except at $j=k$ where $Q_k \reducesto Q'_k$, then
\begin{eqnarray*}
x \vec{Q} & \reducesto & x \vec{Q}\,' \;,  \\
\caseof{x\, \vec{Q}}{i \darrow E_i} & \reducesto &
\caseof{x\, \vec{Q}\,'}{i \darrow E_i} \;.
\end{eqnarray*}
\item[(g4)] If $E_i = E'_i$ except at $i=k$ where $E_k \reducesto E'_k$, then
\[ \caseof{x\, \vec{Q}}{i \darrow E_i} ~\reducesto~
   \caseof{x\, \vec{Q}}{i \darrow E'_i} \;. \]
\end{description}
It is easy to check that this reduction system is sound with respect to the typing rules.
We emphasize that the relation $\reducesto$ is defined on concrete meta-terms,
although it is clear that it also gives rise to a well-defined reduction relation on their
$\alpha$-classes.
An important point to note is that modulo the obvious inclusion of terms into meta-terms,
terms are precisely meta-terms in \emph{normal form}, i.e.\ those that cannot be reduced using $\reducesto$.
(For example, the meta-procedure $\lambda.2$ is in normal form, though the ground meta-term
$(\lambda.2)()$ is not.)
We write $\reducesto^*$ for the reflexive-transitive closure of $\reducesto$.

The above reduction system captures the finitary aspects of evaluation.
In general, however, since terms and meta-terms may be infinitely deep,
evaluation must be seen as an infinite process. 
To account for this infinitary aspect, we use some familiar domain-theoretic ideas.

We write $\sqsubseteq$
for the evident syntactic orderings on concrete meta-procedures and on ground 
meta-terms: thus, $T \sqsubseteq U$ iff $T$ may be obtained from $U$ 
by replacing zero or more ground subterms (perhaps infinitely many) by $\bot$.
It is easy to see that for each $\sigma$, the set of all concrete procedure terms of type $\sigma$
forms a directed-complete partial order under $\sqsubseteq$. 

By a \emph{finite} (concrete) term $t$, we shall mean one generated by the following
grammar, this time construed inductively:
\[ \begin{array}{rrcl}
\mbox{\it Procedures:} &
p,q & ::= & \lambda x_0 \ldots x_{r-1}.\,e \\
\mbox{\it Expressions:} &
d,e & ::= & \bot ~\mid~ n ~\mid~ \ccase~a~\,\oof~ 
            (0 \darrow e_0 \mid \cdots \mid r-1 \darrow e_{r-1}) \\
\mbox{\it Applications:} &
a   & ::= & x\, q_0 \ldots q_{r-1}
\end{array} \]
We regard finite terms as ordinary NSP terms 
by identifying the conditional branching
$(0 \darrow e_0 \mid \cdots \mid r-1 \darrow e_{r-1})$ with 
\[ (0 \darrow e_0 \mid \cdots \mid r-1 \darrow e_{r-1} \mid 
    r \darrow \bot \mid r+1 \darrow \bot \mid \cdots) \;.
\]

We may now explain how a general meta-term $T$ \emph{evaluates} 
to a term $\dang{\!T\!}$.
This will in general be an infinite process,
but we can capture the value of $T$ as the limit of the finite portions that
become visible at finite stages in the reduction.
To this end, for any concrete meta-term $T$ we define
\[ \Downarrow_\fin T ~=~
   \{ t \mbox{~finite~} \mid \exists T'.\; T \reducesto^* T' ~\wedge~ 
   t \sqsubseteq T' \} \;. 
\]
It is not hard to check that for any meta-term $T$, the set $\Downarrow_\fin T$
is directed with respect to $\sqsubseteq$ (see \cite[Proposition~6.1.2]{Longley-Normann}).
We may therefore define 
$\dang{\!T\!}$, the \emph{value} of $T$, to be the ordinary concrete term
\[ \dang{\!T\!} ~=~ \bigsqcup \,(\Downarrow_\fin T) \;. \]
Note in passing that the value $\dang{\!G\!}$ of a ground meta-term $G$
may be either an expression or an application. 
In either case, it is certainly a ground meta-term.
It is also easy to see that
$\dang{\lambda \myvec{x}.E} \,=\, \lambda \myvec{x}.\dang{\!E\!}$,
and that if $T \reducesto^* T'$ then $\dang{\!T\!} \,=\, \dang{\!T'\!}$.

Whilst we have defined our evaluation operation $\dang{\!-\!}$ for concrete meta-terms,
it is clear that this induces a well-defined evaluation operation on their $\alpha$-classes,
and for the most part this is all that we shall need.
We also note that the syntactic ordering $\sqsubseteq$ on concrete terms induces a partial order
$\sqsubseteq$ on their $\alpha$-classes, and that each $\SP(\sigma)$ and $\SP^0(\sigma)$
is directed-complete with respect to this ordering.

In the present paper, an important role will be played by the tracking of variable occurrences
(and sometimes other subterms) through the course of evaluation.
By inspection of the above rules for $\reducesto$, it is easy to see that if $T \reducesto T'$,
then for any occurrence of a (free or bound) variable $x$ within $T'$, we can identify a
unique occurrence of $x$ within $T$ from which it originates (we suppress the formal
definition). The same therefore applies whenever $T \reducesto^* T'$.
In this situation, we may say that the occurrence of $x$ within $T'$ is a \emph{residual} of the
one within $T$, or that the latter is the \emph{origin} of the former.
Note, however, that these relations are relative to a particular reduction path
$T \reducesto^* T'$: there may be other paths from $T$ to $T'$ for which the origin-residual relation
is different.  

Likewise, for any occurrence of $x$ within $\dang{\!T\!}$, we may pick some finite
$t \sqsubseteq \,\dang{\!T\!}$ containing this occurrence, and some $T' \sqsupseteq t$
with $T \reducesto^* T'$; this allows us to identify a unique occurrence of $x$ within $T$
that originates the given occurrence in $\dang{\!T\!}$.
It is routine to check that this occurrence in $T$ will be independent of the choice of
$t$ and $T'$ and of the chosen reduction path $T \reducesto^* T'$;
we therefore have a robust origin-residual relationship between variable occurrences in $T$
and those in $\dang{\!T\!}$.

A fundamental result for NSPs is the \emph{evaluation theorem}, which says that
the result of evaluating a meta-term is unaffected if we choose to evaluate certain subterms
`in advance':

\begin{thm}[Evaluation theorem]  \label{eval-thm}
If $C[-_0,-_1,\ldots]$ is any meta-term context with countably
many holes and $C[T_0,T_1,\ldots]$ is well-formed, then
\[ \dang{C[T_0,T_1,\ldots]} ~=~ 
   \dang{C[\dang{\!T_0\!},\dang{\!T_1\!},\ldots]}. \]
\end{thm}

The proof of this is logically elementary but administratively complex: 
see \cite[Section~6.1.2]{Longley-Normann}.

One further piece of machinery will be useful: 
the notion of \emph{hereditary $\eta$-expansion}, which enables us to convert 
a variable $x$ into a procedure term (written $x^\eta$).
Using this, the restriction that variables may appear only at the head of applications
can be seen to be inessential: e.g.\ the `illegal term' $f x$ may be replaced by the legal term $f x^\eta$.
The definition of $x^\eta$ is by recursion on the type of $x$: 
if $x:\sigma_0,\ldots,\sigma_{r-1} \arrow \nat$, then
\[ x^\eta ~=~ \lambda z_0^{\sigma_0} \ldots z_{r-1}^{\sigma_{r-1}} . \;
   \caseof{x z_0^\eta \ldots z_{r-1}^\eta}{i \darrow i} \;.  \]
In particular, if $x:\nat$ then $x^\eta = \lambda.\,\caseof{x}{i \darrow i}$.
The following useful properties of $\eta$-expansion are proved in 
\cite[Lemma~6.1.14]{Longley-Normann}:

\begin{lem}  \label{eta-properties}
$\dang{x^\eta \vec{q}} \;=\, \caseof{x \vec{q}}{i \darrow i}$, and
$\dang{\lambda \vec{y}.\,p \vec{y}\,^\eta} \;=\, p$.
\end{lem}
The sets $\SP(\sigma)$ may now be made into a total applicative
structure $\SP$ by defining
\[ (\lambda x_0 \cdots x_r.e) \cdot q ~=~ 
    \lambda x_1 \cdots x_r.\dang{e[x_0 \mapsto q]} \;. \]
Clearly the sets $\SP^0(\sigma)$ are closed under this application 
operation, so we also obtain an applicative substructure $\SP^0$ of $\SP$.
It is easy to check that application in $\SP$ is monotone and continuous
with respect to $\sqsubseteq$.
It is also shown in \cite[Section~6.1.3]{Longley-Normann} that both $\SP$ and $\SP^0$
are typed \emph{$\lambda$-algebras}: 
that is, they admit a compositional interpretation of typed $\lambda$-terms that
validates $\beta$-equality. 
(The relevant interpretation of pure $\lambda$-terms is in fact given by three of the
clauses from the interpretation of $\PCF^\Omega$ as defined in Subsection~\ref{subsec-interp} below.)

\subsection{Interpretation of PCF in \texorpdfstring{$\SP^0$}{SP0}}  \label{subsec-interp}

A central role will be played by certain procedures 
$Y_\sigma \in \SP^0((\sigma\arrow\sigma)\arrow\sigma)$
which we use to interpret the $\PCF$ constants $Y_\sigma$
(the overloading of notation will do no harm in practice).
If $\sigma = \sigma_0,\ldots,\sigma_{r-1} \arrow \nat$,
we define $Y_\sigma = \lambda g^{\sigma\arrow\sigma}. F_\sigma[g]$,
where $F_\sigma[g]$ is specified corecursively up to $\alpha$-equivalence by:
\[ F_\sigma[g] ~=_\alpha~ \lambda x_0^{\sigma_0} \ldots x_{r-1}^{\sigma_{r-1}}.\;
                              \caseof{g\;(F_\sigma[g])\; x_0^\eta \cdots x_{r-1}^\eta}{i \darrow i} \;. \]
(A concrete representative of $F_\sigma[g]$ satisfying the no-hiding condition will of course feature 
a different choice of bound variables $x_0,\ldots,x_{r-1}$ at each level.)
We may now give the standard interpretation of $\PCF^\Omega$ in $\SP$.
To each $\PCF^\Omega$ term $\Gamma \vdash M:\sigma$ we associate
a procedure-in-environment $\Gamma \vdash \sem{M}^{\SP}_\Gamma : \sigma$
(denoted henceforth by $\sem{M}_\Gamma$)
inductively as follows:
\begin{eqnarray*}
\sem{x^\sigma}_\Gamma & = & x^{\eta} \\
\sem{\num{n}}_\Gamma & = & \lambda.n \\
\sem{\suc}_\Gamma & = & \lambda x.\,\caseof{x}{i \darrow i+1} \\
\sem{\pre}_\Gamma & = & \lambda x.\,\caseof{x}{0 \darrow 0 \mid i+1 \darrow i} \\
\sem{\ifzero}_\Gamma & = & 
\lambda xyz.\,\ccase~{x}~\oof~(0 \darrow \caseof{y}{j \darrow j} \\
& & ~~~~~~~~~~~~~~~\mid~i+1 \darrow \caseof{z}{j \darrow j}) \\
\sem{Y_\sigma}_\Gamma & = & Y_\sigma \\
\sem{C_f}_\Gamma & = & \lambda x.\,\caseof{x}{i \darrow f(i)} \\
\sem{\lambda x^\sigma.M}_\Gamma & = & \lambda x.\sem{M}_{\Gamma,x} \\
\sem{MN}_\Gamma & = & \sem{M}_\Gamma \cdot \sem{N}_\Gamma
\end{eqnarray*}
(In the clause for $C_f$, we interpret $f(i)$ as $\bot$ whenever $f(i)$ is undefined.)

The following key property of $\sem{-}^\SP$ is shown as Theorem~7.1.16 in \cite{Longley-Normann}:

\begin{thm}[Adequacy]   \label{SP0-adequacy}
For any closed $\PCF^\Omega$ term $M:\nat$,
we have $M \reducesto^* \num{n}$ iff $\sem{M} = \lambda.n$.
\end{thm}

We may now clarify the relationship between $\SP^0$ and $\SF$.
First, there is a natural `observational equivalence' relation $\approx$ on each $\SP^0(\sigma)$,
defined by
\begin{eqnarray*}
  q \approx q' & \mbox{iff} & \forall r \in \SP^0(\sigma\arrow\nat).~~ r \cdot q  ~=~ r \cdot q' \;. 
\end{eqnarray*}
It is not hard to see that if $p \approx p' \in \SP^0(\sigma\arrow\tau)$ and $q \approx q' \in \SP^0(\sigma)$
then $p \cdot q \approx p' \cdot q \approx p' \cdot q' \in \SP^0(\tau)$.
Explicitly, the first of these equivalences holds because for any $r \in \SP^0(\tau\arrow\nat)$ we have
(using Lemma~\ref{eta-properties}) that
\[ r \cdot (p \cdot q) ~=~ (\lambda x.r(\lambda \vec{z}.xq\vec{z}^{\,\eta})) \cdot p 
   ~=~ (\lambda x.r(\lambda \vec{z}.xq\vec{z}^{\,\eta})) \cdot p' ~=~ r \cdot (p' \cdot q) \;, \]
while the second equivalence holds because for any $r \in \SP^0(\tau\arrow\nat)$ we have
\[ r \cdot (p' \cdot q) ~=~ (\lambda y.r(\lambda \vec{z}.p'y^\eta \vec{z}^{\,\eta})) \cdot q 
   ~=~ (\lambda y.r(\lambda \vec{z}.p'y^\eta \vec{z}^{\,\eta})) \cdot q' ~=~ r \cdot (p' \cdot q') \;. \]
We thus obtain a well-defined applicative structure $\SP^0 /\!\approx$ as a quotient of $\SP^0$;
we write $\theta : \SP^0 \arrow \SP^0 /\!\approx$ for the quotient map.

It turns out that up to isomorphism, this structure $\SP^0 /\!\approx$ is none other than $\SF$.
Indeed, in \cite{Longley-Normann} this was taken as the definition of $\SF$, and the
characterization as the closed term model of $\PCF^\Omega$ modulo observational equivalence
proved as a consequence.
In order to fill out the picture a little more,
we will here exhibit the equivalence of these two definitions as a consequence of 
the following non-trivial fact, given as Corollary~7.1.34 in \cite{Longley-Normann}:%
\footnote{What is actually shown in \cite{Longley-Normann} is that every element of $\SP^0$ is denotable
on the nose in a language $\PCF + \mathit{byval}$, with a certain choice of denotation for the
constant $\mathit{byval}$. 
Since the latter satisfies
$\sem{\mathit{byval}} \approx \sem{\lambda f^{\nat\arrow\nat}x^\nat.\,\ifzero\;x\,(fx)(fx)}$,
the present Theorem~\ref{SP0-completeness-thm} follows easily.}

\begin{thm}  \label{SP0-completeness-thm}
For every $p \in \SP^0(\sigma)$, there is a closed $\PCF^\Omega$ term $M: \sigma$ such that
$\sem{M} \approx p$.
\end{thm}

\begin{prop}  \label{SF-char}
$(\SP^0/\!\approx) \iso\SF$, via an isomorphism that identifies $\theta(\sem{M}^\SP)$ 
with $\sem{M}^\SF$ for any closed $\PCF^\Omega$ term $M$.
\end{prop}

\begin{proof}
For any element $x \in \SP^0/\!\approx$, we may take $p \in \SP^0$ with $\theta(p)=x$,
then by Theorem~\ref{SP0-completeness-thm} take $M$ closed with $\sem{M} \approx p$;
we then have that $\theta(\sem{M}) = x$.
In this way, each $\SP^0(\sigma)/\!\approx$ corresponds bijectively to the set of 
closed $\PCF^\Omega$ terms $M:\sigma$ modulo some equivalence relation $\sim$. 

Recall now that $\simeq$ denotes observational equivalence in $\PCF^\Omega$.
To see that $\sim \,\subseteq\, \simeq$, suppose $M \sim M'$, and let $C[-]$ be any suitable observing
context of $\PCF^\Omega$. By the compositionality of $\sem{-}^\SP$, we obtain some $\sem{C} \in \SP^0$
such that $\sem{C[M]} = \sem{C} \cdot \sem{M}$ and similarly for $M'$. 
But $M \sim M'$ means that $\sem{M} \approx \sem{M'}$, whence by the definition of $\approx$
we conclude that $\sem{C[M]} = \sem{C[M']}$ at type $\nat$.
So by Theorem~\ref{SP0-adequacy} we have 
$C[M] \reducesto^* \num{n}$ iff $C[M'] \reducesto^* \num{n}$.
Since $C[-]$ was arbitrary, we have shown that $M \simeq M'$.

To see that $\simeq \,\subseteq\, \sim$, suppose $M \simeq M' : \sigma$. 
It will suffice to show that $\sem{M} \approx \sem{M'}$.
So suppose $r \in \SP^0(\sigma\arrow\nat)$, and 
using Theorem~\ref{SP0-completeness-thm}, take a $\PCF^\Omega$ term $R$ such that
$\sem{R} \approx r$.
Then $r \cdot \sem{M} = \sem{R} \cdot {M} = \sem{RM}$ at type $\nat$, and similarly for $M'$.
But since $M \simeq M'$, we have $RM \reducesto^* n$ iff $RM' \reducesto^* n$,
whence $\sem{RM} = \sem{RM'}$ by Theorem~\ref{SP0-adequacy}.
Thus $r \cdot \sem{M} = r \cdot \sem{M'}$, and we have shown $\sem{M} \approx \sem{M'}$.

Since each $\SF(\sigma)$ consists of closed $\PCF^\Omega$ terms $M:\sigma$ modulo $\simeq$,
we have established a bijection $(\SP^0(\sigma) /\!\approx) \iso \SF(\sigma)$ for each $\sigma$.
Moreover, both $\sem{-}^\SP$ and $\theta$ respect application, so it follows that 
$(\SP^0 /\!\approx) \iso \SF$, and it is immediate by construction that $\theta(\sem{M}^\SP)$
is identified with $\sem{M}^\SF$.
\end{proof}

As we have already seen in Subsection~\ref{subsec-PCF}, 
Milner's context lemma for $\PCF^\Omega$ implies that
$\SF$ is extensional.
From this and Proposition~\ref{SF-char}, we may now read off the following
useful characterization of the equivalence $\approx$:

\begin{lem}[NSP context lemma]  \label{nsp-eq-context-lemma}
Suppose $p,p' \in \SP^0(\sigma_0,\ldots,\sigma_{r-1} \arrow \nat)$. 
Then $p \approx p'$ iff
\[ \forall q_0 \in \SP^0(\sigma_0), \ldots, q_{r-1} \in \SP^0(\sigma_{r-1}). ~~
   p \cdot q_0 \cdot \ldots \cdot q_{r-1} ~=~ p' \cdot q_0 \cdot \ldots \cdot q_{r-1} \;.  \]
\end{lem}

We shall also make use of the \emph{observational ordering} on $\SF$
and the associated preorder on $\SP$.
Let $\sqsubseteq$ be the usual information ordering on $\SF(\nat) \iso \N_\bot$,
and let us endow each $\SF(\sigma)$ with the partial order $\preceq$ defined by
\begin{eqnarray*}
    x \preceq x' & \mbox{iff} & \forall h \in \SF(\sigma\arrow\nat),\, n \in \N.~~
    h \cdot x ~\sqsubseteq~ h \cdot x' \;.  
\end{eqnarray*}
It is easy to see that the application operations $\cdot$ are monotone with respect to $\preceq$.
Moreover, Milner's context lemma also exists in an \emph{inequational} form which says, in effect, that
if $f,f' \in \SF(\sigma_0,\ldots,\sigma_{r-1}\arrow\nat)$ then
\begin{eqnarray*}
   f \preceq f' & \mbox{iff} & \forall y_0 \in \SF(\sigma_0),\ldots,y_{r-1} \in \SF(\sigma_{r-1}).~~
   f \cdot y_0 \cdot \ldots \cdot y_{r-1} ~\sqsubseteq f' \cdot y_0 \cdot \ldots \cdot y_{r-1} \;. 
\end{eqnarray*}
Thus, if elements of $\SF(\sigma\arrow\tau)$ are considered as functions $\SF(\sigma) \arrow \SF(\tau)$,
the partial order $\preceq$ coincides with the pointwise partial order on functions.

We write $\preceq$ also for the preorder on each $\SP^0(\sigma)$
induced by $\preceq$ on $\SF$: that is, $p \preceq p'$ iff $\theta(p) \preceq \theta(p')$.
Furthermore, we extend the use of the notations $\approx,\,\preceq$ in a natural way to open terms
(in the same environment), and indeed to meta-terms: 
e.g.\ we may write $\vec{x} \vdash P \preceq P'$ to mean
$\dang{\lambda \vec{x}.P} \;\preceq\; \dang{\lambda \vec{x}.P'}$.

Clearly $p \approx p'$ iff $p \preceq p' \preceq p$.
The following is also now immediate:

\begin{lem}  \label{nsp-ineq-context-lemma}
Suppose $p,p' \in \SP^0(\sigma)$ where $\sigma = \sigma_0,\ldots,\sigma_{r-1} \arrow \nat$. 
Then the following are equivalent:

\begin{enumerate}[label=(\roman*)]
\item $p \preceq p'$.
\item \( \forall r \in \SP^0(\sigma\arrow\nat).~ r \cdot p \sqsubseteq r \cdot p' \).
\item \( \forall q_0 \in \SP^0(\sigma_0), \ldots, q_{r-1} \in \SP^0(\sigma_{r-1}). ~~
       p \cdot q_0 \cdot \ldots \cdot q_{r-1} \,\sqsubseteq\, p' \cdot q_0 \cdot \ldots \cdot q_{r-1} \).
\end{enumerate}
\end{lem}

\noindent
We conclude this subsection by reformulating some of the major milestones in our proof
using the notation now available.
Specifically, in Sections~\ref{sec-denotations} and \ref{sec-Y-k+1} we will show the following:

\begin{thm}  \label{SP0-main-thm}
For any $k \geq 1$, the element $\sem{Y_{k+1}} \in \SP^0$ is not $\PCF^\Omega_k$-denotable
(whence neither is  $\sem{Y_{0 \arrow (k+1)}}$).
\end{thm}

In Section~\ref{sec-extensional} we will go on
to show that no $Z \approx \sem{Y_{0 \arrow (k+1)}}$ can be $\PCF^\Omega_k$-denotable,
establishing Theorem~\ref{SF-main-thm}(i).
In Section~\ref{sec-pure-type} we will resort to a more refined version of our methods 
to show the same for $Y_{k+1}$;
this will establish Theorem~\ref{SF-main-thm}(ii).

\subsection{The embeddability hierarchy}  \label{subsec-embed}

The following result will play a crucial role in this paper:

\begin{thm}[Strictness of embeddability hierarchy]  \label{no-retraction-thm}
In $\SF$, no type $\pure{k+1}$ can be a \emph{pseudo-retract}
of any finite product $\Pi_i \sigma_i$ where each $\sigma_i$ 
is of level $\leq k$. 
More formally, if $z$ is a variable of type $\pure{k+1}$ and each $x_i$ a variable of type $\sigma_i$,
there cannot exist procedures
\[  z \vdash t_i : \sigma_i \;, \hspace*{2.0em} \vec{x} \vdash r : \pure{k+1}  \]
such that $z \vdash {r[\vec{x} \mapsto \vec{t}\,]} \,\succeq z^\eta$.
\end{thm}

If in the above setting we had $z \vdash {r[\vec{x} \mapsto \vec{t}\,]} \,\approx z^\eta$,
we would call $\pure{k+1}$ a \emph{retract} of $\Pi_i \sigma_i$.
In Appendix A we will show that the notions of retract and pseudo-retract actually coincide,
since $z \vdash p \succeq z^\eta$ implies $z \vdash p \approx z^\eta$.
However, this fact will not be needed for the main results of this paper.

In our statement of Theorem~\ref{no-retraction-thm},
we have referred informally to a product $\Pi_i \sigma_i$ which we have
not precisely defined (although the formal statement of the theorem gives 
everything that is officially necessary).
One may readily make precise sense of this product notation within the
\emph{Karoubi envelope} $\KKar(\SF)$ as studied in \cite[Chapter~4]{Longley-Normann}:
for instance, it is not hard to show that any finite product of level $\leq k$ types can
be constructed as a retract of the pure type $\pure{k+1}$.
In the present paper, however, references to product types may be taken to be
purely informal and motivational.

The proof of Theorem~\ref{no-retraction-thm} appears in
\cite[Section~7.7]{Longley-Normann}, but because of its crucial role in the paper 
we reprise it here with some minor stylistic improvements.

\begin{proof}
By induction on $k$. 
For the case $k=0$, we note that 
$\nat\arrow\nat$ cannot be a pseudo-retract of any $\nat^r$, since (for example)
the set of maximal elements in $\SF(\nat\arrow\nat)$ is of larger cardinality than
the set of all elements of $\SF(\nat)^r$.
(Alternatively, one may note that $\nat\arrow\nat$ is not a \emph{retract} of $\nat^r$,
since the former contains strictly ascending chains of length $r+2$ while the latter does not;
then use the method of Appendix~A in the easy case $k=1$ to show that 
any pseudo-retraction of the relevant type would be a retraction.)

Now assume the result for $k-1$, and suppose 
for contradiction that $z \vdash t_i$ and $\vec{x} \vdash r$ exhibit $\pure{k+1}$
as a pseudo-retract of $\Pi_i \sigma_i$ where each $\sigma_i$ is of level $\leq k$.
Let $v =\, \dang{r[\vec{x} \mapsto \vec{t}\,]}$, 
so that $z \vdash v \succeq z^\eta$,
whence $\dang{v[z \mapsto u]}\, \succeq u$ for any $u \in \SP^0(k+1)$.
We first check that any $v$ with this latter property must have the syntactic form 
$\lambda f^k.\,\caseof{zp}{\cdots}$ for some $p$ of type $\pure{k}$.
Indeed, it is clear that $v$ does not have the form 
$\lambda f.n$ or $\lambda f.\bot$, and the only other alternative form
is $\lambda f.\,\caseof{fp'}{\cdots}$. 
In that case, however, we would have 
\[ \dang{v[z \mapsto \lambda w^k.0]} \cdot\; (\lambda y^{k-1}.\bot) ~=~ \bot \;, \]
contradicting
$\dang{v[z \mapsto \lambda w^k.0]} \cdot\; (\lambda y^{k-1}.\bot) \,\succeq\, (\lambda w.0)(\lambda y.\bot) = 0$.

We now focus on the subterm $p$ in $v = \lambda f^k.\,\caseof{zp}{\cdots}$.
The general direction of our argument will be to show that
$\lambda f^k.p$ represents a function of type $\pure{k} \arrow \pure{k}$
that dominates the identity, and that moreover our construction
of $v$ as $\dang{r[\vec{x} \mapsto \vec{t}\,]}$ can be used to split this into functions
$\pure{k} \arrow \Pi_j \rho_j$ and $\Pi_j \rho_j \arrow \pure{k}$ where the
$\rho_j$ are of level $\leq k-1$, contradicting the induction hypothesis.
An apparent obstacle to this plan is that $z$ as well as $f$ may appear
free in $p$; however, it turns out that we still obtain all the properties
we need if we specialize $z$ here (somewhat arbitrarily) to $\lambda w.0$.

Specifically, we claim that $\lambda f.\!\dang{p[z \mapsto \lambda w.0]} \,\succeq \id_k$.
By Lemma~\ref{nsp-ineq-context-lemma}, it will suffice to show that
$\dang{p[f \mapsto q, z \mapsto \lambda w.0]} \,\succeq q$ for any $q \in \SP^0(k)$.
The idea is that if it is not, then (ignoring the presence of $z$ in $p$ for now)
we may specialize $z$ to some $u$ that will detect the difference
between $p[f \mapsto q]$ and $q$, so that the subterm `$zp$' within $v$ will yield $\bot$,
contradicting that $z \vdash v \succeq z^\eta$.
We can even allow for the presence of $z$ in $p$ by a suitably careful choice of $u$.

Again by Lemma~\ref{nsp-ineq-context-lemma}, it suffices to show that
$\dang{p[f \mapsto q, z \mapsto \lambda w.0]} \cdot\; s \succeq q \cdot s$ for any $s$.
So suppose $q \cdot s = \lambda.n$ whereas 
$\dang{p[f \mapsto q, z \mapsto \lambda w.0]} \cdot\; s \neq \lambda.n$
for some $s \in \SP^0(k-1)$ and $n \in \N$.
Take $u = \lambda g.\,\caseof{gs}{n \darrow 0}$,
so that $u \cdot q' = \bot$ whenever $q' \cdot s \neq \lambda.n$.
Then $u \preceq \lambda w.0$ by Lemma~\ref{nsp-ineq-context-lemma},
so we have $\dang{p[f \mapsto q, z \mapsto u]} \cdot\; s \neq \lambda.n$
since $\lambda.n$ is maximal in $\SP^0(\nat)$.
By the definition of $u$, it follows that
$\dang{(zp)[f \mapsto q, z \mapsto u]}\, = \bot$, whence
$\dang{v[z \mapsto u]} \cdot\, q = \bot$, whereas $u \cdot q=0$,
contradicting $\dang{v[z \mapsto u]}\, \succeq u$.
This completes the proof that
$\lambda f.\!\dang{p[z \mapsto \lambda w.0]} \; \succeq \id_{k}$.

Next, we show how to split the function represented by this procedure
so as to go through some $\Pi_j \rho_j$ as above.
Since $\dang{r[\vec{x} \mapsto \vec{t}\,]} \,= \lambda f.\,\caseof{zp}{\cdots}$, we have that
$r[\vec{x}\mapsto\vec{t}\,]$ reduces in finitely many steps to a head normal form
$\lambda f.\,\caseof{zP}{\cdots}$ where $\dang{\!P\!}\,=p$.
By working backward through this reduction sequence, we may locate
the ancestor within $r[\vec{x}\mapsto\vec{t}\,]$ of this head occurrence of $z$.
Since $z$ does not appear free in $r$, this occurs within some $t_i$, and clearly it must 
appear as the head of some subterm $\caseof{zp'}{\cdots}$.
Now since $t_i$ has type $\sigma_i$ of level $\leq k$,
and $z:\pure{k+1}$ is its only free variable, 
it is easy to see that all \emph{bound}
variables within $t_i$ have pure types of level $<k$.
Let $x'_0,x'_1,\ldots$ denote the finitely many bound variables that are in scope
at the relevant occurrence of $zp'$, and suppose each $x'_j$ has type
$\rho_j$ of level $<k$.
By considering the form of the head reduction sequence
$r[\vec{x}\mapsto\vec{t}\,] \reducesto^*_h \lambda f.\,\caseof{zP}{\cdots}$,
we now see that $P$ has the form $p'[\vec{x}\,' \mapsto \vec{T}]$
where each $T_j: \rho_j$ contains at most $f$ and $z$ free.%
\footnote{The reader wishing to see a more formal justification for this step may consult the
proof of Lemma~\ref{g-lemma-1}(i) below.}

Writing $^*$ for the substitution $[z \mapsto \lambda w.0]$,
define procedures
\[ f^k \vdash t'_j  =\,  \dang{T_j^*} \,: \rho_j \;,  \hspace*{2.0em}
\vec{x}\,' \vdash r'  =\,  \dang{p'^*} \,: \pure{k} \;. \]
Then $\dang{r'[\vec{x}\,' \mapsto \vec{t}\,'\,]}$ coincides with the term 
$\dang{\lambda f.\,P^*} \,= \lambda f.\!\dang{\!p^*\!}$,
which dominates the identity as shown above.
Thus $\pure{k}$ is a pseudo-retract of $\Pi_j \rho_j$,
which contradicts the induction hypothesis.
So $\pure{k+1}$ is not a pseudo-retract of $\Pi_i \sigma_i$ after all,
and the proof is complete.
\end{proof}

As an aside, we remark that for several extensions of $\PCF$ studied in the literature,
the situation is completely different, in that the corresponding fully abstract and universal models
possess a \emph{universal} simple type $\upsilon$ of which all simple types are retracts.
It follows easily in these cases that one can indeed bound the type levels of recursion operators
without loss of expressivity. For example:

\begin{itemize}
\item In the language $\PCF+\por+\eexists$ considered by Plotkin \cite{LCF-considered},
the type $\nat\arrow\nat$ is universal, and the proof of this shows that every program in this
language is observationally equivalent to one in $\PCF_1 + \por+\eexists$.
(This latter fact was already noted in \cite{LCF-considered}.)
\item In $\PCF+\catch$ (a slight strengthening of Cartwright and Felleisen's language SPCF
\cite{Cartwright-Felleisen}), the type $\nat\arrow\nat$ is again universal, and again
the sublanguage $\PCF_1 + \catch$ has the same expressive power.
\item In the language $\PCF+H$ of Longley \cite{SRF}, the type $(\nat\arrow\nat)\arrow\nat$
is universal, but even here, all constants $Y_\sigma$ with $\lev(\sigma)>1$ are dispensable.
\end{itemize}

Further details of each the above scenarios may be found in \cite{Longley-Normann}.
These facts may offer some indication of why a `cheap' proof of our present results 
in the setting of pure $\PCF$ is not to be expected.%
\footnote{That PCF manifests greater structural complexity than many stronger languages
is also a moral of Loader's undecidability theorem for finitary PCF \cite{Loader-PCF}.
However, the complexity we explore here seems quite orthogonal to that exhibited by Loader:
we are concerned purely with `infinitary' aspects of definability, the entire finitary structure
being already represented by our $\PCF_0$.}

\subsection{Other sublanguages of PCF}  \label{subsec-power-pcf1}

Our main theorems establish a hierarchy of languages $\PCF_1 < \PCF_2 < \cdots\,$.
Before proceeding further, however, we pause to clarify the relationship between $\PCF_0$ and $\PCF_1$,
and also to survey some of the interesting territory that lies between them, 
in order to situate our theorems within a wider picture.

On the one hand, $\PCF_0$ is a rather uninteresting language.
As regards the elements of $\SF$ that it denotes, it is equivalent in expressivity to $\PCF_\bot$,
a variant of $\PCF_0$ in which we replace $Y_0$ by a constant $\bot$ (denoting $\bot \in \SF(\nat)$).
This is clear since $Y_0$ and $\bot$ are interdefinable: we have $\bot = Y_0(\lambda x.x)$,
and it is easy to see that $Y_0 = \lambda f. f\bot$ (in $\SF$).
By a syntactic analysis of the possible normal forms of type $\pure{1}$ in $\PCF_\bot$,
one can show that these are very weak languages that do not even define addition.

However, such an analysis is unnecessary for our purposes,
since there are more interesting languages that clearly subsume $\PCF_0$
but are known to be weaker than $\PCF_1$.
For instance, Berger \cite{Berger-min-recursion} considered the language $\sysT_0 +\mmin$,
where $\sysT_0$ (a fragment of G\"odel's System~T) is the $\lambda$-calculus with
first-order primitive recursion over the natural numbers:
\[  \num{0} ~: \nat \;,  \hspace*{2.0em} \suc ~: \N \arrow \N \;, \hspace*{2.0em}
    \rec_0 ~: \nat \arrow (\nat \arrow \nat \arrow \nat) \arrow (\nat \arrow \nat) \;, \]
and $\mmin$ is the classical \emph{minimization} (i.e.\ unbounded search)
operator of type $(\nat\arrow\nat)\arrow\nat$.
On the one hand, it is an easy exercise to define $\bot$ in $\sysT_0 + \mmin$,
and to define both $\rec_0$ and $\mmin$ in $\PCF_1$.
On the other hand, Berger showed that the $\PCF_1$-definable functional
$\Phi_0: (\nat\arrow\nat\arrow\nat) \arrow (\nat \arrow \nat)$ given by
\[ \Phi_0\;g\;n ~=~ g\;n\;(\Phi_0\;g\;(n+1))  \;, \]
is not expressible in $\sysT_0 + \mmin$.%
\footnote{Berger actually considered denotability in the Scott model, 
but his argument applies equally to $\SF$.}
As already indicated in Section~\ref{sec-intro},
this functional and its higher-type analogues will play a crucial role in the present paper.

This situation is revisited in \cite[Section~6.3]{Longley-Normann} from the perspective of substructures
of $\SP^0$. It is shown that $\sysT_0 + \mmin$, and indeed the whole of $\sysT + \mmin$, 
can be modelled within 
the substructure $\SP^{0,\lwf}$ of \emph{left-well-founded} procedures, whereas the above functional 
$\Phi_0$ is not representable by any such procedure; thus $\Phi_0$ is not expressible in $\sysT + \mmin$.
(The reader may wish to study these results and proofs before proceeding further, since they provide
simpler instances of the basic method that we will use in this paper.)
At third order, there are even `hereditarily total' functionals definable in $\PCF_1$ but not
by higher-type iterators, one example being the well-known \emph{bar recursion} operator
(see \cite{Longley-bar-recursion}).

Even weaker than $\sysT_0 + \mmin$ is the language of \emph{(strict) Kleene primitive recursion
plus minimization}, denoted by $\Klex^\smallmin$ in \cite{Longley-Normann};
this again subsumes $\PCF_0$.
It is shown in \cite{Longley-Normann} that the computational power of $\Klex^\smallmin$
coincides with that of computable \emph{left-bounded} procedures;
this is used to show, for example, that even $\rec_0$ is not computable in $\Klex^\smallmin$.
We find it reasonable to regard left-bounded procedures
as embodying the weakest higher-order computability notion of natural interest that is still Turing complete.

\section{Sequential procedures for \texorpdfstring{$\PCF_k$}{PCF k} terms}   \label{sec-denotations}

For the remainder of the paper, we take $k$ to be some fixed natural number greater than $0$.

In this section we give a direct inductive characterization of the $\PCF^\Omega_k$-denotable 
elements of $\SP$ by making explicit how our interpretation works for terms of $\PCF^\Omega_k$.
The first point to observe is that we may restrict attention to $\PCF^\Omega_k$ terms in 
\emph{long $\beta\eta$-normal form}: that is, terms in $\beta$-normal form in which every
variable or constant $z$ of type $\sigma_0,\ldots,\sigma_{r-1} \arrow \nat$ is fully applied
(i.e.\ appears at the head of a subterm $z N_0 \ldots N_{r-1}$ of type $\nat$).
Moreover, an inductive characterization of the class of such terms is easily given.

\begin{prop}  \label{long-beta-eta-prop} \hfill
\begin{enumerate}[label=(\roman*)]
\item A procedure $\Gamma \vdash p : \sigma$ is denotable by a $\PCF^\Omega_k$ term 
$\Gamma \vdash M: \sigma$ iff it is denotable by one in long $\beta\eta$-normal form.
\item The class of long $\beta\eta$-normal forms of $\PCF^\Omega_k$ is inductively generated by 
the following clauses:
\begin{enumerate}[label=(\arabic*)]
\item If $\Gamma \vdash N_i : \sigma_i$ is a normal form for each $i<r$ 
and $x^{\sigma_0,\ldots,\sigma_{r-1} \arrow \nat} \in \Gamma$, 
then $\Gamma \vdash x N_0 \ldots N_{r-1} : \nat$ is a normal form (note that $r$ may be $0$ here).
\item If $\Gamma, x^\sigma \vdash M : \tau$ is a normal form then so is $\Gamma \vdash \lambda x.M : \sigma \arrow \tau$.
\item The numeric literals $\Gamma \vdash \num{n} : \nat$ are normal forms.
\item If $\Gamma \vdash M : \nat$ is a normal form then so are $\Gamma \vdash \suc\;M : \nat$,
$\Gamma \vdash \pre\;M : \nat$ and $\Gamma \vdash C_f\,M : \nat$ for any $f: \N \parrow \N$.
\item If $\Gamma \vdash M : \nat$, $\Gamma \vdash N : \nat$ and $\Gamma \vdash P : \nat$ are
normal forms, then so is $\Gamma \vdash \ifzero\;M\;N\;P : \nat$.
\item If $\sigma = \sigma_0,\ldots,\sigma_{r-1} \arrow \nat$ is of level $\leq k$
and $\Gamma \vdash M : \sigma\arrow\sigma$ and $\Gamma \vdash N_i : \sigma_i$ are normal forms,
then $\Gamma \vdash Y_\sigma M N_0 \ldots N_{r-1} : \nat$ is a normal form.
\end{enumerate}
\end{enumerate}
\end{prop}

\begin{proof} \hfill

\begin{enumerate}[label=(\roman*)]
\item It is a well-known property of simply typed $\lambda$-calculi that every term $M$ 
is $\beta\eta$-equivalent to one in long $\beta\eta$-normal form: 
indeed, we may first compute the $\beta$-normal form of $M$
and then repeatedly apply the $\eta$-rule to expand any subterms that are not already fully applied.
Moreover, it is shown in \cite[Theorem~6.1.18]{Longley-Normann} that $\SP$ is a $\lambda\eta$-algebra,
so that if $\Gamma \vdash M =_{\beta\eta} M'$ then $\sem{M}_\Gamma = \sem{M'}_\Gamma$
in $\SP$. This establishes the claim.
\item This is clear from the fact that no application may be headed by a $\lambda$-abstraction and that
all occurrences of variables and constants must be fully applied.
\qedhere
\end{enumerate}
\end{proof}

\noindent
It follows that the class of $\PCF^\Omega_k$-denotable procedures may be generated inductively
by a set of clauses that mirror the above formation rules for long $\beta\eta$-normal $\PCF^\Omega_k$ terms.
We now consider each of these formation rules in turn in order to spell out the corresponding
operation at the level of NSPs. In Section~\ref{sec-Y-k+1} we will show that these operations cannot give rise
to \emph{$k\!+\!1$-spinal} procedures, from which it will follow that no $\PCF^\Omega_k$-denotable 
procedure can be $k\!+\!1$-spinal.

For the first three formation rules, the effect on NSPs is easily described:

\begin{prop}  \label{NSP-nf-prop} \hfill

\begin{enumerate}[label=(\roman*)]
\item If $\Gamma \vdash x N_0 \ldots N_{r-1} : \nat$ in $\PCF^\Omega$, then
\[ \sem{x N_0 \ldots N_{r-1}}_\Gamma ~=~ 
   \caseof{x \sem{N_0}_\Gamma \cdots \sem{N_{r-1}}_\Gamma}{j \darrow j} \;. \]
\item If $\Gamma \vdash \lambda x.M : \sigma \arrow \tau$ in $\PCF^\Omega$, then
$\sem{\lambda x.M}_\Gamma = \lambda x.\,\sem{M}_{\Gamma,x}$.
\item $\sem{\num{n}}_\Gamma = \lambda.n$.
\end{enumerate}
\end{prop}

\begin{proof} 
Part (i) is easy using the definition of $\sem{-}$ and Lemma~\ref{eta-properties},
and parts (ii) and (iii) are part of the definition of $\sem{-}$.
\end{proof}

As regards the formation rules for $\suc$, $\pre$, $C_f$ and $\ifzero$,
the situation is again fairly straightforward, although a little more machinery is needed:

\begin{defi}  \label{rightward-leaf-def} \hfill
\begin{enumerate}[label=(\roman*)]
\item The set of \emph{rightward} (occurrences of) \emph{numeral leaves} within
a term $t$ is defined inductively by means of the following clauses:
\begin{enumerate}[label=(\arabic*)]
\item A term $n$ is a rightward numeral leaf within itself.
\item Every rightward numeral leaf within $e$ is also one within $\lambda \vec{x}.e$.
\item Every rightward numeral leaf in each $e_i$ is also one in $\caseof{a}{i \darrow e_i}$.
\end{enumerate}
\item If $t$ is a term and $e_i$ an expression for each $i$, 
let $t[i \mapsto e_i]$ denote the result of replacing each rightward leaf occurrence $i$
in $t$ by the corresponding $e_i$.
\end{enumerate}
\end{defi}

\begin{lem}  \label{rightward-leaf-prop}
$\dang{\caseof{d}{i \darrow e_i}} \;= d[i \mapsto e_i]$
for any expressions $d,e_i$.
\end{lem}

\begin{proof}
For each $c \in \N$, define a `truncation' operation $-^{(c)}$ on expressions as follows:
\begin{eqnarray*} 
    n^{(c)} ~=~ n \;, ~~~~  \bot^{(c)} & = & \bot \;, \\
    \caseof{a}{i \darrow e_i}^{(0)} & = & \bot \;, \\
    \caseof{a}{i \darrow e_i}^{(c+1)} & = & \caseof{a}{i \darrow e_i^{(c)}} \;. 
\end{eqnarray*}
Then clearly $d = \bigsqcup_c d^{(c)}$ and 
$d[i \mapsto e_i] = \bigsqcup_c d^{(c)}[i \mapsto e_i]$.
Moreover, we may show by induction on $c$ that
\[ \dang{\caseof{d^{(c)}}{i \darrow e_i}} ~=~ d^{(c)}[i \mapsto e_i] \;. \]
The case $c=0$ is trivial since $d^{(0)}$ can only have the form $n$ or $\bot$.
For the induction step, the situation for $d=n,\bot$ is trivial, so let us suppose
$d = \caseof{a}{j \darrow f_j}$. Then
\begin{eqnarray*}
&  & \dang{\caseof{d^{(c+1)}}{i \darrow e_i}} \\
&=& \dang{\caseof{(\caseof{a}{j \darrow f_j^{(c)}})}{i \darrow e_i}} \\
&=& \caseof{a}{j \darrow\, \dang{\caseof{f_j^{(c)}}{i \darrow e_i}}} \\
&=& \caseof{a}{j \darrow (f_j^{(c)}[i \mapsto e_i])} \mbox{~~by the induction hypothesis} \\
&=& (\caseof{a}{j \darrow f_j^{(c)}})[i \mapsto e_i] \\
&=& d^{(c+1)}[i \mapsto e_i] \;.
\end{eqnarray*}
Since $\dang{\!-\!}$ is continuous, the proposition follows by taking the supremum over $c$.
\end{proof}

From this lemma we may now read off the operations on NSPs that correspond to clauses~4 and 5 
of Proposition~\ref{long-beta-eta-prop}(ii):

\begin{prop}  \label{NSP-arith-prop} \hfill
\begin{enumerate}[label=(\roman*)]
\item If $\Gamma \vdash M : \nat$ in $\PCF^\Omega$, then 
$\sem{C_f\,M}_\Gamma =  \sem{M}_\Gamma [i \mapsto f(i)]$
(understanding $f(i)$ to be $\bot$ when $i \not\in \dom\;f$); similarly for $\suc$ and $\pre$.
\item If $\Gamma \vdash M : \nat$, $\Gamma \vdash N : \nat$ and $\Gamma \vdash P : \nat$, then
$\sem{\ifzero\;M\;N\;P}_\Gamma = \sem{M}_\Gamma [0 \mapsto d,\, i+1 \mapsto e]$
where $\sem{N}_\Gamma = \lambda.d$ and $\sem{P}_\Gamma = \lambda.e$.
\end{enumerate}
\end{prop}

\begin{proof} \hfill
\begin{enumerate}[label=(\roman*)]
\item The definition of $\sem{-}$ yields 
\[ \sem{C_f\,M}_\Gamma ~=~
   \dang{\lambda.\,\caseof{\sem{M}_\Gamma}{i \darrow f(i)}} \, , \] 
and by Lemma~\ref{rightward-leaf-prop} this evaluates to $\sem{M}_\Gamma [i \mapsto f(i)]$.
Likewise for $\suc$ and $\pre$.
\item The definition of $\sem{-}$ yields 
\[ \sem{\ifzero\;M\;N\;P}_\Gamma ~=~ 
   \dang{\lambda.\,\caseof{\sem{M}_\Gamma}{0 \darrow d \mid i+1 \darrow e}} \, , \] 
and by Lemma~\ref{rightward-leaf-prop} this evaluates to
$\sem{M}_\Gamma [0 \mapsto d,\, i+1 \mapsto e]$.
\qedhere
\end{enumerate}
\end{proof}

\noindent
It remains to consider the formation rule involving $Y_\sigma$.
It will be convenient to regard the NSP for $Y M N_0 \ldots N_{r-1}$ as a result of
plugging some simpler NSPs together, in the sense indicated by the following definition.
Here and later, we shall follow the convention that Greek capitals $\Gamma,\Delta$ denote arbitrary
environments, while Roman capitals $Z,X,V$ denote lists of variables of type level $\leq k$.
(Of course, the idea of plugging can be formulated without any restrictions on types,
but we wish to emphasize at the outset that only pluggings at level $\leq k$
will feature in our proof.)

\begin{defi}[Plugging]  \label{plugging-def}
Suppose given the following data:
\begin{itemize}
\item a variable environment $\Gamma$,
\item a finite list $Z$ of `plugging variables' $z$ of level $\leq k$, 
disjoint from $\Gamma$,
\item a \emph{root expression} $\Gamma,Z \vdash e$,
\item a substitution $\xi$ assigning to each $z^\sigma \in Z$ a procedure $\Gamma,Z \vdash \xi(z) : \sigma$.
\end{itemize}
\noindent
In this situation, we define the \emph{($k$-)plugging} $\Pi_{\Gamma,Z}(e,\xi)$ 
(often abbreviated to $\Pi(e,\xi)$) 
to be the meta-term obtained from $e$ by repeatedly expanding variables $z \in Z$ to $\xi(z)$.
To formalize this, let $T^\circ$ denote the meta-term obtained from $T$ by replacing
each ground type subterm $z \vec{Q}$ (where $z \in Z$) by $\bot$.
We may now define, up to $\alpha$-equivalence,
\begin{eqnarray*} 
   \Pi^0(e,\xi) & = & e \;, \\
   \Pi^{m+1}(e,\xi) & = & \Pi^m(e,\xi)[z \mapsto \xi(z) \mbox{~for all $z \in Z$}] \;, \\
   \Pi(e,\xi) & = & \bigsqcup_m \Pi^m(e,\xi)^\circ \;, 
\end{eqnarray*}
where $\bigsqcup$ denotes supremum with respect to the syntactic order on meta-terms.
\end{defi}

It is easy to see that $\Pi_{\Gamma,Z}(e,\xi)$ is well-typed in environment $\Gamma$.
Note that some renaming of bound variables will typically be necessary
in order to realize $\Pi_{\Gamma,Z}(e,\xi)$ as a concrete term conforming to the no-variable-hiding condition;
we will not need to fix on any one particular way of doing this.

The operation on NSPs corresponding to clause~6 of Proposition~\ref{long-beta-eta-prop}(ii)
may now be described as follows:

\begin{prop}  \label{NSP-Y-prop}
Suppose that $\sigma = \sigma_0,\ldots,\sigma_{r-1} \arrow \nat$ is of level $\leq k$
and that $\Gamma \vdash Y_\sigma M N_0 \ldots N_{r-1}$ in $\PCF^\Omega_k$,
where $\sem{M}_\Gamma = \lambda z^\sigma.p = \lambda z^{\sigma} x_0^{\sigma_0} \cdots x_{r-1}^{\sigma_{r-1}}.\,e$ 
and $\sem{N_i}_\Gamma = q_i$ for each $i$.
Then 
\[ \sem{Y_\sigma M N_0 \ldots N_{r-1}}_\Gamma ~=~ \lambda. \dang{\Pi_{\Gamma,Z}(e,\xi)}
\]
where $Z = z,x_0,\ldots,x_{r-1}$, $\xi(z) = p$, and
$\xi(x_i) = q_i$ for each $i$.
\end{prop}

\begin{proof}
Note that in this instance of plugging, the repeated substitutions are needed only for the sake
of the term $\xi(z)$ which may contain $z$ free---only a single substitution step is needed for the
plugging variables $x_i$, since the $q_i$ contain no free variables from $Z$.
We may thus rewrite $\Pi_{\Gamma,Z}(e,\xi)$ as 
\[ \Pi_{\Gamma,\vec{x},Z'}(e,\xi')\;[\vec{x}\mapsto{q}] \,, \] 
where $Z'=\{z\}$ and $\xi'(z)=p$.
The proposition will therefore follow easily (with the help of Theorem~\ref{eval-thm}) once we know that
\[ \sem{Y_\sigma M}_\Gamma ~=~ 
   \lambda \vec{x}.\, \dang{\Pi_{\Gamma,\vec{x},Z'}(e,\xi')} \;. \]
To see this, write $Y_\sigma = \lambda g.F_\sigma[g]$ where 
$F_\sigma[g] = \lambda \vec{x}.\,\caseof{g\,(F_\sigma[g])\,\vec{x}^{\,\eta}}{i \darrow i}$ 
as at the start of Section~\ref{subsec-interp}.
Then clearly 
\[ \sem{Y_\sigma M}_\Gamma ~=~ (\lambda g. F_\sigma[g]) \cdot (\lambda z.p) ~=~
   \dang{F_\sigma[\lambda z.p]}  \;. \]
Here the meta-term $F_\sigma[\lambda z.p]$ is specified corecursively (up to $\alpha$-equivalence) by
\[ F_\sigma[\lambda z.p] ~=~ \lambda \vec{x}.\,\caseof{(\lambda z. p)\,(F_\sigma[\lambda z.p])\,\vec{x}^{\,\eta}}{i \darrow i} \]
That is, $F_\sigma[\lambda z.p]$ coincides with the meta-term $G = \bigsqcup_m G^m$, where
\[  G^0 ~=~ \bot_\sigma \;, ~~~~~~
    G^{m+1} ~=~ \lambda \vec{x}.\,\caseof{(\lambda z.p)\,G^m\,\vec{x}^{\,\eta}}{i \darrow i} \;. \]
We may now compare this with the meta-term $H = \bigsqcup_m H^m$, where
\[  H^0 ~=~ \bot_\sigma \;, ~~~~~~
    H^{m+1} ~=~ \lambda \vec{x}.\, e [z \mapsto H^m] \;. \]
Noting that $(\lambda z.p)\,G^m\,\vec{x}^{\,\eta} \reducesto e[z \mapsto G^m,\,\vec{x} \mapsto \vec{x}^{\,\eta}]$,
we have by Lemmas~\ref{eta-properties} and \ref{rightward-leaf-prop} that
\[ \dang{G^{m+1}} ~=~ \dang{\lambda \vec{x}.\, e[z \mapsto G^m]}  \]
whence by Theorem~\ref{eval-thm} and an easy induction we have $\dang{G^m} \,=\, \dang{H^m}$ 
for all $m$. Hence $\dang{\!G\!} \,=\, \dang{\!H\!}$.

Moreover, it is immediate from the definition that $H$ coincides with the meta-term
$\lambda \vec{x}.\, \Pi_{\Gamma,\vec{x},Z'}(e,\xi')$ mentioned earlier.
We thus have
\[ \sem{Y_\sigma M}_\Gamma ~=~ \dang{F_\sigma[\lambda z.p]} ~=~ \dang{\!G\!} ~=~ \dang{\!H\!} 
   ~=~ \lambda \vec{x}.\, \dang{\Pi_{\Gamma,\vec{x},Z'}(e,\xi')} \]
and the proof is complete.
(We have glossed over some fine details of variable renaming here, but these are easily attended to.)
\end{proof}

Combining Propositions~\ref{NSP-nf-prop}, \ref{NSP-arith-prop} and \ref{NSP-Y-prop} 
with Proposition~\ref{long-beta-eta-prop}, the results of this section may be summarized as follows.

\begin{thm}   \label{denotable-inductive-thm}
The class of $\PCF^\Omega_k$-denotable procedures-in-environment $\Gamma \vdash p$
is the class generated inductively by the following rules:
\begin{enumerate}[label=(\arabic*)]
\item If $\Gamma \vdash q_i$ is denotable for each $i<r$ and $x \in \Gamma$, then
\[ \Gamma \vdash \lambda.\,\caseof{x q_0 \ldots q_{r-1}}{j \darrow j} \] is denotable.
\item If $\Gamma,x \vdash p$ is denotable, then $\Gamma \vdash \lambda x.p$ is denotable.
\item Each $\Gamma \vdash \lambda.n$ is denotable.
\item If $\Gamma \vdash p$ is denotable and $f : \N \parrow \N$, then
$\Gamma \vdash p[i \mapsto f(i)]$ is denotable. (The constructions for $\suc$ and $\pre$
are special cases of this).
\item If $\Gamma \vdash p$,  $\Gamma \vdash \lambda.d$ and $\Gamma \vdash \lambda.e$ are denotable,
then $\Gamma \vdash p[0 \mapsto d, i+1 \mapsto e]$ is denotable.
\item If $\Gamma \vdash \lambda z^\sigma x_0^{\sigma_0} \cdots x_{r-1}^{\sigma_{r-1}}.e$ is denotable
where $\sigma = \sigma_0,\ldots,\sigma_{r-1}\arrow\nat$ is of level $\leq k$,
and $\Gamma \vdash q_i:\sigma_i$ is denotable for each $i<r$, then
\[ \Gamma \vdash \lambda. \dang{\Pi_{\Gamma,Z}(e,\xi)} \] 
is denotable, where $Z = z,\vec{x}$ is disjoint from $\Gamma$, 
$\xi(z) = \lambda \vec{x}.e$, and $\xi(x_i)=q_i$ for each $i$.
\end{enumerate}
\end{thm}

\noindent
To conclude this section,
we introduce a useful constraint on NSPs which, although not satisfied by all $\PCF^\Omega_k$-denotable
procedures, will hold for all those that we will need to consider in the course of our main proofs.
As we shall see, this constraint will interact well with the inductive rules just presented.

Referring back to the examples in Section~\ref{sec-intro}, we see that the recursive definitions of both $Y_{k+1}$ and $\Phi_{k+1}$ involved a variable $g$ of type level $k+2$.
It is therefore natural that our analysis will involve the consideration of terms in which such a variable $g$
appears free. However, it will turn out that apart from this one designated variable, our terms need never
involve any other variables of level $>k$, and this has a pleasant simplifying effect on our arguments.
This motivates the following definition:

\begin{defi}  \label{regular-def}
Suppose $g$ is a variable of type level $k+2$. 
\begin{enumerate}[label=(\roman*)]
\item An environment $\Gamma$ is \emph{($g$-)regular}
if $\Gamma$ contains $g$ but all other variables in $\Gamma$ are of type level $\leq k$.
\item A meta-term $T$ is regular if all free and bound variables within $T$ are of level $\leq k$,
except possibly for free occurrences of $g$.
\item A meta-term-in-environment $\Gamma \vdash T$ is regular if both $\Gamma, T$ are regular.
\end{enumerate}
\end{defi}

\noindent
There is a useful alternative characterization of regularity in the case of normal forms:

\begin{prop}  \label{regular-term-prop}
A term-in-environment $\Gamma \vdash t$ is regular iff $\Gamma$ is regular and $t$ is
not a procedure of type level $\geq k+2$.
\end{prop}

\begin{proof}
The left-to-right implication is trivial, since a procedure of level $\geq k+2$ would have the form
$\lambda \vec{x}.\cdots$ where at least one of the $x_i$ was of level $\geq k+1$.
For the converse, suppose $\Gamma$ is regular and $t$ is not a procedure of level $\geq k+2$.
Then $t$ contains no free variables of level $\geq k$ other than $g$, so we just need to show
that all variables bound by a $\lambda$-abstraction within $t$ are of level $\leq k$.
Suppose not, and suppose that $\lambda \vec{x}.e$ is some outermost subterm of $t$
with $\lev(\vec{x}) > k$. Then $\lambda \vec{x}.e$ cannot be the whole of $t$, 
since $t$ would then be a procedure of level $>k+1$. 
Since $t$ is a normal form, the subterm $\lambda \vec{x}.e$ (of level $>k+1$)
must therefore occur as an argument to some variable $w$
of level $> k+2$. But this is impossible, since $\Gamma$ contains no
such variables, nor can such a $w$ be bound within $t$, since the relevant subterm
$\lambda \vec{w}.d$ would then properly contain $\lambda \vec{x}.e$, contradicting
the choice of the latter.
\end{proof}

Let us now consider how the inductive clauses of Theorem~\ref{denotable-inductive-thm} may be used to 
generate \emph{regular} procedures-in-environment $\Gamma \vdash t$.
The following gives a useful property of derivations involving these clauses:

\begin{prop}  \label{regular-generation-prop}
If $\Gamma \vdash t$ is regular and $\PCF^\Omega_k$-denotable,
then any inductive generation of the denotability of $\Gamma \vdash t$ via the clauses of
Theorem~\ref{denotable-inductive-thm} will consist entirely of regular procedures-in-environment.
\end{prop}

\begin{proof}
It suffices to observe that for each of the six inductive clauses (regarded as rules), if the conclusion
is a regular procedure-in-environment then so are each of the premises.
For clause 1, this is clearly the case because the $q_i$ are subterms of the procedure in the conclusion.
For clause 2, we note that if $\lambda x.p$ is regular then so is $p$, and moreover $x$ has
level $\leq k$ so that $\Gamma,x$ is regular.
Clauses 3 and 4 are trivially handled.
For clause 5, we not that if $\Gamma \vdash p[0 \vdash d, i+1 \mapsto e]$ is regular
then $\Gamma \vdash p$, $\Gamma \vdash \lambda.d$ and $\Gamma \vdash \lambda.e$
are immediately regular by Proposition~\ref{regular-term-prop}
(regardless of whether any leaves $0$ or $i+1$ appear in $p$).
Likewise, for clause 6, we note that under the given hypotheses, both 
$\lambda z \vec{x}.e$ and each $q_i$ are of level $\leq k+1$; hence if $\Gamma$ is regular
then immediately $\Gamma \vdash \lambda z \vec{x}.e$ and $\Gamma \vdash q_i$ are
regular by Proposition~\ref{regular-term-prop}.
\end{proof}

In particular, let us consider again the construction of the procedure $Y_{k+1}$ as
$\lambda g.F_{k+1}[g]$, where
\[ F_{k+1}[g] ~=~ \lambda x^k.\;\caseof{g\,(F_{k+1}[g])\,x^\eta}{i \darrow i} \;. \]
It is clear by inspection that $g \vdash F_{k+1}[g]$ is regular;
hence, if it were $\PCF^\Omega_k$-denotable, then Proposition~\ref{regular-generation-prop} would apply.
We shall show, however, that a purely regular derivation via the clauses of 
Theorem~\ref{denotable-inductive-thm} cannot generate `spinal' terms such as $F_{k+1}[g]$;
hence $g \vdash F_{k+1}[g]$ is not $\PCF^\Omega_k$-denotable.
This will immediately imply that $Y_{k+1}$ itself is not $\PCF^\Omega_k$-denotable
(Theorem~\ref{SP0-main-thm}),
since the only means of generating non-nullary $\lambda$-abstractions is via clause 2 of Theorem~\ref{denotable-inductive-thm}.

\section{\texorpdfstring{$\PCF^\Omega_k$-denotable}{PCF Omega k denotable} procedures are non-spinal}  \label{sec-Y-k+1}

In this section, we will introduce the crucial notion of a \emph{($k\!+\!1$-)spinal term}, and
will show that the clauses of Theorem~\ref{denotable-inductive-thm} (in the regular case)
are unable to generate spinal terms from non-spinal ones. 
Since the procedure $F_{k+1}[g]$ will be easily seen to be spinal, this will 
establish Theorem~\ref{SP0-main-thm}.

More specifically, we will actually introduce the notion of a \emph{$g$-spinal term}, 
where $g$ is a free variable which we treat as fixed throughout our discussion.
We shall do this first for the case 
\[ g ~:~ (k+1) \arrow (k+1) \] 
as appropriate for the analysis of $F_{k+1}[g]$ and hence of $Y_{k+1}$.
Later we will also consider a minor variation for $g$ of type $\nat \arrow (k+1) \arrow (k+1)$,
as appropriate to the definition of $\Phi_{k+1}$ given in Section~\ref{sec-intro}.
In both cases, we shall be able to dispense with the term `$k\!+\!1$-spinal', since the type level $k+1$ 
may be read off from the type of $g$.

Some initial intuition for the concept of spinality was given in Section~\ref{sec-intro}.
We now attempt to provide some further motivation by examining a little more closely
the crucial difference between $Y_{k+1}$ and $Y_k$ (say) that we are trying to capture.

The most obvious difference between these procedures is that
$Y_{k+1}$ involves an infinite sequence of nested calls to a variable $g: (k+1) \arrow (k+1)$,
whereas $Y_k$ does not.
One's first thought might therefore be to try and show that no procedure
involving an infinite nesting of this kind can be constructed using the means at our disposal
corresponding to $\PCF^\Omega_k$ terms.

As it stands, however, this is not the case.
Suppose, for example, that $\up_k: {k} \arrow {k+1}$ and
$\down_k: {k+1} \arrow {k}$ are $\PCF_0$ terms defining a
standard retraction ${k} \lhd {k+1}$. 
More specifically, let us inductively define
\[ \begin{array}{rclcrcl}
\up_0   & = & \lambda x^0.\lambda z^0.x \;, & &
\down_0 & = & \lambda y^1.\,y\,\num{0} \;, \\
\up_{k+1}   & = & \lambda x^{k+1}.\lambda z^{k+1}.\,x(\down_k\;z) \;, & &
\down_{k+1} & = & \lambda y^{k+2}.\lambda w^{k}.\,y(\up_k\;w) \;.
\end{array} \]
Now consider the $\PCF_k$ program 
\[ Z_{k+1} ~=~ 
   \lambda g: (k+1) \arrow (k+1).\; \up_k\,(Y_k\,(\down_k \circ g \circ \up_k)) \;. \]
This is essentially just a representation of $Y_k$ modulo our encoding of type $k$ in type $k+1$.
A simple calculation shows that the NSPs for $Y_{k+1}$ and $Z_{k+1}$ are
superficially very similar in form, both involving an infinite sequence
of nested calls to $g: (k+1) \arrow (k+1)$.
(These NSPs are shown schematically in Figure~1 for the case $k=2$.)
We will therefore need to identify some more subtle property of NSPs that differentiates
between $Y_{k+1}$ and $Z_{k+1}$.

\begin{figure}  \label{Y_Z_NSPs}
\begin{center}
\includegraphics[scale=.60]{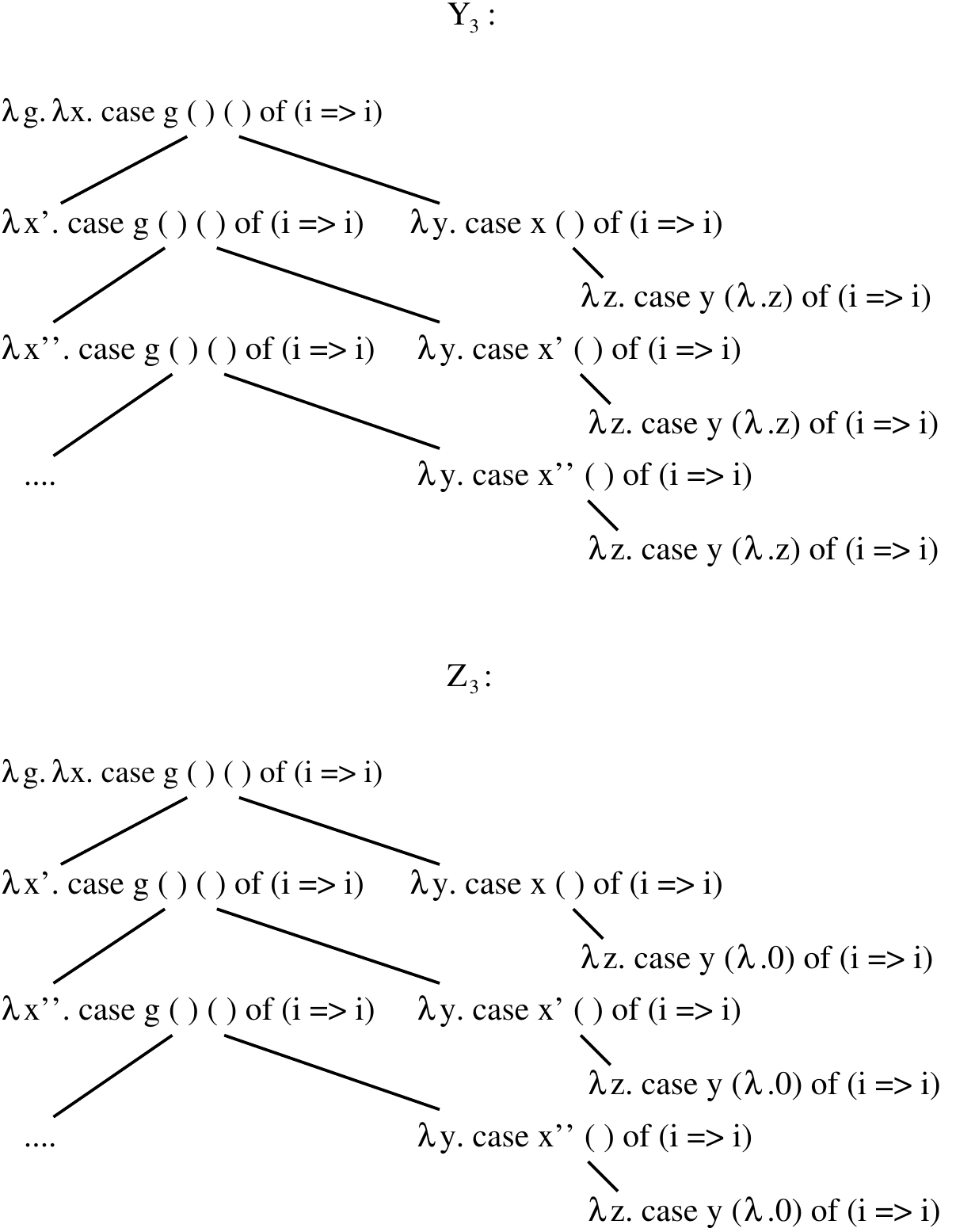}
\end{center}
\caption{The NSPs for $Y_3$ and $Z_3$. Here $\lambda.z$ abbreviates
$\lambda.\,\caseof{z}{i \darrow i}$.}

\end{figure}

The intuitive idea will be that in the NSP for $Z_{k+1}$, the full potency of 
$g$ as a variable of type ${k+1} \arrow {k+1}$ is not exploited,
since both the input and output of $g$ are `funnelled' through the simpler type ${k}$. 
Such funnelling will inevitably entail some loss of information, 
as Theorem~\ref{no-retraction-thm} tells us that the type ${k}$ cannot
fully represent the structure of the type ${k+1}$.
A useful mental picture here is that of a $(k+1)$-dimensional space being `flattened' down to a $k$-dimensional one.
Broadly speaking, then, we shall want to define a $g$-spinal term to be one containing an infinite sequence
of nested calls to $g$ but with no essential `flattening' of the arguments.
It will then be the case that $Y_{k+1}$ is $g$-spinal, but $Z_{k+1}$ is not.

We now approach the formal definition of a $g$-spinal term, 
generalizing the structure exhibited by the terms $F_{k+1}[g]$.
To get our bearings, let us examine the form of these terms one more time.
Note that $F_{k+1}[g]$ has the form $\lambda x^k. H[g,x]$, where
\[  H[g,x] ~=~ \caseof{g(\lambda x'.H[g,x']) x^\eta}{\cdots} \;. \]
Spinal expressions of this kind, in which the topmost $g$ of the spine appears at the very head of
the expression, will be referred to as \emph{head-spinal}.
In fact, we shall say that $H[g,x]$ is head-spinal with respect to the variable $x$, since as noted above,
it is significant here that $x$ is passed to $g$ with no `flattening' (in the form of the procedure $x^\eta$).
As a first attempt, then, one might hazard that we should define a concept of head-spinality relative to
a type $k$ variable coinductively as follows: an expression $e$ is $g$-head-spinal w.r.t.\ $x$ if it is of the
form
\[ \caseof{g(\lambda x'.e')x^\eta}{\cdots} \]
where $e'$ is itself $g$-head-spinal w.r.t.\ $x$.

In fact, in order for the set of non-spinal terms to have appropriate closure properties,
we shall need to relax this definition in two ways.
Firstly, we allow $\lambda x'.e'$ to be replaced by $\lambda x'.E[e']$ where $E[-]$ is any expression context:
that is, we allow the head-spinal subterm $e'$ to appear at positions other than the head of this procedure.
Secondly, we allow $x^\eta$ to be replaced by a procedure term $o$ that can be \emph{specialized} to
(something close to) $x^\eta$: the intuition is that any such $o$ will embody the whole content of $x$ 
with no flattening.

This leads us, at last, to the following definition. Note that this makes reference to the technical notion of
an \emph{$x,V$-closed substitution}, the explanation of which we shall defer to Definition~\ref{xV-closed-def} below. This entails that the notion of head-spinality needs to be defined relative to a certain set $V$ of
variables as well as a type $k$ variable $x$.
We shall adopt the convention that any environment denoted by $\Gamma$ will be $g$-regular
(and hence will contain $g$); recall that Roman letters such as $V,X,Z$ 
always denote lists of variables of level $\leq k$ (which may also contribute to the
environments we consider).

\begin{defi}[Spinal terms]  \label{spinal-def}
Suppose $g$ has type $(k+1) \arrow (k+1)$.
Suppose $\Gamma \vdash e$ is $g$-regular, 
and that $x^k \in \Gamma$ and $V \subseteq \Gamma$.
\begin{enumerate}[label=(\roman*)]
\item In this situation, we coinductively declare $e$ to be \emph{$g$-head-spinal with respect to $x,V$} 
iff $e$ has the form
\[  \caseof{g(\lambda {x'}. E[e']) o}{\cdots} \]
where $E[-]$ is an expression context, and
\begin{enumerate}[label=(\arabic*)]
\item for some $x,\!V$-closed substitution $^\circ$ covering the free variables of $o$ other than $x$,
we have $o^\circ \succeq x^\eta$,
\item $e'$ is $g$-head-spinal with respect to $x',V'$, where $V'$ is the local variable environment for $E[-]$.
(Clearly $e'$ will automatically be $g$-regular in some $\Gamma'$ that contains both $x'$ and $V'$.)
\end{enumerate}
In other words, we take `$e$ is $g$-head-spinal w.r.t.\ $x,V$' to be the largest relation
that satisfies the above statement.

\item In the above setting, we may also refer to the application $g(\lambda {x'}. E[e']) o$ itself
as $g$-head-spinal w.r.t.\ $x,V$. 

\item We say a term $t$ is \emph{$g$-spinal} if it contains a subexpression that is $g$-head-spinal
w.r.t.\ some $x,V$.
\end{enumerate}
\end{defi}

\noindent
Whilst this definition makes use of local variable environments which in principle pertain to concrete terms,
it is easily seen that the notion of $g$-spinal term is $\alpha$-invariant.
Since we are taking $g$ to be fixed throughout the discussion, we will usually omit mention of it
and speak simply of spinal and head-spinal terms, and of regular (meta-)terms and environments.

In condition~1, one might have expected to see
$o^\circ \approx x^\eta$, but it turns that the argument goes through most smoothly
with $\succeq$ in place of $\approx$.
In Appendix~A we will see that $o^\circ \succeq x^\eta$
is actually equivalent to $o^\circ \approx x^\eta$, 
although this is somewhat non-trivial to show and is not needed for our main proof.

It remains to define the notion of an $x,V$-closed substitution.
Suppose that $^\circ = [\vec{w} \mapsto \vec{r}\,]$ is some substitution proposed for use in
condition 1 of Definition~\ref{spinal-def}(i).
Since we are wishing to compare $o^\circ$ with $x^\eta$, it is natural to require that the $\vec{r}$
contain no free variables other than $x$.
However, what we want to ensure here is intuitively that the whole unflattened content of $x$ is present
in $o$ itself rather than simply being introduced by the substitution.
This can be ensured if we allow $x$ as a free variable only in procedures $r_i$ of type level $<k$:
such procedures can only introduce `flattened' images of $x$, since the $x$ is here being funnelled through
a type of level $\leq k-1$.

For technical reasons, we furthermore need to restrict such occurrences of $x$ to those $r_i$
substituted for variables $w_i$ in a certain set $V$, 
which in practice will consist of variables
bound between one spinal occurrence of $g$ and the next 
(as can be seen from the specification of $V'$ above).
The necessity for the set $V$ is admittedly difficult to motivate at this point:
it is simply what the details of the proof seem to demand
(see the last page of the proof of Lemma~\ref{g-lemma-2}).

\begin{defi}  \label{xV-closed-def}
If $x$ is a variable of type $k$ and $V$ a set of variables, 
a substitution $^\circ = [\vec{w} \mapsto \vec{r}\,]$ is called
\emph{$x,\!V$-closed} if the $r_i$ contain no free variables, except that if $w_i \in V$
and $\lev(w_i)<k$ then $r_i$ may contain $x$ free.
\end{defi}

It is worth remarking that if we were only interested in
showing the non-definability of $Y_{k+1}$ as an element of $\SP^0$, one
could do without the notion of $x,\!V$-closedness altogether, 
and more simply require in Definition~\ref{spinal-def} that $^\circ$ is closed 
(and moreover that $\dang{\!o^\circ\!}\, = x^\eta$ on the nose).
The weaker definition we have given is designed with the proof of non-definability 
in $\SF$ in mind: we will be able to show in Section~\ref{sec-extensional} that every (simple) procedure 
representing the functional $\Phi_{k+1} \in \SF$ is spinal in this weaker sense.%
\footnote{It can be shown using Theorem~\ref{no-retraction-thm} that if $o^\circ \succeq x^\eta$
where $^\circ$ is $x,\!V$-closed, then at least one $x$ in $\dang{\!o^\circ\!}$ must originate from
$o$ rather than from $^\circ$.
We have not actually settled the question of whether there are procedures $o$ such that
$o^\circ \succeq x^\eta$ for some $x,\!V$-closed $^\circ$ but not for any closed $^\circ$;
fortunately this is not necessary for the purpose of our proof.}

We now digress briefly to explain the small modification of this machinery 
that we will need in Section~\ref{sec-extensional}.
Since our purpose there will be to analyse the functional $\Phi_{k+1}$ which we defined
in Section~\ref{sec-intro}, we shall be working in a setup in which
the global variable $g$ has the slightly different type ${0}\arrow(k+1)\arrow(k+1)$.
In this setting, we may vary the above definition
by coinductively declaring $e$ to be $g$-head-spinal w.r.t.\ $x,V$ iff $e$ has the form
\[  \caseof{gb(\lambda x'. E[e']) o}{\cdots} \]
where $b$ is a procedure term of type ${0}$ and conditions~1 and 2 above are also satisfied.
Subject to this adjustment, all the results and proofs of the present section go through
in this modified setting, with the extra argument $b$ playing no active role.
For the remainder of this section, we shall work with
a global variable $g$ of the simpler type $(k+1) \arrow (k+1)$, on the understanding that
the extra arguments $b$ can be inserted where needed to make formal
sense of the material in the modified setting.
We do not expect that any confusion will arise from this.

Clearly $g \vdash F_{k+1}[g]$ is spinal. 
The main result of this section will be that every $\PCF^\Omega_k$-denotable procedure
$\Gamma \vdash p$ is non-spinal (Theorem~\ref{no-gremlin-thm}).
We shall establish this by induction on the generation of denotable terms
as in Theorem~\ref{denotable-inductive-thm}, the only challenging case being the one for rule~6,
which involves plugging.
Here we require some technical machinery whose purpose is to show that 
if the result of a plugging operation is spinal,
then a spinal structure must already have been present in one of the components of the plugging:
there is no way to `assemble' a spinal structure from material in non-spinal fragments.

The core of the proof will consist of some lemmas developing the machinery 
necessary for tackling rule 6.
We start with some technical but essentially straightforward facts concerning evaluation and 
the tracking of subterms and variable substitutions. 

\begin{lem} \label{g-lemma-1}
Suppose that
\[ \Gamma \;\vdash\; \dang{K[d]} \;=_\alpha\; K'[c] \]
where $K[-],K'[-]$ are concrete meta-term contexts with local environments $\vec{v},\vec{v}\,'$ respectively,
and $\Gamma,\vec{v} \vdash d = \caseof{gpq}{\cdots}$,
$\Gamma,\vec{v}\,' \vdash c = \caseof{gp'q'}{\cdots}$ are concrete expressions.
Suppose also that:
\begin{enumerate}[label=(\arabic*)]
\item $\Gamma \vdash K[d]$ is regular,
\item in the evaluation above, the head $g$ of $c$ originates from that of $d$.
\end{enumerate}
Then:
\begin{enumerate}[label=(\roman*)]
\item There is a substitution $^\dag = [\vec{v}\mapsto\vec{s}\,]$ of level $\leq k$
arising from the $\beta$-reductions in the above evaluation,
with $\Gamma,\vec{v}\,' \vdash \vec{s}$ regular, such that
$\Gamma,\vec{v}\,' \vdash gp'q' =_\alpha \,\dang{\!(gpq)^\dag\!}$,
whence $\dang{\!d^\dag\!}$ is of form $\caseof{gp'q'}{\cdots}$ up to $=_\alpha$.%
\footnote{Note that although both $\dang{\!d^\dag\!}$ and $c$ have the form $\caseof{gp'q'}{\cdots}$,
they will in general have different case-branches, for instance when $K[-]$ is of the form
$\caseof{-}{\cdots}$.}

\item If furthermore $c$ is head-spinal w.r.t.\ some $x,V$,
then also $\dang{d^\dag}$ is head-spinal w.r.t.\ $x,V$.

\item If $K[-]$ contains no $\beta$-redexes $P \vec{Q}$ with $P$ of type level $k+1$,
then $^\dag$ is \emph{trivial for level $k$ variables}: 
that is, there is an injection $\iota$ mapping each level $k$ variable $v_i \in \vec{v}$ to
a variable $\iota(v_i) \in \vec{v}\,'$ such that $s_i = \iota(v_i)^\eta$.
\end{enumerate}
\end{lem}

\noindent
In reference to part~(iii), recall that substitutions $v \mapsto v^\eta$ have no effect on the meaning
of a term, as established by Lemma~\ref{eta-properties}.
Note that the environments $\vec{v},\vec{v}\,'$, and hence the injection $\iota$, will in general
depend on the concrete choice of $K[d]$ and $K'[c]$. However, for the purpose of proving the theorem,
it is clearly harmless to assume that $K'[c]$ is, on the nose, the concrete term obtained by evaluating $K[d]$.
In this case, we will see from the proof below that each $\iota(v_i)$ will be either $v_i$ itself or a renaming
of $v_i$ arising from the evaluation.

\begin{proof} \hfill
\begin{enumerate}[label=(\roman*)]
\item We first formulate a suitable property of terms that is preserved under all individual reduction steps.
Let $K[-],d,p,q$ and $\vec{v}$ be fixed as above, and suppose that
\[ K^0[\caseof{gP^0Q^0}{\cdots}] ~\reducesto~ K^1[\caseof{gP^1Q^1}{\cdots}] \]
via a single reduction step,
where the $g$ on the right originates from the one on the left, 
and moreover $K^0,P^0,Q^0$ enjoy the following properties 
(we write $\vec{v}\,^0$ for the local environment for $K^0[-]$):
\begin{enumerate}[label=(\arabic*)]
\item $\Gamma \vdash K^0[\caseof{gP^0Q^0}{\cdots}]$ is regular.
\item There exists a substitution $^{\dag0} = [\vec{v}\mapsto\vec{s}\,^0]$
(with $\Gamma,\vec{v}\,^0 \vdash \vec{s}\,^0$ regular) 
such that $\dang{gP^0Q^0} \,=_\alpha\, \dang{(gpq)^{\dag 0}}$.
\end{enumerate}

We claim that $K^1,P^1,Q^1$ enjoy these same properties w.r.t.\ the
local environment $\vec{v}\,^1$ for $K^1[-]$.
For property~1, clearly $K^1[\caseof{gP^1Q^1}{\cdots}]$ cannot contain
variables of level $>k$ other than $g$, because $K^0[\caseof{gP^0Q^0}{\cdots}]$ does not.
For property~2, we define the required substitution $^{\dag1} = [\vec{v}\mapsto\vec{s}\,^1]$
by cases according to the nature of the reduction step:
\begin{itemize}
\item If the subexpression $\caseof{gP^0Q^0}{\cdots}$ is unaffected by the
reduction (so that $P^0=P^1$ and $Q^0=Q^1$),
or if the reduction  is internal to $P^0,Q^0$ or to the rightward portion $(\cdots)$,
or if the reduction has the form
\begin{eqnarray*}
\caseof{(\caseof{gP^0Q^0}{i \darrow E^0_i})}{j \darrow F_j} & \reducesto & \\
\caseof{gP^0Q^0}{i \darrow \caseof{E^0_i}{j \darrow F_j}} 
\end{eqnarray*}
then the conclusion is immediate, noting that $\vec{v}\,^1 = \vec{v}\,^0$ 
and taking $^{\dag 1}=\,^{\dag 0}$.
\item If the reduction is for a $\beta$-redex $(\lambda \vec{x}.E)\vec{R}$
where the indicated subexpression $\caseof{gP^0Q^0}{\cdots}$ lies within some $R_i$,
we may again take $^{\dag 1}$ to be $^{\dag 0}$, with obvious adjustments to compensate for
any renaming of bound variables within $R_i$ or $\vec{s}\,^0$.
In this case $\vec{v}\,^1$ may contain more variables than $\vec{v}\,^0$,
but we will still have that $\Gamma,\vec{v}\,^1 \vdash \vec{s}\,^1$
once these renamings have been effected.
\item If the reduction is for a $\beta$-redex $(\lambda \vec{x}.E)\vec{R}$
where $\caseof{gP^0Q^0}{\cdots}$ lies within $E$,
then $P^1 = P^0[\vec{x} \mapsto \vec{R}\,']$ and $Q^1 = Q^0[\vec{x} \mapsto \vec{R}\,']$
for some $\vec{R}\,' =_\alpha \vec{R}$.
In this case, the local environment $\vec{v}\,^1$ for $K^1[-]$ will be $\vec{v}\,^0 - \vec{x}$
(perhaps modulo renamings of the $v^0_i$),
so that the conclusion follows if we take $^{\dag 1} = [\vec{v}\mapsto\vec{s}\,^1]$ 
where $s_i^1 = \dang{s_i^0 [\vec{x}\mapsto\vec{R}\,']}$ for each $i$
(modulo the same renamings).
Note here that $\Gamma,\vec{v}\,^1 \vdash \vec{s}\,^1$ is regular since $\vec{R}$ is regular
by condition~1 of the hypothesis.
\end{itemize}

Now in the situation of the lemma we will have some 
finite reduction sequence
\[  K[\caseof{gpq}{\cdots}] ~\reducesto^*~ K''[\caseof{gP'Q'}{\cdots}] \;, \]
where, intuitively, $K''[-]$ is fully evaluated down as far as the hole.
More formally, there is a finite normal-form context $t[-] \sqsubseteq K''[-]$
containing the hole in $K''[-]$ such that $t[-] \sqsubseteq K'[-]$;
from this we may also see that $\dang{\!P'\!}\,=p'$, $\dang{\!Q'\!}\,=q'$
and $\dang{\!K''[-]\!}\,=K'[-]$. 
Moreover, we now see that $K,p,q$ themselves trivially satisfy the above invariants
if we take $^\dag = [\vec{v} \mapsto \vec{v}\,^\eta]$ (Lemma~\ref{eta-properties} is used here).
We therefore infer by iterating the argument above that $K'',P',Q'$ also satisfy these invariants with respect to 
some $^\dag = [\vec{v} \mapsto \vec{s}\,]$ with $\Gamma,\vec{v}\,' \vdash \vec{s}$ regular.
(The environment $\Gamma,\vec{v}\,'$ is correct here, as $K'[-],K''[-]$ have the same
local environment.)
We now have $gp'q' =\, \dang{gP'Q'} \,=_\alpha\, \dang{(gpq)^\dag}$.
That the $\vec{v}$ are of level $\leq k$ is automatic, because $K[d]$ is regular.
It also follows immediately that $\dang{\!d^\dag\!}$ has the stated form.

\item If $c$ is head-spinal w.r.t.\ $x,V$, then we see from Definition~\ref{spinal-def} that $gp'q'$ 
and hence $\dang{d^\dag} \,= \caseof{gp'q'}{\cdots}$ are head-spinal w.r.t.\ $x,V$.

\item From the proof of (i), we see that in the reduction of $K[\caseof{gpq}{\cdots}]$
to $K''[\caseof{gP'Q'}{\cdots}]$, any $v_i \in \vec{v}$ 
can be tracked through the local environments for the intermediate contexts $K^0[-],K^1[-],\ldots$
until (if ever) it is a substitution variable for a $\beta$-reduction.
For those $v_i$ that never serve as such a variable, it is clear from the construction that
$v_i$ gives rise to some variable $\iota(v_i) \in \vec{v}\,'$ (either $v_i$ itself or a renaming thereof),
and that $s_i = v_i^\dag = \iota(v_i)^\eta$.
We wish to show that all $v_i \in \vec{v}$ of level $k$ are in this category.

Recalling that $\vec{v}$ is the local environment for $K[-]$,
any $v_i \in \vec{v}$ of level $k$ is bound by the leading $\lambda$ of some
meta-procedure $P$ within $K[-]$ of level $k+1$. 
By hypothesis, this $P$ does not occur in operator position; nor can it occur
as an argument to another $\lambda$-abstraction within $K[-]$, 
since this would require a bound variable of level $\geq k+1$.
It must therefore occur as a level $k+1$ argument to $g$,
so that we have a subterm $g(\lambda v_i.E[-])\cdots$.
But this form of subterms is stable under reductions, since $g$ is a global variable;
it follows easily that this subterm has a residual $g(\lambda v_i'.E'[-])\cdots$ in each of the intermediate
reducts (where $v_i'$ is either $v_i$ or a renaming thereof),
and thus that $v_i$ and renamings thereof never serve as substitution variables for $\beta$-reductions.
\qedhere
\end{enumerate}
\end{proof}

\noindent
Thus, in the setting of the above lemma, 
if $c$ is head-spinal then $d$ can be specialized and evaluated to yield a head-spinal term via the substitution
$[\vec{v} \mapsto \vec{s}\,]$.
However, we wish to show more, namely that in this setting, $d$ itself is already a spinal term,
so that the $\vec{s}$ make no essential contribution to the spinal structure.
(This will give what we need in order to show that $k$-pluggings cannot manufacture spinal terms
out of non-spinal ones.) 
This is shown by the next lemma, whose proof forms the
most complex and demanding part of the entire argument.
The main challenge will be to show that all the head-spinal occurrences of $g$ in
$\dang{d[\vec{v} \mapsto \vec{s}\,]}$ originate from $d$ rather than from $\vec{s}$.
The reader is advised that great care is needed as regards which variables can appear free where,
and for this reason we shall make a habit of explicitly recording the variable environment 
for practically every term or meta-term that we mention.

\begin{lem}  \label{g-lemma-2}
Suppose we have regular terms
\[ \Gamma,\vec{v} \;\vdash~ d ~=~ \caseof{gpq}{\cdots} \;, ~~~~~~
   \Gamma,\vec{v}\,' \vdash \vec{s} \;,  ~~~~~~
   \lev(\vec{v}), \lev(\vec{v}\,') \leq k \;, \] 
where $\Gamma,\vec{v}\,' \vdash\, \dang{d[\vec{v}\mapsto\vec{s}\,]}$ 
is head-spinal with respect to some $x,V$.
Then $d$ itself is spinal.
\end{lem}

\begin{proof}
We begin with some informal intuition.
The term $\dang{d[\vec{v}\mapsto\vec{s}\,]}$, being head-spinal, will be of the form
\[ \Gamma,\vec{v}\,' \;\vdash~ t ~=~ 
    \caseof{g\,(\lambda x'.\,E[\caseof{gF'o'}{\cdots}])\,o}{\cdots} \;, \]
where ${o'}^\circ \succeq {x'}^\eta$ for some $^\circ$ (and likewise for $o$ and $x$).
Here the head $g$ of $t$ clearly originates from that of $d$;
likewise, the $\lambda x'$ originates from the leading $\lambda$ of $p$ within $d$, 
rather than from $\vec{s}$.
Suppose, however, that the second displayed spinal occurrence of $g$ in $t$
originated from some $s_i$ rather than from $d$. 
In order to form the application of this $g$ to $o'$, 
the whole content of ${x'}^\eta$ would in effect need to be passed in to $s_i$
when $d$ and $\vec{s}$ are combined. But this is impossible, since the arguments to
$s_i$ are of level $<k$, so by Theorem~\ref{no-retraction-thm} 
we cannot funnel the whole of ${x'}^\eta$ through them:
that is, the interface between $d$ and $\vec{s}$ is too narrow 
for the necessary interaction to occur.
(The situation is made slightly more complex by the fact that 
some components of $^\circ$ might also involve $x'$, but the same idea applies.)
It follows that the second spinal $g$ in $t$ originates from $d$ after all.
By iterating this argument, we can deduce that all the spinal occurrences of $g$,
and indeed the entire spinal structure, comes from $d$.

We now proceed to the formal proof.
By renaming variables if necessary,
we may assume for clarity that the same variable is never bound in two places within the entire
list of terms $d,\vec{s}$,
and that all bound variables within $d,\vec{s}$ are distinct from those of $\vec{v}$ and $\vec{v}\,'$.

Let $^\dag = [\vec{v}\mapsto\vec{s}\,]$,
and let us write the subterm $p$ appearing within $d$ as $\Gamma,\vec{v} \vdash \lambda {x'}^k.e$,
where $\Gamma,\vec{v},x' \vdash e$ is regular. Then
\[ \dang{d^\dag} ~=~ \caseof{g\,(\lambda x'.\!\dang{e^\dag})\dang{q^\dag}}{\cdots} \;, \]
and since $\dang{\!d^\dag\!}$ is head-spinal by hypothesis,
$\dang{\!{e}^\dag\!}$ will be some term 
$\Gamma,x',\vec{v}\,' \vdash E[c]$, where
$\Gamma,x',\vec{v}\,',\vec{y}\,' \vdash c~=~ \caseof{gF'o'}{\cdots}$ is itself head-spinal 
with respect to $x'$ and $\vec{y}\,'$.
(Here $\vec{y}\,'$ denotes the local environment for $E[-]$, so that
$\Gamma,x',\vec{v}\,',\vec{y}\,'$ contains no repetitions.)
We will first show that the head $g$ of $c$ comes from $e$ rather than from $^\dag$;
we will later show that the same argument can be repeated for 
lower spinal occurrences of $g$.

\emph{Claim 1: In the evaluation $\dang{\!e^\dag\!} \,= E[c]$, the head $g$ of $c$
originates from $e$.}

\emph{Proof of Claim 1:} Suppose for contradiction that the head $g$ of $c$ originates from some
substituted occurrence of an $s_i$ within ${e}^\dag$, 
say as indicated by ${e}^\dag = D[s_i]$ and $s_i = L[d']$, 
where $\Gamma,x',\vec{v}\,' \vdash D[-]$, $\Gamma,\vec{v}\,' \vdash s_i$,
and $\Gamma,\vec{v}\,',\vec{z} \vdash d' = \caseof{gp'q'}{\cdots}$.
(Here $\vec{z}$ is the local variable environment for $L[-]$;
note that $\vec{z}$ is disjoint from $\Gamma,x',\vec{v}\,'$, but may well overlap with $\vec{y}\,'$.)
Then
\[ \Gamma,x',\vec{v}\,' \;\vdash~
   \dang{\!e^\dag\!} ~=~ \dang{D[L[d']]} ~=~ E[c] \;, \]
where the head $g$ in $d'$ is the origin of the head $g$ in $c$.
We will use this to show that a head-spinal term may be obtained from $d'$ 
via a substitution of level $< k$;
this will provide the bottleneck through which ${x'}^\eta$ is unable to pass.

We first note that the above situation satisfies the conditions of Lemma~\ref{g-lemma-1}, 
where we take the $\Gamma,K,d,K',c$ of the lemma 
to be respectively $(\Gamma,x',\vec{v}\,'),$ $D[L[-]],d',E,c$.
Condition~1 of the lemma holds because $\Gamma,\vec{v},x' \vdash e$ and 
$\Gamma,\vec{v}\,' \vdash \vec{s}$ are clearly regular, 
whence so is $\Gamma,x',\vec{v} \vdash e^\dag = D[L[d']]$;
condition~2 is immediate in the present setup.

We conclude that there is a substitution $[\vec{y}\mapsto\vec{t}\,]$ 
(called $[\vec{v}\mapsto\vec{s}\,]$ in the statement of Lemma~\ref{g-lemma-1})
with $\vec{y}$ the local environment for $D[L[-]]$ 
and $\Gamma,x',\vec{v}\,',\vec{y}\,' \vdash \vec{t}$
(recalling that $\vec{y}\,'$ are the local variables for $E[-]$),
such that $\dang{d'[\vec{y}\mapsto\vec{t}\,]}$ is head-spinal
and indeed of the form $\caseof{gF'o'}{\cdots}$ with $F',o'$ as above.
Furthermore, the only $\beta$-redexes in $e^\dag$ are those arising from the
substitution $^\dag$, with some $s_j$ of level $k$ as operator.
There are therefore no $\beta$-redexes in $e^\dag$ with a substitution variable of level $k$,
so by Lemma~\ref{g-lemma-1}(iii), the substitution $[\vec{y} \mapsto \vec{t}\,]$ 
is trivial for variables of level $k$. 
Note also that $\vec{y}$ (the environment for $D[L[-]]$) subsumes $\vec{z}$
(the environment for $L[-]$); it is disjoint from $\Gamma,x,\vec{v}\,'$ but may well overlap with $\vec{y}\,'$.

Let us now split the substitution $[\vec{y} \mapsto \vec{t}\,]$ as
$[\vec{y}\,^+ \mapsto \vec{t}\,^+, \, \vec{y}\,^- \mapsto \vec{t}\,^-]$,
where $\vec{y}\,^+$ consists of the variables in $\vec{y}$ of level $k$, 
and $\vec{y}\,^-$ consists of those of level $<k$.
As we have noted, the substitution for $\vec{y}\,^+$ is trivial:
that is, there is a mapping associating with each $y_j \in \vec{y}\,^+$ a variable $\iota(y_j) \in \vec{y}\,'$
such that $t_j = \iota(y_j)^\eta$.

Taking stock, we have that
\begin{eqnarray*} 
   \Gamma,x',\vec{v}\,',\vec{y}\,'  & \vdash &
   \dang{d'[\vec{y} \mapsto \vec{t}\,]} ~=~ \caseof{gF'o'}{\cdots} \;,  \\
   \Gamma,\vec{v}\,',\vec{z} & \vdash & d' = \caseof{gp'q'}{\cdots} \;, \\
    \Gamma,x',\vec{v}\,',\vec{y}\,' & \vdash & \vec{t} \;,
\end{eqnarray*}
where $[\vec{y}\mapsto\vec{t}\,]$ is trivial for level $k$ variables, 
and $gF'o'$ is head-spinal w.r.t.\ $x,\vec{y}\,'$.
From this we may read off that
\[ \Gamma,x',\vec{v}\,',\vec{y}\,' \;\vdash\; \dang{q'[\vec{y} \mapsto \vec{t}\,]} ~=~ o' \;. \]
Since $\vec{y}$ subsumes $\vec{z}$,
we may henceforth regard $q'$ as a term in environment $\Gamma,\vec{v}\,',\vec{y}$.
(This is compatible with the no-variable-hiding condition:
our conventions ensure that the variables of $\vec{y}-\vec{z}$ come from $d$ rather than $s_i$
and so do not appear bound in $q'$.)
We may harmlessly write $q'[\vec{y} \mapsto \vec{t}\,]$ as above, 
even though there are variables of $\vec{y}$ that cannot appear in $q'$.

Since $x'$ does not occur free in $q'$, each free occurrence of $x'$ in $o'$ above must originate from some $t_j \in \vec{t}$, 
which must furthermore have some type $\rho_j$ of level $<k$, since if $t_j$ had level $k$ then 
we would have $t_j = \iota(y_j)^\eta$ which does not contain $x'$ free. 
In fact, we may decompose the substitution $[\vec{y} \mapsto \vec{t}\,]$ as
$[\vec{y}\,^+ \mapsto \iota(\vec{y}\,^+)^\eta]$ followed by $[\vec{y}\,^- \mapsto \vec{t}\,^-]$,
since none of variables of $\vec{y}\,^-$ appear free in the $\iota(y_j)^\eta$ for $y_j \in \vec{y}\,^+$.
Setting $q'^* =\, \dang{q[\vec{y}\,^+ \mapsto \iota(\vec{y}\,^+)^\eta]}$ (so that $q'^*$ is just $q'$
with the variables in $\vec{y}\,^+$ rewritten via $\iota$),
we therefore have $\dang{q'^*[\vec{y}\,^- \mapsto \vec{t}\,^-]}\, = o'$.
Thus:
\begin{eqnarray*}
\Gamma,\vec{v}\,',\iota(\vec{y}\,^+),\vec{y}\,^- & \vdash & q'^* : \pure{k} \;, \\
\Gamma,x,\vec{v}\,',\vec{y}\,' & \vdash & t_j : \rho_j \mbox{~~~for $t_j \in \vec{t}\,^-$} \;, \\
\Gamma,x,\vec{v}\,',\vec{y}\,' & \vdash & \dang{q'^* [\vec{y}\,^- \mapsto \vec{t}\,^-]} ~=~ o' \;.
\end{eqnarray*}

Since $o'^\circ \succeq {x'}^\eta$ for a suitable $x',\vec{y}\,'$-closed substitution $^\circ$
(as part of the fact that $\dang{\!d^\dag\!}$ is head-spinal),
the above already comes close to exhibiting $\pure{k}$ as a pseudo-retract of a level $<k$ product type, 
contradicting Theorem~\ref{no-retraction-thm}.
To complete the argument, we must take account of the effect of $^\circ$,
which we here write as $[\vec{w}\mapsto\vec{r}\,]$
(we may assume that $\vec{w}$ is exactly $\Gamma,\vec{v}\,',\vec{y}\,'$). 
Reordering our variables, we may now write $x,\vec{w} \vdash \vec{t}\,^-$.

Next, let us split $^\circ$ into two independent parts: a substitution $[\vec{w}\,^+\mapsto\vec{r}\,^+]$ 
covering the variables in $\Gamma,\vec{v}\,',\vec{y}\,'$ of level $\geq k$, 
and $[\vec{w}\,^-\mapsto\vec{r}\,^-]$ covering those of level $<k$.
Since $^\circ$ is $x',\vec{y}\,'$-closed, we have $\vdash \vec{r}\,^+$ and $x' \vdash \vec{r}\,^-$.
Now set ${q'}^\wr =\, \dang{{q'^*}[\vec{w}\,^+\mapsto\vec{r}\,^+]}$, 
so that $\vec{u}\,^- \vdash {q'}^\wr$ 
where $\vec{u}\,^-$ consists of the variables of $\Gamma,\vec{v}\,',\vec{y}$ of level $<k$.
The idea is that $\vec{u}\,^- \vdash {q'}^\wr$ may now serve as one half of a suitable pseudo-retraction.
For the other half, let $[\vec{u}\,^- \mapsto \vec{a}\,^-]$ denote the effect of the substitution
$[\vec{y}\,^- \mapsto \vec{t}\,^-]$ followed by $^\circ =[\vec{w} \ \mapsto \vec{r}\,]$
(the order is important here as $\vec{y}\,^-$ and $\vec{w}$ may overlap).
Since $\vec{u}\,^- \subseteq \vec{y} \cup \vec{w}$ and $x,\vec{w} \vdash \vec{t}\,^-$ 
and $x \vdash \vec{r}$,
this substitution does indeed cover at least the variables of $\vec{u}\,^-$ and we have $x \vdash \vec{a}\,^-$.
We may now verify that  $\vec{u}\,^- \vdash {q'}^\wr$ and $x \vdash \vec{a}\,^-$ 
constitute a pseudo-retraction as follows:

\begin{eqnarray*}
x' & \vdash & \dang{{q'}^\wr [\vec{u}\,^- \mapsto \vec{a}\,^-]} \\
   & = & \dang{q'^* [\vec{w}\,^+ \mapsto \vec{r}\,^+] [\vec{y}\,^- \mapsto \vec{t}\,^-] [\vec{w} \ \mapsto \vec{r}\,]} \\
   & = & \dang{(q'^* [\vec{y}\,^- \mapsto \vec{t}\,^-]) [\vec{w} \ \mapsto \vec{r}\,] } \\
   & = & \dang{{o'}^\circ} ~\succeq~ {x'}^\eta \;.
\end{eqnarray*}
As regards the second equation here, the first substitution $[\vec{w}\,^+ \mapsto \vec{r}\,^+]$ may be safely
omitted as $\vec{w}\,^+$ and $\vec{y}\,^-$ are disjoint and the terms $\vec{r}\,^+$ do not contain
any of the $\vec{w}\,^+$ or $\vec{y}\,^-$ free.
We therefore have $\pure{k}$ as a pseudo-retract of a product of level $<k$ types.
This contradicts Theorem~\ref{no-retraction-thm}, so the proof of Claim 1 is complete.

We may therefore suppose that in the evaluation $\dang{\!e^\dag\!} = E[c]$,
the originating occurrence of the head $g$ in $c$
is as indicated by $\Gamma,x',\vec{v} \vdash e = C[d']$,
where $\Gamma,x',\vec{v},\vec{v}\,'' \vdash d' =\caseof{gp'q'}{\cdots}$.
(Here $\vec{v}\,''$ is the local environment for $C[-]$.
The symbols $d',p',q'$ are available for recycling now that the proof of Claim~1 is complete.)

In order to continue our analysis to greater depth, note that we may write
\[ \Gamma,x',\vec{v}\,' \vdash~
    \dang{\!e^\dag\!} ~=~ \dang{(\lambda \vec{v}.\,C[d'])\vec{s}} ~=~ E[c] \;, \] 
where $c$ is head-spinal w.r.t.\ $x',\vec{y}\,'$, and the head $g$ of $d'$ is the origin of the head $g$ of $c$.
(Recall that $\vec{y}\,'$ is the local environment for $E[-]$ and that 
$\Gamma,x',\vec{v} \vdash e = C[d']$ is regular, whence $\lev(\vec{v}\,'') \leq k$.)

We claim that once again we are in the situation of Lemma~\ref{g-lemma-1},
taking $\Gamma,K,d,K',c$ of the lemma to be respectively
$(\Gamma,x',\vec{v}\,')$, $(\lambda \vec{v}.\,C[-])\vec{s}$, $d', E, c$.
Condition~2 of the lemma is immediate in the present setup;
for condition~1, we again note that
$\Gamma,x',\vec{v}\,' \vdash C[d'] = e$ and $\Gamma,\vec{v}\,' \vdash \vec{s}$ are regular,
so by Proposition~\ref{regular-term-prop} contain no bound variables of level $>k$;
hence the same is true for $(\lambda \vec{v}.\,C[d'])\vec{s}$.
Applying Lemma~\ref{g-lemma-1}, we obtain a substitution 
$^{\dag'} = [\vec{v}\,^+ \mapsto \vec{s}\,^+]$ of level $\leq k$
(with $\vec{v}\,^+ = \vec{v},\vec{v}\,''$),
where $\Gamma,x',\vec{v}\,',\vec{v}\,^+ \vdash d'$ 
and $\Gamma,x',\vec{v}\,',\vec{y}\,' \vdash \vec{s}\,^+$ are regular, such that 
\[ \Gamma,x',\vec{v}\,',\vec{y}\,' \;\vdash\; \dang{d'[\vec{v}\,^+\!\mapsto\!\vec{s}\,^+]} \]
is head-spinal w.r.t.\ $x',\vec{y}\,'$, and indeed of the form $\caseof{gF'o'}{\cdots}$ with $F',o'$ as above.
(We may in fact write just $\Gamma,x',\vec{v}\,^+ \vdash d'$
since, by assumption, the variables of $\vec{v}\,'$ do not overlap with the free or bound variables of $d$
so do not appear in $d$.)
We may also read off that $\dang{(p')^{\dag'}} \,= F'$ and $\dang{(q')^{\dag'}} \,= o'$.
As regards the substitution $^{\dag'} =  [\vec{v}\,^+ \mapsto \vec{s}\,^+]$, it is clear that this
extends $ [\vec{v} \mapsto \vec{s}\,]$ since the evaluation of $(\lambda \vec{v}.\,C[d'])\vec{s}$
starts by $\beta$-reducing this term. 
Moreover, the argument of Lemma~\ref{g-lemma-1}(iii) shows that $^{\dag'}$ is trivial for
any level $k$ variables in $\vec{v}\,''$, as $C[-]$ is in normal form.

We are now back precisely where we started, in the sense that $d',\vec{v}\,^+,\vec{s}\,^+$
themselves satisfy the hypotheses of Lemma~\ref{g-lemma-2}, 
with $(\Gamma,x')$ now playing the role of $\Gamma$
and $(\vec{v}\,',\vec{y}\,')$ that of $\vec{v}\,'$.
Explicitly, we have regular terms
\[  \Gamma,x,\vec{v}\,^+ \vdash d' = \caseof{gp'q'}{\cdots} \;, ~~~~~~~~
    \Gamma,x,\vec{v}\,',\vec{y}\,' \vdash \vec{s}\,^+  \]
(so that $\lev(\vec{v}\,^+), \lev(\vec{v}\,',\vec{y}\,') \leq k$) 
where $\Gamma,x,\vec{v}\,',\vec{y}\,' \vdash \dang{d' [\vec{v}\,^+ \mapsto \vec{s}\,^+]}$
is head-spinal w.r.t.\ $x',\vec{y}\,'$.
We can therefore iterate the whole of the above argument to obtain an infinite descending chain of subterms
\[ \begin{array}{rrlrl}
\Gamma,\vec{v} \vdash &
d = & \caseof{gpq}{\cdots} \;, & 
p = & \lambda x'.\,C[d'] \;, \\
\Gamma,\vec{v},x',\vec{v}\,'' \vdash &
d' = & \caseof{gp'q'}{\cdots} \;, & 
p' = & \lambda x''.\,C'[d''] \;, \\
\Gamma,\vec{v},x',\vec{v}\,'',x'',\vec{v}\,'''' \vdash &
d'' = & \caseof{gp''q''}{\cdots} \;, &
p'' = & \lambda x'''.\,C''[d'''] \;, \\
& \cdots & & \cdots &
\end{array} \]
along with associated substitutions $^\dag$, $^{\dag'}$, $^{\dag''}, \ldots$
applicable to $d,d',d'',\ldots$ respectively, such that 
$\dang{\!p^\dag\!}$, $\dang{(p')^{\dag'}\!}$, $\dang{(p'')^{\dag''}\!}, \ldots$
coincide with the successive procedure subterms $F,F',F'',\ldots$ 
from the spine of the original term $\dang{\!d^\dag\!}$,
and likewise $\dang{\!q^\dag\!}$, $\dang{(q')^{\dag'}\!}$, $\dang{(q'')^{\dag''}\!}, \ldots$
coincide with $o,o',o''\ldots$.

We cannot quite conclude that $d$ is head-spinal, because the critical $x$ in 
$q^\dag$ might originate not from $q$ but from a level $k$ term in $\vec{s}$ (for example).
However, we can show that this problem does not arise for $q',q'',\ldots$, essentially because
$x',x'',\ldots$ are bound locally within $p$. 
We will in fact show that $d'$ is head-spinal w.r.t.\ $x',\vec{v}\,''$,
where $\vec{v}\,''$ is the local environment for $C[-]$; 
this will imply that $d$ is spinal.
In the light of Definition~\ref{spinal-def}, 
it will be sufficient to show that $(q')^{\circ'} \succeq {x'}^\eta$ for some $x',\vec{v}\,''$-closed
specialization $^{\circ'}$ covering the free variables of $q'$ except $x'$
(namely those of $\Gamma,\vec{v},\vec{v}\,''$);
the same argument will then obviously apply also to $q'',q''',\ldots$.

Recall that $\Gamma,\vec{v},\vec{v}\,'',x' \vdash q'$ 
and $\Gamma,\vec{v}\,',\vec{y}\,',x' \vdash o'$.
Since $gF'o'$ is head-spinal w.r.t.\ $x,\vec{y}\,'$,
we may as before take $^\circ = [\vec{w}\mapsto\vec{r}\,]$
$x',\vec{y}\,'$-closed such that ${o'}^\circ \succeq {x'}^\eta$,
where $\vec{w} = \Gamma,\vec{v}\,',\vec{y}\,'$ and $x' \vdash \vec{r}$.
Now define
\[ ^{\circ'} ~=~ [\vec{v} \mapsto \vec{s}\,^\circ, \; 
                          \vec{v}\,'' \mapsto (\vec{s}\,'')^\circ, \; 
                          \vec{w}\,^\Gamma\mapsto\vec{r}\,^\Gamma] \]
(where we write $\vec{s}\,^+$ as $\vec{s},\vec{s}\,''$, 
and $\vec{w}\,^\Gamma \mapsto \vec{r}\,^\Gamma$ denotes the restriction of $^\circ$ to $\Gamma$).
This covers the free variables of $q'$ except $x'$, and we have
$x' \vdash \vec{s}\,^\circ, (\vec{s}\,'')^\circ, \vec{r}\,^\Gamma$ because 
$\vec{w} \vdash \vec{s}, \vec{s}\,''$ and $x \vdash \vec{r}$.
Moreover, we have
\[  q'^{\circ'} ~=~ (q' [\vec{v}\,^+ \mapsto (\vec{s}\,^+)^\circ])[\vec{w}\,^\Gamma \mapsto \vec{r}\,^\Gamma] ~=~ (q'^{\dag'})^\circ ~\approx~ o^\circ ~\succeq~ x'^\eta \]
since $(\vec{s}\,^+)^\circ$ contains no free variables except $x$.
To check that $^{\circ'}$ is $x',\vec{v}\,''$-closed, it remains to show that
that $u^{\circ'}$ may contain $x'$ free only when $u \in \vec{v},''$ and $u$ is of level $<k$.
(Indeed, it is because of the possibility of $x'$ occurring free in these terms that 
the machinery of $x,V$-closed substitutions is necessary at all.)
The remaining cases are handled as follows:
\begin{itemize}
\item The terms $\vec{s}$ exist in environment $\Gamma,\vec{v}\,'$, so do not involve $x'$
or any of the variables of $\vec{y}\,'$. 
Since $^\circ$ is $x',\vec{y}\,'$-closed, it follows that the terms $\vec{s}\,^\circ$ do not involve $x'$.
\item For any variables $v \in \vec{v}\,''$ of level $k$, we have $v^{\dag'} = v^\eta$ which contains
no free variables of level $<k$, so that $(v^{\dag'})^\circ$ cannot involve $x'$.
\item The $\vec{r}\,^\Gamma$ cannot involve $x'$, since $^\circ$ is $x',\vec{y}\,'$-closed
and $\Gamma$ is disjoint from $\vec{y}\,'$.
\end{itemize}
This completes the verification that $d$ is spinal.
\end{proof}

From the above lemma we may immediately conclude, for example, 
that in the setting of Lemma~\ref{g-lemma-1}(ii) the term $d$ is spinal.

We are now ready for the main result of this section:

\begin{thm}  \label{no-gremlin-thm}
Every $\PCF^\Omega_k$-denotable procedure $\Gamma \vdash p$ is non-$g$-spinal
where $g: (k+1) \arrow (k+1)$.
\end{thm}

\begin{proof}
In the light of Section~\ref{sec-denotations}, it will suffice to show that the 
clauses of Theorem~\ref{denotable-inductive-thm}
cannot generate spinal terms from non-spinal ones.
For clauses 1--5 this is very easily seen.
For clause 6, it will be sufficient to show that non-spinal terms are closed under $k$-plugging,
and it is here that the machinery of Lemmas~\ref{g-lemma-1} and \ref{g-lemma-2} comes into play.

Suppose that $t =\, \dang{\Pi_{\Gamma,Z}(e,\xi)}$ where $\Gamma,Z \vdash e$
and $\Gamma,Z \vdash \xi(z)$ for each $z \in Z$.
For later convenience, to each $z_i \in Z$ let us associate the procedure
$\Gamma \vdash r_i =\, \dang{\Pi_{\Gamma,Z}(\xi(z_i),\xi)}$;
it is then clear from the definition of plugging and the evaluation theorem
that $t =\, \dang{e[\vec{z}\mapsto\vec{r}\,]}$
and that $r_i =\, \dang{\xi(z_i)[\vec{z}\mapsto\vec{r}\,]}$ for each $i$.

It will be natural to frame the argument contrapositively.
Suppose that $t$ is spinal, i.e.\ $t$ contains some head-spinal expression $c$ 
at position $K[-]$. We shall focus on the head occurrence of $g$ in $c$.
Clearly this occurrence must originate from one of the ingredients
$\Gamma,Z \vdash e$ or $\Gamma,Z \vdash \xi(z)$ of the plugging $\Pi_{\Gamma,Z}(e,\xi)$;
let us denote this ingredient by $\Gamma,Z \vdash t_0$.
We will show that $t_0$ itself is spinal.

Suppose that the relevant occurrence of $g$ in $t_0$ is as indicated by
\[ \Gamma,Z \;\vdash\; t_0 ~=~ L[d] \;,  \hspace*{3.0em}
    \Gamma,Z,\vec{v} \;\vdash\; d ~=~ \caseof{g p q}{\cdots} \;, \]
where $\vec{v}$ is the local environment for $L[-]$.
Writing $^\star$ for $[\vec{z}\mapsto\vec{r}\,]$, we have that
$\dang{\Pi(t_0,\xi)} \,=\, \dang{t_0^\star}$; 
and if $C[-]$ is the context encapsulating the remainder of the plugging $\Pi(e,\xi)$, 
then we may write
\[ \Gamma \vdash~ K[c] ~=~ t ~=~ \dang{C[\Pi(t_0,\xi)]} ~=~ \dang{C[t_0^\star]} 
   ~=~ \dang{C[L^\star[\dang{\!d^\star\!}]]} \;, \]
where 
\[ \dang{\!d^\star\!} ~=~ 
   \caseof{g \dang{\!p^\star\!} \dang{\!q^\star\!}\!}{\cdots} \;. \]
We claim that we are in the situation of Lemma~\ref{g-lemma-1},
taking $\Gamma,K,d,K',c$ of the lemma to be respectively
$\Gamma,C[L^\star[-]],\dang{\!\!d^\star\!\!},K,c$.
Condition~1 of the lemma holds because $C[t_0^\star]$ is constructed by substitution
from normal-form terms of level $\leq k$,
and condition~2 is immediate in the present setup.

By Lemma~\ref{g-lemma-1}, 
we may therefore conclude that
for a suitable substitution $[\vec{v} \mapsto \vec{s}\,]$
with $\Gamma, \vec{v}\,' \vdash \vec{s}$ regular, 
$\Gamma \vdash\, \dang{\dang{\!d^\star\!}[\vec{v}\mapsto\vec{s}\,]}$ is head-spinal.
(Note that the local variables of $C[-]$ do not appear in $d^\star$, because $\Gamma,Z \vdash t_0$
and $\Gamma \vdash \vec{r}$.)
Equivalently, we may say that $\dang{d[\vec{v}\,^+\!\mapsto\vec{s}\,^+]}$ 
is head-spinal, where
\[ [\vec{v}\,^+\mapsto\vec{s}\,^+] ~=~ 
   [\vec{z}\mapsto\vec{r}, ~\vec{v}\mapsto\vec{s}\,]
\;, \]
so that $\Gamma, \vec{v}\,' \vdash \vec{s}\,^+$ 
and $\lev(\vec{v}\,^+),\lev(\vec{v}\,') \leq k$.
(Note that the $\vec{z}$ do not appear free in $\vec{s}$, nor the $\vec{v}$ in $\vec{r}$.)

Since $\Gamma,Z,V \vdash d$ and $\Gamma,\vec{v}\,' \vdash \vec{s}\,^+$ are regular,
we are in the situation of Lemma~\ref{g-lemma-2}, so may conclude
that $d$ itself is spinal, and hence that $t_0$ is spinal.
We have thus shown that $k$-plugging cannot assemble spinal terms from non-spinal ones,
and this completes the proof.
\end{proof}

In particular, since the procedure $g \vdash F_{k+1}[g]$ mentioned at the start of the section
is spinal, we may conclude that this procedure is not $\PCF^\Omega_k$-denotable,
and hence neither is the procedure $Y_{k+1}$. This establishes Theorem~\ref{SP0-main-thm}.

We conclude the section by mentioning some minor variations on Theorem~\ref{no-gremlin-thm}
that we will require below.
First, as already indicated, the whole of the above proof goes through for the modified notion 
of spinal term appropriate to a variable $g: 0 \arrow (k+1) \arrow (k+1)$.
Secondly, the theorem also holds for an innocuous extension of
$\PCF^\Omega_k$ with a constant $\byval : (\nat\arrow\nat) \arrow \nat \arrow \nat$,
whose denotation in $\SP^0$ we take to be 
\[ \lambda fx.\;\caseof{x}{i \darrow \caseof{fi}{j \darrow j}} \;. \]
To see that the proof of Theorem~\ref{no-gremlin-thm} goes through in the presence of $\byval$,
it suffices simply to add an extra clause to the inductive proof noting that the procedure for $\byval$
is non-spinal.
This mild extension will allow a significant simplification of the
forms of procedures that we need to consider in Section~\ref{sec-extensional}.%
\footnote{The operator $\byval$ plays a major role in \cite[Section~7.1]{Longley-Normann},
where it is shown that every element of $\SP^0$ is denotable in $\PCF^\Omega + \byval$.
The sense in which it is innocuous is that its denotation in $\SF$ coincides with that of
$\lambda fx.\,\ifzero\,x\,(fx)(fx)$; thus $\byval$ adds nothing to the expressivity of $\PCF^\Omega_k$
as regards $\SF$.}

\section{Non-definability in the extensional model}  \label{sec-extensional}

To obtain corresponding non-definability results for $\SF$ rather than $\SP^0$,
one must show not only that the canonical procedures $Y_\tau$ considered above are not
$\PCF^\Omega_k$-denotable, but also that no extensionally equivalent procedures are.
It is easy to see that there are indeed many other procedures $Z$
with the same extension as $Y_\tau$.
To give a trivial example, we may present the canonical procedure $Y_{k+1}$ as
$\lambda gx.C[g,x]$, where
\[ C[g,x] ~=~ \caseof{A[g,x]}{i \darrow i} \;,  \hspace*{3.0em}
   A[g,x] ~=~ g(\lambda x'.C[g,x'])x^\eta \;. \]
However, another candidate for the fixed point operator is
\[  Z_0 ~=~ \lambda gx.\,\caseof{A[g,x]}{i \darrow {C[g,x]}} \;. \]
Intuitively, this computes the desired value twice, discarding the first result.

As a slightly more subtle example, consider the procedure
\[  Z_1 ~=~ \lambda gx.\,\caseof{g(\lambda x'.\caseof{A[g,x]}{i \darrow {C[g,x']}}) x^\eta}{k \darrow k} \;. \]
Here, within the $\lambda x'$ subterm, we have smuggled in a repetition of the 
top-level computation $A[g,x]$ before proceeding to evaluate what is really required.
The effect is that
$\lambda x'.\caseof{A[g,x]}{i \darrow C[g,x']}$ may be 
extensionally below $\lambda x'.C[g,x']$,
and this may indeed affect the result when $g$ is applied.
However, this can only happen when $Y_{k+1}gx$ is undefined anyway,
so it is easy to see that $Z_1$ as a whole will have the same extension as $Y_{k+1}$.

Yet another way to construct procedures extensionally equivalent to $Y_{k+1}$
is to vary the subterms of the form $x^\eta$ (where $x$ has type $k$).
For instance, in the case $k=1$, we could replace $x^\eta$ by an
extensionally equivalent term such as
\[ X_0 ~=~ \lambda y^0.\,
   x(\lambda.\,\caseof{y}{0 \darrow \caseof{x(\lambda.0)}{j \darrow 0} \mid i+1 \darrow i+1}) \;. \]
This is different in character from the previous examples: 
rather than simply repeating the computation of $xy^\eta$, we are performing the specific
computation $x(\lambda.0)$ which we can see to be harmless given that
this point in the tree has been reached.
Clearly, such `time-wasting' tricks as the above may be combined and elaborated 
to yield more complex examples of procedures equivalent to $Y$.

However, all of the above are rather innocuous variations
and do not really yield a fundamentally different method of computing fixed points.
For example, the bodies of both $Z_0,Z_1$ are still head-spinal terms, 
and it is essentially the spines that are really computing the desired fixed point
by the canonical method.
This suggests that we should try to show that every procedure extensionally equivalent
to $Y_{k+1}$ is spinal;
from this it would follow easily by Theorem~\ref{no-gremlin-thm} that
the fixed point functional $Y_{k+1}$ in $\SF$ 
is not denotable in $\PCF^\Omega_k$.

Unfortunately, we are currently unable to show this in the case of $Y_{k+1}$:
indeed, the syntactic analysis of procedures $Z \approx Y_{k+1}$
appears to present considerable technical difficulties.
We shall establish the result for $Y_{0 \arrow (k+1)}$, 
although even here, it is simplest to concentrate not on $Y_{0 \arrow (k+1)}$ itself, 
but on a certain functional that is readily definable from it,
namely the functional $\Phi_{k+1}$ introduced in Section~\ref{sec-intro}.
Nonetheless, the above examples of `time-wasting' procedures illustrate some of the 
situations that our proof will need to deal with, and they may help to motivate
some of the technical machinery that follows.

Specifically, within $\PCF_{k+1}$, let us define
\begin{eqnarray*}
\Phi_{k+1} & : & ({0} \arrow (k+1) \arrow (k+1)) 
                         \arrow ({0} \arrow (k+1))  \\
\Phi_{k+1}\,g^{0\arrow(k+1)\arrow(k+1)} & = & 
   Y_{{0}\arrow(k+1)}\, (\lambda f^{0\arrow(k+1)}.\lambda n.\,g\,n\,(f(\suc\;n))) \;,
\end{eqnarray*}
so that informally
\[ \Phi_{k+1}\,g\,n ~=~ g\,n\,(g\,(n+1)\,(g\,(n+2)\, (\cdots))) \;. \]
For the rest of this section we will write $\Phi_{k+1}$ simply as $\Phi$.
For each $n \in \N$, let $g \vdash \Phi^n[g] = \Phi\,g\,\num{n} : {k+1}$,
and let $p_n \in \SP(k+1)$ be the canonical NSP for $\Phi^n[g]$
(that is, the one arising from the above $\PCF$ definition
via the standard interpretation in $\SP^0$).
These procedures may be defined simultaneously by:
\[ g~ \vdash~ p_n ~=~ \lambda x^k.\,\caseof{g\,(\lambda.n)\,p_{n+1}\,x^\eta}{i \darrow i}
   ~:~ k+1 \;. \]

By a syntactic analysis of the possible forms of (simple) procedures $g \vdash q \approx p_n$,
we will show that any such $q$ is necessarily spinal.
Here we have in mind the modified notion of spinal term that is applicable to
terms involving a global variable $g : \rho$, where $\rho = {0} \arrow (k+1) \arrow (k+1)$
(see the explanation following Definition~\ref{spinal-def}).
Using Theorem~\ref{no-gremlin-thm} (adapted to this modified setting), 
it will then be easy to conclude that within $\SF$, the element
$\sem{\lambda g.\,\Phi^0[g]}$, and hence $Y_{{0}\arrow(k+1)}$ itself, 
is not $\PCF^\Omega_k$-denotable in $\SF$.

To show that any $q \approx p_n$ is head-spinal, our approach will be as follows.
First, we show that any such $q$ must broadly resemble $p_n$ in at least its top-level
structure, in that $q$ must have the form $\lambda x.\,\caseof{garo}{\cdots}$,
where the arguments $a,r,o$ are closely related to the corresponding
arguments $(\lambda.n), p_{n+1}, x^\eta$ occurring within $p_n$.
We do this by showing that if $q$ were to deviate in any way from this prescribed form, 
we would be able to cook up procedures $G \in \SP^0(\rho)$ and $X \in \SP^0(k)$ 
manifesting an extensional difference between $q$ and $p_n$, i.e.\ such that
$q[g \mapsto G]X \not\approx p_n[g \mapsto G]X$.
(Contrary to our usual convention, we will here use the uppercase letters $G,X$ to range over
normal-form closed procedures that may be substituted for $g,x$ respectively.)
In particular, we shall establish a sufficiently close relationship between $r$ and $p_{n+1}$
that the same analysis can then be iterated to arbitrary depth, showing that $q$ has
a spinal structure as required.

The main complication is that $r$ need not superficially resemble $p_{n+1}$,
since within $r$, the crucial application of $g$ that effectively computes the value of $p_{n+1}$
may be preceded by other `time-wasting' applications of $g$ (the idea is illustrated by the
example $Z_1$ above). However, it turns out that such time-wasting subterms
$g a^1 r^1 o^1$ must be of a certain kind if the extensional behaviour $q \approx p_n$
is not to be jeopardized: in particular, the first argument $a^1$ must evaluate to some 
$i < n$. (As in the example of $Z_1$, the idea is that if the subterm $g a^1 r^1 o^1$
merely repeats some `outer' evaluation, it will make no overall difference to the extension
if the evaluation of this subterm does not terminate.)
In order to formulate the relationship between $r$ and $p_{n+1}$, we therefore need a
means of skipping past such time-wasting applications in order to reach the application of $g$
that does the real work.
We achieve this with the help of a \emph{masking} operator $\mu_{n,n'}$,
which (for any $n \leq n'$) overrides the behaviour of $g$ on numerical arguments $n \leq i < n'$
with a trivial behaviour returning the dummy value $0$.

We now proceed to our formal development.
As a brief comment on notation, we recall from Section~\ref{sec-background}
that the relations $\approx$ and $\preceq$ of observational equivalence and inequality
make sense not just for elements of $\SP^0$ but for arbitrary meta-terms (including applications), closed or otherwise. 
Throughout this section, for typographical convenience, 
we will tend to express the required relationships mostly at the level of meta-terms,
writing for instance $pq \approx \lambda.n$ rather than the equivalent $\dang{pq}\,=\lambda.n$
or $p \cdot q = \lambda.n$.
We shall also perpetrate other mild abuses of notation, such as writing 
a procedure $\lambda.n$ simply as $n$ (except for special emphasis),
$\lambda \vec{x}.\bot$ as $\bot$, $x^\eta$ as $x$,
a meta-expression $\caseof{A}{i \darrow i}$ just as $A$,
and abbreviating a substitution $[g \mapsto G,\, x \mapsto X]$ to $[G,X]$.

We shall say that $G \in \SP^0(0\arrow(k+1)\arrow(k+1))$ is \emph{strict} 
if $G\bot ro \approx \bot$ for any $r,o$. Clearly, $G$ is strict iff
$G \approx \lambda izx.\,\caseof{i}{j \darrow G (\lambda.j) z^\eta x^\eta}$.
In connection with meta-terms with free variable $g$, we shall write
$T \approx^\mydot T'$ to mean that $T[g \mapsto G] \approx T'[g \mapsto G]$
for all \emph{strict} $G$; the notation $\preceq^\mydot$ is used similarly.
We shall actually analyse the syntactic forms of procedures $g \vdash q$ based on
the assumption that $q \succeq^\mydot p_n$,
where $p_n$ is the canonical procedure for $\Phi^n[g]$ as above.

We shall say a procedure $g \vdash q$ is \emph{simple} if for every application
$garo$ appearing within $q$, the first argument $a$ is just a numeral $\lambda.n$.
The following observation simplifies our analysis of terms considerably;
it uses the operator $\byval$ and its NSP interpretation,
as introduced at the end of Section~\ref{sec-Y-k+1}.

\begin{prop}  \label{simple-prop}
If there is a procedure $g \vdash q \succeq^\mydot p_n$-denotable in $\PCF^\Omega_k$,
then there is a simple procedure $g \vdash q' \succeq^\mydot p_n$
denotable in $\PCF^\Omega_k + \byval$.
\end{prop}

\begin{proof}
Suppose $g \vdash q$ is $\PCF^\Omega_k$-denotable where $q \succeq^\mydot p_n$, and write 
\[ S[g] ~=~ \lambda izx.\,\caseof{i}{j \darrow g(\lambda.j)z^\eta x^\eta} \;. \]
It is easy to see that
\[ S[g] ~=~ \sem{\lambda izx.\,\byval\,(\lambda j. gjzx)\,i\,}_g \;. \]
Take $g \vdash q' =\, \dang{q[g \mapsto S[g]]}$,
so that $q'$ is denotable in $\PCF^\Omega_k + \byval$.
Then $q \approx^\mydot q'$ since $S[G] \approx G$ for all strict $G$, so 
$q' \succeq^\mydot p_n$. Finally, $q'$ is clearly simple: every occurrence of $g$ within
$q[g \mapsto S[g]]$ has a numeral as its first argument, so the same will be true of
$\dang{q[g \mapsto S[g]]}$.
\end{proof}

For any $n \leq n'$, let us define the \emph{masking} $\mu_{n,n'}(g)$ of $g$
to be the following procedure term (here $\rho = 0 \arrow (k+1) \arrow (k+1)$):
\[ g^\rho ~\vdash~ \mu_{n,n'}(g) ~=~ 
   \lambda izx.\, 
   \caseof{i}{n \darrow 0 \mid \cdots \mid n'-1 \darrow 0 \mid - \darrow gizx} ~:~ \rho \;. 
\]
(The wildcard symbol `$-$' covers all branch indices not covered by the preceding clauses.)
We may also write $\mu_{n,n'}(P)$  for $\mu_{n,n'}(g)[g \mapsto P]$ if $P$ is any
meta-procedure of type $\rho$.
We write $\mu_{n,n+1}$ simply as $\mu_n$; note also that $\mu_{n,n}(g) \approx^\mydot g$.
Clearly $\mu_n(\mu_{n+1}(\cdots(\mu_{n'-1} (g)) \cdots)) \approx \mu_{n,n'}(g)$.

We shall say that a closed meta-term $\vdash P : \rho$ is \emph{trivial at $n$} if 
$P(\lambda.n)\bot\bot \approx 0$.
Note that $\mu_{n,n'}(G)$ is trivial at each of $n,\ldots,n'-1$ for any closed $G$;
indeed, $G$ is trivial at $n,\ldots,n'-1$ iff $G \succeq^\mydot \mu_{n,n'}(G)$.

The following lemma now implements our syntactic analysis of the top-level structure of 
simple procedures $q \succeq^\mydot p_n$.

\begin{lem}  \label{Qn-lemma}
Suppose $g \vdash q$ is simple and $q \succeq^\mydot p_n$. Then $q$ has form 
$\lambda x^k.\,\caseof{g a r o}{\cdots}$, where:
\begin{enumerate}[label=(\arabic*)]
\item $a = \lambda.n$,
\item $o[g \mapsto G] \succeq x^\eta$ whenever $\vdash G$ is trivial at $n$,
\item $r[g \mapsto \mu_n(g),\, x \mapsto X] \succeq^\mydot p_{n+1}$ for any $X$.
\end{enumerate}
\end{lem}

\begin{proof}
Suppose $q = \lambda x^k.e$.
Clearly $e$ is not constant since $q \succeq p_n$;
and if $e$ had head variable $x$, we would have
$q[g \mapsto \lambda izx.\,\caseof{i}{- \darrow 0}](\lambda w.\bot) = \bot$, 
whereas
$p_n[g \mapsto \lambda izx.\,\caseof{i}{- \darrow 0}](\lambda w.\bot) = 0$,
contrary to $q \succeq^\mydot p_n$.
So $e$ has form $\caseof{g a r o}{\cdots}$.

For claim~(1), we have $a = \lambda.m$ for some $m \in \N$
because $q$ is simple. Suppose that $m \neq n$,
and consider
\[ G' ~=~ \lambda izx.\,\caseof{i}{n \darrow 0 \mid - \darrow \bot} \;. \]
Then for any $X$, clearly $q[G']X \approx \bot$, whereas
$p_n[G']X \approx G'(\lambda.n)(\cdots)X \approx 0$,
contradicting $q \succeq^\mydot p_n$. Thus $a = \lambda.n$.

For claim~(2), suppose that $G (\lambda.n) \bot \bot \approx 0$
but not $o[G] \succeq x^\eta$;
then we may take $X \in \SP^0({k})$ and $u \in \SP^0({k-1})$
such that $Xu \approx l \in \N$ but $o[G,X]u \not\approx l$. Now define
\[ G' ~=~ \lambda izx.\,\caseof{i}{j \darrow \caseof{xu}{l \darrow Gjzx \mid - \darrow \bot}} \;. \]
Then $G' \preceq G$, so $o^* u \not\approx l$ where $^* = [G',X]$;
hence $G' a^* r^* o^* \approx \bot$ and so $q[G']X \approx \bot$.
On the other hand, we have
\begin{eqnarray*}
p_n[G']X & \approx & \caseof{G'(\lambda.n)p_{n+1}^*X}{i \darrow i} \\
& \approx & \caseof{Xu}{l \darrow G(\lambda.n)p_{n+1}^*X} ~\approx~ 0 \;,
\end{eqnarray*}
contradicting $q \succeq^\mydot p_n$ (note that $G'$ is strict). Thus $o[G] \succeq x^\eta$.

For claim~(3), suppose that $p_{n+1}[G]X' \approx l$ for some strict $G \in \SP^0(\rho)$
and $X' \in \SP^0(k)$. 
We wish to show that $r[\mu_n(G), X]X' \approx l$ for any $X$.
Suppose not, and consider
\[ G' ~=~ 
\lambda izx.\,\caseof{i}{n \darrow \caseof{zX'}{l \darrow 0 \mid - \darrow \bot} \mid - \darrow Gizx} \;. \]
Then $G' \preceq \mu_n(G)$, so $r[G',X]X' \not\approx l$.
Moreover, since $a=\lambda.n$ by claim~1, 
we see that $G' a^* r^* o^* \approx \bot$, where $^* = [G',X]$,
so that $q[G']X \approx \bot$.
On the other hand, we have
\begin{eqnarray*}
p_n[G']X & \approx & \caseof{G'(\lambda.n)p_{n+1}^*X}{i \darrow i} \\
& \approx & \caseof{p_{n+1}^*X'}{l \darrow 0} \;.
\end{eqnarray*}
Here, since $p_{n+1}$ does not contain $x$ free, we have $p_{n+1}^* = p_{n+1}[G']$.
But it is easy to see that $p_{n+1}[G'] \approx p_{n+1}[G]$, since
every occurrence of $g$ within $p_{n+1}$ is applied to $\lambda.n'$ for some $n'>n$,
and for all such $n'$ we have $G'(\lambda.n') \approx G(\lambda.n')$.
(Since $p_{n+1}$ contains infinitely many applications of $g$, an appeal to continuity is
formally required here.)
But $p_{n+1}[G]X' \approx l$ by assumption; thus $p_{n+1}^*X' \approx l$,
allowing us to complete the proof that $p_n[G']X \approx 0$.
Once again, this contradicts $q \succeq^\mydot p_n$, so claim~3 is established.
\end{proof}

In the light of Appendix~A, one may strengthen claim~2 of the above lemma 
by writing $o[g \mapsto G] \approx x^\eta$.
This gives a fuller picture of the possible forms of terms $q \approx p_n$,
but is not needed for showing that such $q$ are spinal.

We have now almost completed a circle, in the sense that claim~3 tells us that the procedure
$\dang{r[g \mapsto \mu_n(g), x \mapsto X]}$ itself satisfies the hypothesis for $q$
(with $n+1$ in place of $n$).
However, there still remains a small mismatch between the hypothesis and the conclusion,
in that claim~3 concerns not $r$ itself but rather $\dang{r[g \mapsto \mu_n(g)]}$. 
(In the light of claim~3, the variable $x$ may be safely ignored here.)
This mismatch is repaired by the following lemma, which lends itself to iteration to any depth.
Note the entry of a term context $E[-]$ here.

\begin{lem}  \label{q-structure-lemma}
Suppose $g \vdash q$ is simple and 
$q[g \mapsto \mu_{n,n'}(g)] \succeq^\mydot p_{n'}$.
Then $q$ has the form $\lambda x^k.\,E[\caseof{garo}{\cdots}]$, where:
\begin{enumerate}[label=(\arabic*)]
\item $E[-]$ has empty local variable environment,
\item $a = \lambda.n'$,
\item $o[g \mapsto G] \succeq x^\eta$ whenever $\vdash G$ is trivial at $n,\cdots,n'$,
\item $r[g \mapsto \mu_{n,n'+1}(g),\,x \mapsto X] \succeq^\mydot p_{n'+1}$ 
for any $X$.
\end{enumerate}
\end{lem}

\begin{proof}
Let $q' =\, \dang{q[g \mapsto \mu_{n,n'}(g)]}$.
Under the above hypotheses, we have by Lemma~\ref{Qn-lemma} that $q'$ is of the form 
$\lambda x.\,\caseof{ga'r'o'}{\cdots}$, where $a'=\lambda.n'$,
$o'[g \mapsto G] \succeq x^\eta$ whenever $G$ is trivial at $n'$,
and $r'[g \mapsto \mu_{n'}(g)] \succeq^\mydot p_{n'+1}$.
Write $q$ as $\lambda x.\,E[\caseof{garo}{\cdots}]$ where the displayed
occurrence of $g$ originates the head $g$ of $q'$ via the substitution
$^\dag = [g \mapsto \mu_{n,n'}(g)]$.

Suppose that the hole in $E[-]$ appeared within an abstraction $\lambda y.-$;
then the hole in $E[g \mapsto \mu_{n,n'}(g)][-]$ would likewise appear within such an abstraction.
Moreover, the evaluation of $q[g \mapsto \mu_{n,n'}(g)]$ consists simply of the contraction of
certain expressions $\mu_{n,n'}(g)(\lambda.m)r''o''$ to either $0$ or $g(\lambda.m)r''o''$,
followed by reductions $\caseof{0}{i \darrow e_i} \reducesto e_0$;
thus, any residue in $q'$ of the critical $g$ identified above will likewise appear underneath $\lambda y$.
But this is impossible, because the head $g$ of $q'$ is a residue of this $g$ by assumption.
This establishes condition~(1).

In the light of this, by Lemma~\ref{g-lemma-1}(i) we have
$a' \approx a^\dag$, $o' \approx o^\dag$ and $r' \approx r^\dag$.
But since $q$ is simple, $a$ is a numeral, so $a=\lambda.n'$, giving condition~(2).
For condition~(3), suppose $G$ is trivial at $n,\ldots,n'$.
Then $G \succeq \mu_{n,n'+1}(G)$, so that
\[ o[g \mapsto G] ~\succeq~ o[g \mapsto \mu_{n,n'}(\mu_{n'}(G))]
   ~\approx~ o^\dag[g \mapsto \mu_{n'}(G)] ~\approx~ o'[g \mapsto \mu_{n'}(G)]
   ~\succeq~ x^\eta  \;, \]
since $\mu_{n'}(G)$ is trivial at $n'$.
Condition~(4) also holds since 
$r'[g \mapsto \mu_{n'}(g)] \approx r^\dag[g \mapsto \mu_{n'}(g)] \approx r[g \mapsto \mu_{n,n'+1}(g)]$, where
$r'[g \mapsto \mu_{n'}(g)] \succeq^\mydot p_{n'+1}$.
\end{proof}

\begin{cor}  \label{spinal-cor}
If $g \vdash q$ is simple and $q \succeq^\mydot p_n$, then $q$ is spinal
(in the modified sense).
\end{cor}

\begin{proof}
Since condition~3 of the above lemma matches its hypotheses, 
starting from the assumption that 
$q \approx^\mydot q[g \mapsto \mu_{n,n}(g)] \succeq^\mydot p_n$,
we may apply the lemma iteratively to obtain a spinal structure as
prescribed by Definition~\ref{spinal-def} 
(subject to the relevant adjustments for $g : 0 \arrow (k+1) \arrow (k+1)$).
Note that at each level, a suitable substitution $^\circ$ will be the closed one that
specializes $g$ to $\lambda izx.0$ (which is trivial at all $n$), and all variables $x^k$
other than the innermost-bound one to $\bot$.
\end{proof}

Thus, if there exists a $\PCF^\Omega_k$-denotable procedure $t \approx \lambda g.p_0$, for instance,
then by Proposition~\ref{simple-prop} there is also a simple such procedure $t'$
denotable in $\PCF^\Omega_k + \byval$,
and by Corollary~\ref{spinal-cor}, this $t'$ will be spinal in the modified sense.
But this contradicts Theorem~\ref{no-gremlin-thm} 
(understood relative to the modified setting, and applied to $\PCF^\Omega_k + \byval$ as indicated
at the end of Section~\ref{sec-Y-k+1}).
We conclude that no $t \approx \lambda g.p_0$ can be $\PCF^\Omega_k$-denotable.
Since the interpretation of $\PCF^\Omega_k$ in $\SF$ factors through $\SP^0$,
this in turn means that within the model $\SF$, the element $\sem{\lambda g.\,\Phi^0[g]}$
is not denotable in $\PCF^\Omega_k$.
On the other hand, this element is obviously denotable
relative to $Y_{{0}\arrow (k+1)} \in \SF$ even in $\PCF_1$,
so the proof of Theorems~\ref{operational-main-thm}(i) and \ref{SF-main-thm}(i) is complete.
This establishes Berger's conjecture, and also suffices for Corollary~\ref{no-finite-basis-cor}.

\section{Extensional non-definability of \texorpdfstring{$Y_{k+1}$}{Y k+1}}  \label{sec-pure-type}

We have so far shown that the element $Y_{0 \arrow (k+1)} \in \SF$ is not $\PCF^\Omega_k$-denotable.
We shall now refine our methods slightly to show that 
even $Y_{k+1} \in \SF$ is not $\PCF^\Omega_k$-denotable. 
Since $\pure{k+1}$ is clearly a $\PCF_0$-definable retract of every level $k+1$ type,
this will establish that no $Y_\sigma \in \SF$ where $\lev(\sigma)=k+1$ is denotable in
$\PCF^\Omega_k$.

The idea is as follows. Within each type level $k \geq 1$, we can stratify the types into 
\emph{sublevels} $(k,l)$ where $l = 1,2,\ldots$, essentially by taking account of the `width' of the type 
as well as its depth. 
We thus obtain a sublanguage $\PCF^\Omega_{k,l}$ of $\PCF^\Omega_k$ by admitting $Y_\sigma$
only for types $\sigma$ of sublevel $(k,l)$ or lower.
We show that, roughly speaking, all our previous methods can be adapted to show that for a given $k$,
the languages $\PCF^\Omega_{k,l}$ for $l = 1,2,\ldots$ form a strict hierarchy.
(This is true as regards definability in $\SP^0$; for $\SF$, we will actually show only that
$\PCF^\Omega_{k,l}$ is strictly weaker than $\PCF^\Omega_{k,l+2}$.)
This more refined hierarchy is of some interest in its own right, and illustrates that the structure of $\SF$
is much richer than that manifested by the pure types.%
\footnote{This contrasts with the situation for extensional \emph{total} type structures over $\N$, for example.
There, under mild conditions, every simple type is isomorphic to a pure type: 
see Theorem~4.2.9 of \cite{Longley-Normann}.}

Any term of $\PCF^\Omega_k$ will be a term of some $\PCF^\Omega_{k,l}$.
Previously we showed only that no such term could define $Y_{0 \arrow (k+1)} \in \SF$;
however, we now see that there is actually plenty of spare headroom between $\PCF^\Omega_{k,l}$
and the pure type $k+1$.
Specifically, there are operators $Y_\sigma$ in $\PCF^\Omega_{k,l+2}$
that are not $\PCF^\Omega_{k,l}$-denotable; and since all such $\sigma$ are of level $\leq k$
and are thus easily seen to be retracts of $\pure{k+1}$,
we may conclude that $Y_{k+1} \in \SF$ is not $\PCF^\Omega_k$-denotable.

The following definition sets out the finer stratification of types that we shall use.

\begin{defi}  \label{width-def} \hfill
\begin{enumerate}[label=(\roman*)]
\item The \emph{width} $w(\sigma)$ of a type $\sigma$ is defined inductively as follows:
\[  w(\nat) ~=~ 0 \;, ~~~~~~~~ 
w(\sigma_0,\ldots,\sigma_{r-1} \arrow \nat) ~=~ \max(r,w(\sigma_0),\ldots,w(\sigma_{r-1})) \;. \]
For $k,l \geq 1$, we say $\sigma$ has \emph{sublevel} $(k,l)$ if $\lev(\sigma)=k$ and $w(\sigma)=l$.
If $\sigma=\nat$, we simply say that $\sigma$ has \emph{sublevel} $0$.
We order sublevels in the obvious way: $0$ is the lowest sublevel,
and $(k,l) < (k',l')$ if either $k<k'$ or $k=k'$ and $l<l'$.

\item For each $k,l$ we define a type $\rho_{k,l}$ by:
\[ \rho_{0,l} ~=~ \nat \;, ~~~~~~~~
   \rho_{k+1,l} ~=~ \rho_{k,l},\ldots,\rho_{k,l} \arrow \nat \mbox{~~($l$ arguments)} \;. \]
When $k,l \geq 1$, we may call $\rho_{k,l}$ the \emph{homogeneous} type of sublevel $(k,l)$.
\end{enumerate}
\end{defi}

\noindent
The following facts are easily established. Here we shall say that $\sigma$ is a \emph{simple retract} of $\tau$
if there is a $\PCF_0$-definable retraction $\sigma \lhd \tau$ within $\SP^0$.

\begin{prop}  \label{sublevel-facts} \hfill
\begin{enumerate}[label=(\roman*)]
\item For $k \geq 1$, every type of sublevel $(k,l)$ or lower is a simple retract of $\rho_{k,l}$.
Hence, for all $k \geq 0$, for every finite list of types $\sigma_i$ of level $\leq k$, there is some $l$ such that
each $\sigma_i$ is a simple retract of $\rho_{k,l}$.

\item Every finite product of level $\leq k$ types is a simple retract of $\pure{k+1}$.

\item If $\sigma$ is a simple retract of $\tau$, then $Y_\sigma$ is $\PCF_0$-definable from 
$Y_\tau$ in $\SP_0$.
\end{enumerate}
\end{prop}

\begin{proof}  \hfill
\begin{enumerate}[label=(\roman*)]
\item The first claim (for $k \geq 1$) is easy by induction on $k$, and the second claim 
(which is trivial when $k=0$) follows easily.

\item By induction on $k$. The case $k=0$ is easy. For $k \geq 1$,
suppose $\sigma_0,\ldots,\sigma_{m-1}$ are level $\leq k$ types, 
where $\sigma_i = \tau_{i0},\ldots,\tau_{i(n_i-1)} \arrow \nat$ for each $i$.
Here the $\tau_{ij}$ are of level at most $k-1$, so by (i), 
we may choose $l$ such that each $\tau_{ij}$ is a simple retract of $\rho_{k-1,l}$.
Taking $n = \max_i n_i$, we then have that each $\sigma_i$ is a simple retract of
$\rho_{k-1,l},\ldots,\rho_{k-1,l} \arrow \nat$ (with $n$ arguments).
The product $\Pi_i \sigma_i$ is therefore a simple retract of the type
$\sigma = \nat,\rho_{k-1,l},\ldots,\rho_{k-1,l} \arrow \nat$.
But by the induction hypothesis, the product of $\nat,\rho_{k-1,l},\ldots,\rho_{k-1,l}$
is a simple retract of $\pure{k}$ whence $\sigma$ itself is a simple retract of $\pure{k+1}$.

\item is left as an easy exercise.
\qedhere
\end{enumerate}
\end{proof}

\noindent
Next, we adapt the proof of Theorem~\ref{no-retraction-thm} to establish the crucial gap
between $\rho_{k,l}$ and $\rho_{k,l+1}$. 
This gives an indication of how our methods may be used to map out the embeddability relation 
between types in finer detail, although we leave an exhaustive investigation of this to future work.

\begin{thm}  \label{kl-no-retraction-thm}
Suppose $k \geq 1$. Within $\SF$, the type $\rho_{k,l+1}$ is not a pseudo-retract of any finite product
of types of sublevel $\leq (k,l)$ or lower.
\end{thm}

\begin{proof}
In view of Proposition~\ref{sublevel-facts}(i), it will suffice to show that $\rho_{k,l+1}$ is not 
a pseudo-retract of a finite power of $\rho_{k,l}$.
We argue by induction on $k$. The arguments for both the base case $k=1$ and the step case $k>1$ 
closely parallel the argument for the step case in Theorem~\ref{no-retraction-thm}, 
so we treat these two cases together as far as possible, omitting details that are easy adaptations
of those in the earlier proof.

Suppose for contradiction that there were procedures
\[ z^\rho \vdash t_i : \sigma  ~~(i<m) \;,  \hspace*{2.0em}
   x_0^{\sigma},\ldots,x_{m-1}^{\sigma} \vdash r : \rho \;, \]
 where $\rho = \rho_{k,l+1}$, $\sigma = \rho_{k,l}$,
 such that $z \vdash r[\vec{x} \mapsto \vec{t}\,] \succeq z^\eta$.
 Let $v =\, \dang{r[\vec{x} \mapsto \vec{t}\,]}$, so that $z \vdash v \succeq z^\eta$.
 As in the proof of Theorem~\ref{no-retraction-thm}, one may show that
 $v$ has the form $\lambda f_0 \ldots f_l.\,\caseof{zp_0 \ldots p_l}{\cdots}$
 where $p_i[z \mapsto \lambda \vec{w}.0] \succeq f_i$ for each $i$. 
Next, we note that $r[\vec{x} \mapsto \vec{t}\,]$ reduces to a head normal form
$\lambda f_0 \ldots f_l.\, \caseof{zP_0\ldots P_l}{\cdots}$ where $\dang{P_i} \,= p_i$ for each $i$;
moreover, the ancestor of the leading $z$ here must lie within some $t_i$, say at the head of
some subterm $z\vec{P}\,'$, where $\vec{P}$ is an instance of $\vec{P}\,'$ via some substitution $^\dag$.

At this point, the arguments for the base case and step case part company.
In the base case $k=1$, we have that $z^\rho \vdash t_i : \sigma$ 
where $\rho = \nat^{l+1} \arrow \nat$ and $\sigma = \nat^l \arrow \nat$;
thus there are no bound variables within $t_i$ except the top-level ones---say $w_0,\ldots,w_{l-1}$,
all of type $\nat$. So in fact $^\dag$ has the form $[\vec{w} \mapsto \vec{W}]$ for certain meta-terms
$f_0^\nat,\ldots,f_l^\nat,z \vdash W_j : \nat$. 
Now consider the terms $f_0,\ldots,f_l \vdash \vec{W}^*$ and $\vec{w} \vdash \vec{P}\,' {}^*$,
writing $^*$ for the substitution $[z \mapsto \lambda w.0]$.
These compose to yield $\vec{f} \, \vdash \, \dang{\vec{P}\,^*} \,=\, \dang{\vec{p}\,^*} \,\succeq f$,
so we have expressed $\nat^{l+1}$ as a pseudo-retract of $\nat^l$ within $\SF$.
As already noted in the course of the proof of Theorem~\ref{no-retraction-thm}, this is impossible.

For the step case $k>1$, we proceed much as in the proof of Theorem~\ref{no-retraction-thm}, 
using the substitution $^\dag$ to express $\rho_{k-1,l+1}$ as a retract of a finite product of types
of sublevel $(k-1,l)$ or lower, contrary to the induction hypothesis. We leave the remaining details
as an exercise.
\end{proof}

We now outline how the ideas of Sections~\ref{sec-denotations}, \ref{sec-Y-k+1} and \ref{sec-extensional}
may be adapted to show that $Y_{\rho_{k,l+1}} \in \SP^0$ is not $\PCF^\Omega_{k,l}$-denotable.
We assume that $k \geq 2$ for the time being (the case $k=1$ will require special treatment).
The adaptations are mostly quite systematic: the type $\rho_{k,l+1}$ 
now plays the role of $\pure{k+1}$; types of sublevel $\leq (k,l)$ play the role of types of level $\leq k$;
$\rho_{k-1,l+1}$ plays the role of $\pure{k}$; and types of sublevel $\leq (k-1,l)$ play the role of types
of level $<k$. Since the proof we are adapting is quite lengthy, we leave many routine details to be checked
by the reader.

First, the evident adaptation of Definition~\ref{plugging-def} yields a notion of
\emph{$k,l$-plugging} where the plugging variables are required to be of sublevel $\leq (k,l)$,
and we thus obtain an inductive characterization of the $\PCF^\Omega_{k,l}$-denotable procedures
analogous to Theorem~\ref{denotable-inductive-thm}.
We also adapt the notion of \emph{regular} meta-term as follows:

\begin{defi}
Suppose $g : \rho_{k,l+1} \arrow \rho_{k,l+1}$ for $k,l \geq 1$.
An environment $\Gamma$ is \emph{$g$-($k,l$-)regular} if $\Gamma$ contains $g$ but no other variables
of sublevel $> (k,l)$. A meta-term $T$ is $g$-regular if it contains no variables of sublevel $> (k,l)$
except possibly for free occurrences of $g$. We say $\Gamma \vdash T$ is $g$-regular
if both $\Gamma$ and $T$ are $g$-regular.
\end{defi}

Next, we proceed to the ideas of Section~\ref{sec-Y-k+1}.
Our convention here will be that $\Gamma$ ranges over regular environments, and Roman capitals $V,Z$
range over sets of variables of sublevel $\leq (k,l)$. 
The notions of $x,V$-closed substitution and spinal term carry over as follows:

\begin{defi}  \label{kl-spinal-def}
Suppose $g : \rho_{k,l+1} \arrow \rho_{k,l+1}$ where $k \geq 2$ and $l \geq 1$.
\begin{enumerate}[label=(\roman*)]
\item If $\vec{x}$ is a list of variables of type $\rho_{k-1,l+1}$ and $V$ a set of variables of sublevel
$\leq (k,l)$, a substitution $^\circ = [\vec{w} \mapsto \vec{r}\,]$ is called \emph{$\vec{x},V$-closed}
if the $r_i$ contain no free variables, except that if $w_i \in V$ and $w_i$ is of sublevel $\leq (k-1,l)$
then $r_i$ may contain the $\vec{x}$ free.

\item Suppose $\Gamma \vdash e$ is $g$-$k,l$-regular and $\vec{x},V \subseteq \Gamma$.
We coinductively declare $e$ to be \emph{$g$-head-spinal} 
w.r.t.\ $\vec{x},V$ iff $e$ has the form 
$\caseof{g(\lambda \vec{x}\,'.E[e'])\vec{o}}{\cdots}$, where $E[-]$ is an expression context, and 
\begin{enumerate}[label=(\arabic*)]
\item for some $\vec{x},V$-closed specialization $^\circ$ covering the free variables of $\vec{o}$ 
other than those of $\vec{x}$, we have $\vec{o}\,^\circ \succeq \vec{x}^{\,\eta}$,
\item $e'$ is $g$-head-spinal w.r.t.\ $\vec{x}\,',V'$, where $V'$ is the local variable environment for $E[-]$.
\end{enumerate}

\item We say a regular term $\Gamma \vdash t$ is \emph{$g$-spinal} 
if it contains a $g$-head-spinal subexpression w.r.t.\ some $\vec{x},V$.
\end{enumerate}
\end{defi}

\noindent
Lemma~\ref{g-lemma-1} and its proof go through with the above adaptations;
here the local environments $\vec{v}, \vec{v}\,'$ are now of sublevel $\leq (k,l)$, and part~(iii)
of the lemma now states that if $K[-]$ contains no redexes with operator of sublevel $> (k,l)$, then
the substitution $^\dag$ is trivial for variables of sublevel $\geq (k-1,l+1)$.
The crucial Lemma~\ref{g-lemma-2}, which forms the heart of our proof, now translates as follows:

\begin{lem}  \label{kl-lemma-2}
Suppose $g: \rho_{k,l+1} \arrow \rho_{k,l+1}$ and we have $g$-regular terms
\[ \Gamma,\vec{v} \;\vdash~ d ~=~ \caseof{gpq}{\cdots} \;, ~~~~~~
   \Gamma,\vec{v}\,' \vdash \vec{s} \,, \] 
where $\vec{v},\vec{v}\,'$ are of sublevel $\leq (k,l)$, and
$\Gamma,\vec{v}\,' \vdash\, \dang{d[\vec{v}\mapsto\vec{s}\,]}$ 
is $g$-head-spinal with respect to some $\vec{x},V$.
Then $d$ itself is $g$-spinal.
\end{lem}

The entire proof of this lemma translates systematically according to the correspondences we have indicated,
invoking Theorem~\ref{kl-no-retraction-thm} for the fact that
$\rho_{k-1,l+1}$ is not a pseudo-retract of a product of sublevel $\leq (k-1,l)$ types.
The analogue of Theorem~\ref{no-gremlin-thm} now goes through readily, so we obtain:

\begin{thm}
If $k \geq 2$ and $l \geq 1$, every $\PCF^\Omega_{k,l}$-denotable procedure is non-$g$-spinal
where $g: \rho_{k,l+1} \arrow \rho_{k,l+1}$.
\end{thm}

As in our original proof, we will actually use a version of this theorem for the modified notion of 
spinal term that incorporates the extra argument $b$, and for the extension of $\PCF^\Omega_{k,l}$
with the operator $\byval$.

To adapt the material of Section~\ref{sec-extensional}, we now take $g$ to be a variable of type
$0 \arrow \rho_{k,l+1} \arrow \rho_{k,l+1}$, and argue that the $\PCF_{k,l+2}$-denotable element
$\Phi = \lambda g.\, Y_{0 \arrow \rho_{k,l+1}} (\lambda fn.\,g\,n\,(f(\suc\,n)))$ within $\SF$
is not $\PCF^\Omega_{k,l}$-denotable. The proof is a completely routine adaptation of that in
Section~\ref{sec-extensional}. Since $Y_{0 \arrow \rho_{k,l+1}}$ is readily definable from $Y_{k+1}$
by Proposition~\ref{sublevel-facts}, this implies that $Y_{k+1} \in \SF$ is not $\PCF^\Omega_{k,l}$-denotable.
We have thus shown:

\begin{thm}  \hfill
\begin{enumerate}[label=(\roman*)]
\item For $k \geq 2$ and $l \geq 1$, the element $Y_{0 \arrow \rho_{k,l+1}} \in \SF$ is denotable
in $\PCF_{k,l+2}$ but not in $\PCF^\Omega_{k,l}$.

\item For $k \geq 2$, the element $Y_{k+1} \in \SF$ is not denotable in $\PCF^\Omega_k$.
\end{enumerate}
\end{thm}

\noindent
A slightly different approach is needed for the case $k=1$. This is because at level $0$ our only
type is $\nat$, so we are unable to make a distinction between sublevels $l$ and $l+1$.
To establish Lemma~\ref{kl-lemma-2} in this case, we again wish to show that we cannot pass in
the content of the relevant $\vec{x}\,'$ to the relevant $s_i$, but now the idea is to appeal to the
fact that $\vec{x}\,'$ consists of $l+1$ variables of type $\nat$, whereas $s_i$ accepts at most $l$ 
arguments of type $\nat$. (We have already seen that $\nat^{l+1}$ cannot be a retract of $\nat^l$.)
However, we also need to exclude the possibility that the substitution $^\circ$
is being used to import some components of $\vec{x}\,'$.
We can achieve this if we require $^\circ$ to be actually closed rather than just $\vec{x}\,',V'$-closed, 
and it turns out that this is permissible if we also tighten our notion of spinal term slightly, 
essentially to ensure that no intermediate applications
of $g$ appear in between those declared to constitute the spine of the term:

\begin{defi}  \label{strongly-spinal-def}
Suppose $g : \rho_{1,l+1} \arrow \rho_{1,l+1}$ where $l \geq 1$.

Suppose $\Gamma \vdash e$ is $g$-$1,l$-regular and $\vec{x} \subseteq \Gamma$.
We coinductively declare $e$ to be \emph{strongly $g$-head-spinal} w.r.t.\ $\vec{x}$
iff $e$ has the form $\caseof{g(\lambda \vec{x}\,'.E[e'])\vec{o}}{\cdots}$, 
where $E[-]$ is an expression context, and 
\begin{enumerate}[label=(\arabic*)]
\item the hole within $E[-]$ does not itself occur within an application $gpq$,
\item for some closed substitution $^\circ$ covering the free variables of $\vec{o}$ other than
those of $\vec{x}$, we have $\vec{o}\,^\circ \succeq \vec{x}^{\,\eta}$,
\item $e'$ is strongly $g$-head-spinal w.r.t.\ $\vec{x}\,'$.
\end{enumerate}
The notion of strongly $g$-spinal term follows suit.
\end{defi}

The counterpart of Lemma~\ref{g-lemma-1} goes through as expected, although without part~(iii):
the relevant sublevel distinction does not exist at type level $0$, and we cannot conclude that the
substitution in question is trivial for all variables of type $\nat$.
We may now indicate the required changes to Lemma~\ref{kl-lemma-2} and its proof:

\begin{lem}
Suppose $g: \rho_{1,l+1} \arrow \rho_{1,l+1}$ and we have $1,l$-regular terms
\[ \Gamma,\vec{v} \;\vdash~ d ~=~ \caseof{gpq}{\cdots} \;, ~~~~~~
   \Gamma,\vec{v}\,' \vdash \vec{s} \,, \] 
where $\vec{v},\vec{v}\,'$ are of sublevel $\leq (1,l)$, and
$\Gamma,\vec{v}\,' \vdash\, \dang{d[\vec{v}\mapsto\vec{s}\,]}$ 
is strongly $g$-head-spinal with respect to some $\vec{x}$.
Then $d$ itself is strongly $g$-spinal.
\end{lem}

\begin{proof}
The proof of Lemma~\ref{g-lemma-2} up to the end of the proof of Claim~1 adapts straightforwardly, 
and is somewhat simplified by the fact that the substitution $^\circ$ is closed.
As sketched above, the crucial contradiction is provided by the fact that $\nat^{l+1}$ is not a pseudo-retract
of $\nat^l$ in $\SF$.

For the remainder of the proof, the key point to note is that $\vec{v}\,''$ (the local environment for $C[-]$)
is actually empty in this case. This is because $C[-]$ is in normal form and contains no free variables
of level $\geq 2$ except $g$, so any $\lambda$-term containing the hole would need to appear as an
argument to $g$. It would then follow that the hole within $E[-]$ lay within an argument to some occurrence of $g$, as precluded by the definition of strongly spinal term. 

It follows trivially that the $\vec{x}\,',\vec{v}\,''$-closed substitution $^{\circ'}$ constructed at the very end
of the proof is actually closed. Moreover, the spinal structure of $d'$ identified by the proof cannot contain
any intermediate applications of $g$, since these would give rise under evaluation to intermediate applications
of $g$ in the spine of $\dang{d[\vec{v}\mapsto\vec{s}\,]}$ as precluded by 
Definition~\ref{strongly-spinal-def}.
Thus, the identified spinal structure in $d'$ is actually a strongly spinal structure, 
and the argument is complete.
\end{proof}

A trivial adaptation of the proof of Theorem~\ref{no-gremlin-thm} now yields:

\begin{thm}
No $\PCF^\Omega_{1,l}$-denotable procedure can be strongly $g$-spinal
where $g: \rho_{1,l+1} \arrow \rho_{1,l+1}$. 
\end{thm}

As before, this adapts easily to a variable $g$ of type $0 \arrow \rho_{1,l+1} \arrow \rho_{1,l+1}$.
From here on, we again follow the original proof closely. 
The only additional point to note is that in place of Corollary~\ref{spinal-cor}
we now require that any simple $g \vdash q$ with $q \succeq' p_n$ must be \emph{strongly} spinal,
but this is already clear from the proof of Lemma~\ref{q-structure-lemma}.
We therefore have everything we need for:

\begin{thm}  \hfill
\begin{enumerate}[label=(\roman*)]
\item For any $l \geq 1$, $Y_{\nat^{l+2} \arrow \nat} \in \SF$ is
not denotable in $\PCF^\Omega_{1,l}$.

\item $Y_2 \in \SF$ is not denotable in $\PCF^\Omega_1$. 
\end{enumerate}
\end{thm}

\noindent
The proof of Theorems~\ref{operational-main-thm} and \ref{SF-main-thm} is now complete.

\section{Related and future work}  \label{sec-further-work}

\subsection{Other hierarchies of Y-combinators}

There have been a number of previous results from various research traditions showing that in some sense
the power of level $k$ recursions increases strictly with $k$.
Whilst many of these results look tantalizingly similar to ours,
it turns out on inspection that their mathematical substance is quite different, and 
we do not expect any substantial technical connections with our own work to be forthcoming.
Nonetheless, it is interesting to see how strikingly different ideas and methods arising in other contexts 
can lead to superficially similar results.

Previous results to the effect that $Y$-combinators for level $k+1$ are not definable
from those for level $k$ have been obtained by Damm \cite{Damm-IO} and 
Statman \cite{Statman-lambda-Y}. It is convenient to discuss the latter of these first.
Statman works in the setting of the simply typed \emph{$\lambda Y$-calculus}, essentially the
pure $\lambda$-calculus extended with constants $Y_\sigma : (\sigma\arrow\sigma)\arrow\sigma$
and reduction rules $Y_\sigma M \reducesto M(Y_\sigma M)$. 
He gives a succinct proof that $Y_{k+1}$ is not definable from $Y_k$ 
up to computational equality, based on the following idea. 
If $Y_{k+1}$ were definable from $Y_k$, it would follow that 
the recursion equation $Y_{k+1} g = g(Y_{k+1} g)$ could be derived with only finitely many
uses of the equation $Y_k M = M(Y_k M)$ (say $m$ of them).
It would then follow, roughly speaking, that $mn$ recursion unfoldings for $Y_k$ 
would suffice to fuel $n$ recursion unfoldings for $Y_{k+1}$. 
On the other hand, it can be shown that the size of normal-form terms definable 
using $n$ unfoldings of $Y_{k+1}$ (as a function of the size of the starting term) grows faster than
can be accounted for with $mn$ unfoldings of $Y_k$.

The language $\lambda Y$ is seemingly less powerful than $\PCF$,%
\footnote{One might consider translating $\PCF$ into $\lambda Y$ by representing natural numbers as Church numerals; 
however, it appears that predecessor is not $\lambda Y$-definable for this representation.}
although this is perhaps not the most essential difference between Statman's work and ours.
More fundamentally, Statman's method establishes the non-definability only up to computational
equality (that is, the equivalence relation generated by the reduction rules), 
whereas we have been concerned with non-definability modulo observational 
(or extensional) equivalence.
Even for non-denotability in $\SP^0$, an approach along Statman's lines 
would be unlikely to yield much information,
since there is no reason why the number of unfoldings of $Y_k$ required to generate 
the NSP for $Y_{k+1}$ to depth $n$ should not grow dramatically as a function of $n$.

A result very similar to Statman's was obtained earlier in Damm \cite{Damm-IO}, 
but by a much more indirect route as part of a far-reaching investigation of the 
theory of tree languages.
In Damm's setting, programs are \emph{recursion schemes}---%
essentially, families of simultaneous (and possibly mutually recursive) defining
equations in typed $\lambda$-calculus---but in essence these can be considered as terms of $\lambda Y$
relative to some signature consisting of typed constants. 
(Actually, Damm's $\lambda$-terms involve a restriction to \emph{derived types}, 
which has the effect of limiting attention to what are elsewhere called \emph{safe} recursion schemes.)
Any such program can be expanded to an
infinite tree (essentially a kind of B\"ohm tree), and Damm's result (Theorem~9.8 of \cite{Damm-IO}) 
is that if programs are considered up to \emph{tree equality}, 
then safe level $k+1$ recursions give more expressive power than safe level $k$ ones.
Damm's result is thus distinguished from Statman's in two ways: by the restriction to safe recursion schemes,
and by the use of tree equality in place of the stricter computational equality.
This latter point brings Damm's work somewhat closer in spirit to our work:
indeed, in the case of pure $\lambda Y$, tree equality agrees with equality of innocent strategies
if the base type is interpreted as a certain trivial game---or equivalently with equality in a variant of 
our $\SP^0$ with no ground type values.
However, in the case of a signature for $\PCF$, tree equality will still be considerably more fine-grained than
equality in $\SP^0$, let alone in $\SF$, since in effect the $\PCF$ constants are left uninterpreted.
It therefore seems unlikely that an approach to our theorems via Damm's methods is viable.

The strictness of the recursion scheme hierarchy was further investigated by Ong
\cite{Ong-model-checking}, who used innocent game semantics to show that the complexity of
certain model checking problems for trees generated by level $k$ recursions increases strictly with $k$.
It was also shown by Hague, Murawski, Ong and Serre \cite{Collapsible-PDA} 
that the trees generated by level $k$ recursion schemes (with no safety restriction) 
were precisely those that could be generated by \emph{collapsible pushdown automata} of order $k$.
Later, Kartzow and Parys \cite{Kartzow-Parys} used pumping lemma techniques to show that
the collapsible pushdown hierarchy was strict, hence so was the recursion scheme hierarchy
with no safety restriction.
Again, despite the intriguing parallels to our work, these results appear to be manifesting something 
quite different:
in our setting, the power of level $k$ recursion has nothing to do with the `difficulty of computing' the relevant
NSPs in the sense of automata theory, since the full power of Turing machines is required even at level~1.

Also of interest is the work of Jones \cite{Jones-cons-free} from the functional programming community.
Jones's motivation is close to ours in that he seeks to give mathematical substance to the programming
intuition that some (combinations of) language features yield 
`more expressive power' than others.
As Jones notes, an obvious obstacle to obtaining such results is that all programming languages
of practical interest are Turing complete, so that no such expressivity distinctions are visible at the level of the
(first-order) computable functions.
Whereas we have responded to this by considering the situation at higher types, where genuine 
expressivity differences do manifest themselves, Jones investigates the power of (for example)
recursion at different type levels in the context of a restricted language of `read-only' or `cons-free' programs.
Amongst other results, he shows that in such a language, if data values and general recursion of type level 
$k \geq 1$ are admitted, then the computable functions from lists of booleans to booleans are exactly the
{\sc{exp}$^k$\sc{time}} computable ones.
These results inhabit a mathematical territory very different from ours, although yet again,
the basic moral that the power of recursion increases strictly with its type level shines through.

\subsection{Relationship to game semantics}

Next, we comment briefly how our work relates
to the known \emph{game models} of PCF \cite{AJM,Hyland-Ong}.
It is known that these models are in fact isomorphic to our $\SP^0$,
although the equivalence between the game-theoretic definition of application and our own
is mathematically non-trivial (see \cite[Section~7.4]{Longley-Normann}).
This raises the obvious question of whether our proofs could be conducted equally well, or better, in a game-semantical setting. 

Whilst a direct translation is presumably possible, our present impression 
is that the sequential procedure presentation, and our calculus of meta-terms in particular, 
allows one to see the wood for the trees more clearly, 
and also to draw more easily on familiar intuitions from $\lambda$-calculus.
However, a closer look at game-semantical approaches would be needed in order to judge
whether either approach really offers some genuine mathematical or conceptual advantage over the other.

\subsection{Sublanguages of PCF: further questions}  \label{sublang-questions}

We now turn our attention to some potential extensions and generalizations of our work.

So far, we have worked mainly with a coarse stratification of types in terms of their levels, 
although we have illustrated in Section~\ref{sec-pure-type} how finer stratifications are also possible.
Naturally, there is scope for a still more fine-grained analysis of types and the relative strength of their
$Y$-combinators; this is of course closely related to the task of mapping out the embeddability relation 
between types (as in Subsection~\ref{subsec-embed}) in finer detail.

Even at level~1, there is some interest here.
Our analysis in Section~\ref{sec-pure-type} has shown that, for $l \geq 1$,
\begin{itemize}
\item the element $Y_{\nat^{l+1} \arrow \nat} \in \SP^0$ is not $\PCF^\Omega_0$-definable
from $Y_{\nat^l \arrow \nat}$,
\item  the element $Y_{\nat^{l+2} \arrow \nat} \in \SF$ is not $\PCF^\Omega_0$-definable
from $Y_{\nat^l \arrow \nat}$.
\end{itemize}

However, this leaves us with a small gap for $\SF$: e.g.\ we have not shown either that
$Y_{\nat\arrow\nat}$ is strictly weaker than $Y_{\nat^2\arrow\nat}$ or that
$Y_{\nat^2\arrow\nat}$ is strictly weaker than $Y_{\nat^3\arrow\nat}$,
although according to classical logic, at least one of these must be the case.
(This is reminiscent of some well-known situations from complexity theory.)
We expect that a more delicate analysis will allow us to fill this gap.
One can also envisage an even more fine-grained hierarchy obtained by admitting other base types
such as the booleans or unit type.

A closely related task is to obtain analogous results for the \emph{call-by-value} 
interpretation of the simple types (as embodied in Standard~ML, for example).
As is shown in \cite[Section~4.3]{Longley-Normann}, a call-by-value (partial) type structure
$\SF_v$ can be constructed by fairly general means from $\SF$:
here, for example, $\SF_v(\pure{1})$ consists of all partial functions $\N \parrow \N$
rather than (monotone) total functions $\N_\bot \arrow \N_\bot$, 
and $\SF_v(\pure{2})$ consists of partial functions $\N_\bot^\N \parrow \N$.
From known results on the interencodability of call-by-name and call-by-value models
(see \cite[Section~7.2]{Longley-Normann}), it is easy to read off the analogue of 
Corollary~\ref{no-finite-basis-cor} for $\SF_v^\eff$; 
however, more specific results on the relative strengths of various $Y$-combinators
within $\SF_v$ would require some further reworking of our arguments.%

Of course, one can also pose relative definability questions for other elements besides 
$Y$-combinators.
For instance, it is natural to consider the \emph{higher-order primitive recursors} $\rec_\sigma$
of System~T, as well as the closely related \emph{iterators} $\iter_{\sigma\tau}$:
\begin{eqnarray*}
\rec_\sigma & : & 
  \sigma \arrow (\sigma \arrow \nat \arrow \sigma) \arrow \nat \arrow \sigma \;, \\
\iter_{\sigma\tau} & : & (\sigma\arrow (\sigma + \tau)) \arrow \sigma \arrow \tau \;.
\end{eqnarray*}
The idea behind the latter is to embody the behaviour of a \emph{while} construct for
imperative-style loops with state $\sigma$ and exit type $\tau$.%
\footnote{The sum type $\sigma + \tau$ is not officially part of our system, but can 
(for any given $\sigma,\tau$) be represented as a retract of some existing simple type.}

It is shown in \cite[Section~6.3]{Longley-Normann} that each $\rec_\sigma$ may be
interpreted by a \emph{left-well-founded} procedure (cf.\ Subsection~\ref{subsec-power-pcf1}),
and it is not hard to check that the same is true for each $\iter_{\sigma\tau}$.
Furthermore, it is clear that all left-well-founded procedures are non-spinal,
so it requires only the addition of an easy base case in the proof of Theorem~\ref{no-gremlin-thm}
to show that $Y_{k+1}$ cannot be defined 
even in $\PCF_k^\Omega$ extended with all primitive recursors and iterators.
(Actually, we can dispense with the primitive recursors here, 
as it is straightforward to define them from suitable iterators.)

The dual question, roughly speaking, is whether any or all of the recursors $\rec_\sigma$ 
or iterators $\iter_{\sigma\nat}$ for types $\sigma$ of level $k+1$ are definable in $\PCF_k$.
We conjecture that they are not, and that this could be shown by suitably 
choosing a substructure of $\SP^0$ so as to exclude such $\iter_{\sigma\nat}$.
(This would incidentally answer Question~2 of \cite[Section~5]{Berger-min-recursion}.)
One could also look for substructures that more precisely determine the strength of 
various \emph{bar recursion} operators or the \emph{fan functional}.
All in all, our experience leads us to expect that many further substructures of $\SP^0$ 
should be forthcoming, leading to a harvest of non-definability results exhibiting a rich
`degree structure' among $\PCF$-computable functionals.

Another very natural kind of question is the following: given a particular sublanguage $\LLL$
of $\PCF$, what is the \emph{simplest} possible type for an element of $\SF^\eff$ that is 
not denotable in $\LLL$? Or to look at it another way: given a type $\sigma$, what is the
smallest natural sublanguage of $\PCF$ that suffices for defining all elements of 
$\SF^\eff(\sigma)$?
Here the analysis of \cite[Section~7.1]{Longley-Normann} yields several
positive definability results, whilst the analysis of the present paper provides ammunition on the
negative side. The current state of our knowledge is broadly as follows.
As in \cite{Longley-Normann}, we write $\Klexmin$ of the language of \emph{Kleene $\mu$-recursion}:
this is equivalent (in its power to denote elements of $\SF$) 
to $\PCF_0$ extended with a strict primitive recursor for ground type and a strict iterator for ground type, 
but with no form of general recursion.

\begin{itemize}
\item For first-order types $\sigma$, even $\Klexmin$ suffices for defining
all elements of $\SF^\eff(\sigma)$; likewise, the oracle version $\Klexmin^\Omega$
suffices for $\SF(\sigma)$.
\item For second-order types $\sigma$ of the special form $(\nat\arrow\nat)^r \arrow \nat$,
$\Klexmin^\Omega$ still suffices for $\SF(\sigma)$; however, this result is non-constructive,
and $\Klexmin$ does not suffice for $\SF^\eff(\sigma)$.
(We draw here on some recent work of Dag Normann \cite{Normann-sequential-dcpo}.)
\item For general second-order types, $\Klexmin^\Omega$ no longer suffices, but 
the languages $\PCF_1$, $\PCF_1^\Omega$ suffice for $\SF^\eff$, $\SF$ respectively---%
indeed, even the single recursion operator $Y_{\nat\arrow\nat}$ is enough here.
\item For third-order types, we do not know whether $\PCF_1$ suffices (for $\SF^\eff$).
We do know that $\PCF_2$ suffices, and that $Y_{\nat\arrow\nat}$ alone is not enough.
\item For types of level $k \geq 4$, $\PCF_{k-3}$ does not suffice,
but $\PCF_{k-2}$ does.
\end{itemize}
Again, there is scope for a more fine-grained view of the hierarchy of types.

\subsection{Other languages and models}

We have so far concentrated almost entirely on $\PCF$-style sequential computation.
To conclude, we briefly consider which other notions of higher-order computation
are likely to present us with an analogous situation.

As already noted at the end of Subsection~\ref{subsec-embed}, 
several extensions of $\PCF$ studied in the literature present a strikingly different picture:
in these settings, universal types exist quite low down, and as a consequence, 
only $Y$-combinators of low type (along with the
other constants of the language) are required for full definability. 
There is, however, one important language which appears to be more analogous 
to pure $\PCF$ in these respects, namely an extension with \emph{local state} 
(essentially Reynolds's \emph{Idealized Algol}).
This language was studied in \cite{Game-semantics}, where a fully abstract game model 
was provided (consisting of well-bracketed but possibly non-innocent strategies).
Unpublished work by Jim Laird has shown that there is no universal type 
in this setting.
We would conjecture also that the recursion hierarchy for this language is strict,
where we consider expressibility modulo observational equivalence in Idealized Algol.
This would constitute an interesting variation on our present results.

Related questions also arise in connection with \emph{total} functionals.
Consider, for example, the type structure $\Ct$ of total continuous functionals,
standardly constructed as the extensional collapse (relative to $\N$) of 
the Scott model $\PC$ of partial continuous functionals.
It is shown by Normann \cite{Normann2000} that every effective element of $\Ct$
is represented by a \emph{$\PCF$-denotable} element of $\PC$, and the proof actually
shows that $\PCF_1$ suffices here. (The further generalization of these ideas by
Longley \cite{Ubiquity} makes some use of second-order recursions as in $\PCF_2$;
we do not know whether these can be eliminated.) Thus, in this setting,
only recursions of low type are needed to obtain all computable functionals.
Similar remarks apply to the total type structure $\HEO$, obtained as the extensional collapse
of $\PC^\eff$.

On the other hand, one may consider the \emph{Kleene computable} functionals over $\Ct$,
or over the full set-theoretic type structure $\Set$, as defined by the schemes S1--S9.
As explained in \cite[Chapter~6]{Longley-Normann}, sequential procedures can be
seen as abstracting out the algorithmic content common to both $\PCF$-style and
Kleene-style computation (note that Kleene's S9 in some sense does duty for the 
$Y$-combinators of $\PCF$). 
This naturally suggests that our strict hierarchy for $\PCF$ may have an analogue for
the Kleene computable functionals (say over $\Set$ or $\Ct$), where at level $k$ we consider
the evident restriction of S9 to types of level $\leq k$.
We conjecture that this is indeed the case, although the required counterexamples may be
more difficult to find given that we are limited to working with total functionals.

\section*{Acknowledgements}
 
I am grateful to Ulrich Berger, Mart\'{i}n Escard\'{o}, Dag Normann and Alex Simpson
for valuable discussions and correspondence, and to Luke Ong and Colin Stirling for helping me to
navigate the existing literature on $\lambda Y$ and recursion schemes
(as discussed in Section~\ref{sec-further-work}).
Many of the participants in the Domains XII workshop in Cork and the Galop XI workshop in Eindhoven
also offered helpful comments; thanks in particular to Neil Jones for drawing his work to my attention.
Finally, I thank the anonymous referees for their careful work on the paper and their valuable suggestions;
these have resulted in significant improvements in both the overall architecture and the formal details.

\appendix
\section{Super-identity procedures}

In the course of our main proof, we have frequently encountered assertions of the form
$p \succeq x^\eta$ for various procedures $x^k \vdash p$.
Although not necessary for our main argument, it is natural to ask whether there are
any such procedures other than those for which $p \approx x^\eta$.
The following theorem shows that the answer is no: in other words,
no procedure $\lambda x.p : \pure{k} \arrow \pure{k}$  
can extensionally `improve on' the identity function.
We here present this as a result of some interest in its own right,
whose proof is perhaps less trivial than one might have expected.

Recall that $\preceq$ denotes the extensional order on $\SF$,
as well as the associated preorder on $\SP^0$.
Within $\SF$, we will write $f \prec f'$ to mean $f \preceq f'$ but $f \neq f'$;
we also write $f \sharp f'$ to mean that $f,f'$ have no upper bound
with respect to $\preceq$.

We shall call an element of $\SP^0$ \emph{finite} if it is a finite tree once branches
of the form $i \darrow \bot$ have been deleted.
We say an element of $\SF$ is \emph{finite} if it is represented by some finite
element of $\SP^0$.
We write $\SP^{0,\fin}(\sigma), \SF^\fin(\sigma)$ for the set of finite elements
in $\SP^0(\sigma), \SF(\sigma)$ respectively.

\begin{thm}  \label{super-identity-thm}  \hfill
\begin{enumerate}[label=(\roman*)]
\item If $f \in \SF^\fin(k)$ and $f \prec f'$, then there exists $f'' \sharp f'$
with $f \prec f''$.

\item If $\Phi \in \SF(k \arrow k)$ and $\Phi \succeq \id$,
there can be no $f \in \SF(k)$ with $\Phi(f) \succ f$; hence $\Phi = \id$.
\end{enumerate}
\end{thm}

\begin{proof}  \hfill
\begin{enumerate}[label=(\roman*)]
\item The cases $k=0,1$ are easy, so let us assume $k \geq 2$.
Suppose $f \prec f'$ where $f$ is finite.
Then for some $g \in \SF(k-1)$ we have $f(g)=\bot$ but $f'(g)=n \in \N$, say,
and by continuity in $\SP^0$ we may take $g$ here to be finite.
Take $p,q \in \SP^{0,\fin}$ representing $f,g$ respectively;
we may assume that $p,q$ are `pruned' so that every subtree containing no
leaves $m \in \N$ must itself be $\bot$.

\textit{Case 1:} $g(\bot^{k-2}) = a \in \N$.
In this case, we may suppose that $q = \lambda x.a$.
Consider the computation of $p \cdot q$.
Since all calls to $q$ trivially evaluate to $a$, this computation follows the
rightward path through $p$ consisting of branches $a \darrow \cdots$.
But since $p$ is finite and $p \cdot q = \bot$ (because $f(g)=\bot$), 
this path must end in a leaf occurrence of $\bot$ within $p$.
Now extend $p$ to a procedure $p'$ by replacing this leaf occurrence
by some $n' \neq n$; then clearly $p' \cdot q = n'$.
Taking $f''$ to be the function represented by $p'$, we then have
$f \preceq f''$ and $f''(g) = n' \,\sharp\, n = f'(g)$, so $f'' \sharp f'$
(whence also $f'' \neq f$ so $f \prec f''$).

\textit{Case 2:} $g(\bot^{k-2}) = \bot$.
Take $N$ larger than $n$ and all numbers appearing in $p,q$.
Define $p' \sqsupseteq p$ as follows: if $p = \lambda x.\bot$, take $p' = \lambda x.N$,
otherwise obtain $p'$ from $p$ by replacing each case branch $j \darrow \bot$
anywhere within $p$ by $j \darrow N$ whenever $j \leq N$.
Extend $q$ to $q'$ in the same way. 
Note in particular that every $\ccase$ subterm within $p',q'$ will now be equipped
with a branch $N \darrow N$.

Now consider the computation of $p \cdot q$.
Since $p,q$ are finite and $f(g)=\bot$, this evaluates to an occurrence of $\bot$
which originates from $p$ or $q$. Since no numbers $>N$ ever arise in the computation,
this occurrence of $\bot$ cannot be part of a branch $j \darrow \bot$ for $j>N$,
so will have been replaced by $N$ in $p'$ or $q'$.
Now suppose that we head-reduce $pq$ until $\bot$ first appears in head position,
and let $U$ be the resulting meta-term.
Then it is easy to see that $p'q'$ correspondingly reduces to a meta-term $U'$
with $N$ in head position. (Formally, we reason here by induction on the length
of head-reduction sequences not involving the rule for $\caseof{\bot}{\cdots}$.)

We now claim that $p' \cdot q' = N$. Informally, this is because the head
occurrence of $N$ in $U'$ will be propagated to the top level by the case branches 
$N \darrow N$ within both $p'$ and $q'$.
More formally, let us define the set of meta-expressions \emph{led by $N$}
inductively as follows:
\begin{enumerate}[label=(\arabic*)] 
\item $N$ is led by $N$.
\item If $E$ is led by $N$, then so is $\caseof{E}{i \darrow D_i}$.
\end{enumerate}
We say that an NSP meta-term $T$ is \emph{saturated at $N$}
if every $\ccase$ subterm within $T$ has a branch $N \darrow E$ 
where $E$ is led by $N$.
Clearly $p'q'$ is saturated at $N$, and it is
easy to check that the terms saturated at $N$ are closed under head reductions;
thus $U'$ is saturated at $N$.
But we have also seen that $U'$ has $N$ in head position, so is led by $N$.
Finally, an easy induction on term size shows that
every finite meta-term that is led by $N$ and saturated at $N$ evaluates to $N$ itself.
This shows that $p' \cdot q' = N$.

To conclude, let $f'',g'$ be the functions represented by $p',q'$ respectively,
so that $f \preceq f''$ and $g \preceq g'$.
Then $f'(g')=n$, but $p'' \cdot q = N$ so $f''(g')=N \neq n$, whence $f'' \sharp f'$
(and also $f'' \neq f$ so $f \prec f''$).
 
\item Suppose $\Phi \succeq \id$ and $\Phi(f) \succ f$ for some $f$.
Again by continuity, we may take $f$ to be finite. 
Then by (i), we may take $f'' \succ f$ such that $f'' \,\sharp\, \Phi(f)$.
But this is impossible because $\Phi(f'') \succeq f''$ and $\Phi(f'') \succeq \Phi(f)$.
Thus $\Phi=\id$.
\qedhere
\end{enumerate}
\end{proof}

\noindent
It is easy to see that the above theorem holds with any finite type over $\nat$
in place of $\pure{k}$.
However, it will trivially fail if the unit type $\unit$ is admitted as an additional base type:
e.g.\ the function $(\lambda x.\top) \in \SF(\unit\arrow\unit)$ strictly dominates the identity.
An interesting question is whether the theorem holds for all finite types over 
the type $\bool$ of booleans: note that the above proof fails here
since it requires the base type to be infinite.
For comparison, we mention that in other models of computation,
improvements on the identity are sometimes possible for such types.
For example, if $\sigma=\bool\arrow\bool$, then a functional
of type $\sigma\arrow\sigma$ strictly dominating the identity exists in the Scott model.
Indeed, such a function $J$ can be defined in
$\PCF$ augmented with the parallel conditional $\pif$, e.g.\ as 
\[ J ~=~ \lambda f^\sigma.\lambda x^\bool.\,\mathit{vote}(f(x),f(\ttrue),f(\ffalse)) \;. \]
Here $\mathit{vote}$ is Sazonov's voting function, definable by
\[ \mathit{vote}(x,y,z) ~=~ \pif(x,\pif(y,\ttrue,z),\pif(y,z,\ffalse)) \;. \]
The point is that $J$ will `improve' the function $\lambda x.\,\iif(x,\ttrue,\ttrue)$ to $\lambda x.\ttrue$.
We do not know whether phenomena of this kind can arise within the model $\SF$.
 
\end{document}